\documentclass[twocolumn,trackchanges]{aastex62}
\usepackage{float, bm, graphicx, amsmath, morefloats}
\usepackage{CJK,CJKspace,CJKpunct}
\usepackage{color}
\usepackage{booktabs}
\newcommand{\code}[1]{\texttt{\detokenize{#1}}}

\newcommand{\name}{SN2018gep}

\newcommand{\redshift}{\mbox{$0.03154$}}
\newcommand{\distance}{\mbox{$143$}}

\newcommand{\teff}{\mbox{$T_{\rm eff}$}}
\newcommand{\nickel}{\mbox{$\rm {}^{56}Ni$}}

\newcommand{\mni}{\mbox{$M_{\rm Ni}$}}
\newcommand{\mej}{\mbox{$M_{\rm ej}$}}

\newcommand{\msol}{\mbox{$M_\odot$}}

\newcommand{\trise}{\mbox{$t_{\rm rise}$}}
\newcommand{\tpeak}{\mbox{$t_{\rm peak}$}}
\newcommand{\lpeak}{\mbox{$L_{\rm peak}$}}

\newcommand{\magnitudes}{\mbox{$\rm mag$}}

\newcommand{\lbol}{\mbox{$L_\mathrm{bol}$}}

\newcommand{\days}{\mbox{$\rm day$}}

\newcommand{\pyr}{\mbox{$\rm yr^{-1}$}}
\newcommand{\phr}{\mbox{$\rm hr^{-1}$}}

\newcommand{\psec}{\mbox{$\rm s^{-1}$}}
\newcommand{\gm}{\mbox{$\rm g$}}

\newcommand{\erg}{\mbox{$\rm erg$}}
\newcommand{\kev}{\mbox{$\rm keV$}}

\newcommand{\mpc}{\mbox{$\rm Mpc$}}

\newcommand{\cm}{\mbox{$\rm cm$}}
\newcommand{\km}{\mbox{$\rm km$}}

\newcommand{\ghz}{\mbox{$\rm GHz$}}

\newcommand{\phz}{\mbox{$\rm Hz^{-1}$}}

\newcommand{\chandra}{{\em Chandra}}
\newcommand{\fermi}{{\em Fermi}}
\newcommand{\galex}{{\em GALEX}}

\newcommand{\swift}{{\em Swift}}
\newcommand{\wise}{{\em WISE}}

\newcommand{\degsq}{\mbox{$\rm deg^{2}$}}
\newcommand{\pcmsq}{\mbox{$\rm cm^{-2}$}}
\newcommand{\pcmcub}{\mbox{$\rm cm^{-3}$}}

\newcommand{\kelvin}{\mbox{$\rm K$}}

\newcommand{\package}[1]{\textsc{#1}}

\begin{document}

\title{
Evidence for Late-stage Eruptive Mass-loss in the Progenitor to SN2018gep, a Broad-lined Ic Supernova: \\
Pre-explosion Emission and a Rapidly Rising Luminous Transient}

\author[0000-0002-9017-3567]{Anna Y. Q.~Ho}
\affiliation{Cahill Center for Astrophysics, 
California Institute of Technology, MC 249-17, 
1200 E California Boulevard, Pasadena, CA, 91125, USA}

\author[0000-0003-3461-8661]{Daniel A.~Goldstein}
\altaffiliation{Hubble Fellow}
\affiliation{Cahill Center for Astrophysics, 
California Institute of Technology, MC 249-17, 
1200 E California Boulevard, Pasadena, CA, 91125, USA}

\author[0000-0001-6797-1889]{Steve Schulze}
\affiliation{Department of Particle Physics and Astrophysics, Weizmann Institute of Science, 234 Herzl St, 76100 Rehovot, Israel}

\author{David K.~Khatami}
\affiliation{Department of Astronomy, University of California, Berkeley, CA, 94720}

\author{Daniel A.~Perley}
\affiliation{Astrophysics Research Institute, Liverpool John Moores University, IC2, Liverpool Science Park, 146 Brownlow Hill, Liverpool L3 5RF, UK}

\author{Mattias Ergon}
\affiliation{The Oskar Klein Centre \& Department of Astronomy, Stockholm University, AlbaNova, SE-106 91 Stockholm, Sweden}

\author{Avishay Gal-Yam}
\affiliation{Department of Particle Physics and Astrophysics, Weizmann Institute of Science, 234 Herzl St, 76100 Rehovot, Israel}

\author{Alessandra Corsi}
\affiliation{Department of Physics and Astronomy, Texas Tech University, Box 1051, Lubbock, TX 79409-1051, USA}

\author[0000-0002-8977-1498]{Igor Andreoni}
\affiliation{Cahill Center for Astrophysics, 
California Institute of Technology, MC 249-17, 
1200 E California Boulevard, Pasadena, CA, 91125, USA}

\author{Cristina Barbarino}
\affiliation{The Oskar Klein Centre \& Department of Astronomy, Stockholm University, AlbaNova, SE-106 91 Stockholm, Sweden}

\author[0000-0001-8018-5348]{Eric C. Bellm}
\affiliation{DIRAC Institute, Department of Astronomy, University of Washington, 3910 15th Avenue NE, Seattle, WA 98195, USA} 

\author[0000-0003-0901-1606]{Nadia Blagorodnova}
\affiliation{Department of Astrophysics/IMAPP, Radboud University, Nijmegen, The Netherlands}

\author[0000-0002-7735-5796]{Joe S.~Bright}
\affiliation{Department of Physics, University of Oxford, Denys Wilkinson Building, Keble Road, Oxford OX1 3RH, UK}

\author{E.~Burns}
\affiliation{NASA Postdoctoral Program Fellow, Goddard Space Flight Center, Greenbelt, MD 20771, USA}

\author{S.\ Bradley~Cenko}
\affiliation{Astrophysics Science Division, NASA Goddard Space Flight Center, Mail Code 661, Greenbelt, MD 20771, USA}
\affiliation{Joint Space-Science Institute, University of Maryland, College Park, MD 20742, USA}

\author{Virginia Cunningham}
\affiliation{Astronomy Department, University of Maryland, College Park, MD 20742, USA}

\author[0000-0002-8989-0542]{Kishalay De}
\affiliation{Cahill Center for Astrophysics, 
California Institute of Technology, MC 249-17, 
1200 E California Boulevard, Pasadena, CA, 91125, USA}

\author{Richard Dekany}
\affiliation{Caltech Optical Observatories, California Institute of Technology, MC 11-17, 1200 E. California Blvd., Pasadena, CA  91125 USA
}

\author{Alison Dugas}
\affiliation{Cahill Center for Astrophysics, 
California Institute of Technology, MC 249-17, 
1200 E California Boulevard, Pasadena, CA, 91125, USA}

\author{Rob P.~Fender}
\affiliation{Department of Physics, University of Oxford, Denys Wilkinson Building, Keble Road, Oxford OX1 3RH, UK}

\author{Claes Fransson}
\affiliation{The Oskar Klein Centre \& Department of Astronomy, Stockholm University, AlbaNova, SE-106 91 Stockholm, Sweden}

\author{Christoffer Fremling}
\affiliation{Division of Physics, Mathematics and Astronomy, California Institute of Technology, Pasadena, CA 91125, USA}

\author[0000-0002-0587-7042]{Adam Goldstein}
\affiliation{Science and Technology Institute, Universities Space Research Association, Huntsville, AL 35805, USA}

\author{Matthew J.\ Graham}
\affiliation{Division of Physics, Mathematics and Astronomy, California Institute of Technology, Pasadena, CA 91125, USA}

\author{David Hale}
\affiliation{Caltech Optical Observatories, California Institute of Technology, MC 11-17, 1200 E. California Blvd., Pasadena, CA  91125 USA
}

\author{Assaf Horesh}
\affiliation{Racah Institute of Physics, Hebrew University, Jerusalem 91904, Israel}

\author{Tiara Hung}
\affiliation{Department of Astronomy and Astrophysics, University of California, Santa Cruz, CA 95064, USA}

\author{Mansi M.~Kasliwal}
\affiliation{Cahill Center for Astrophysics, 
California Institute of Technology, MC 249-17, 
1200 E California Boulevard, Pasadena, CA, 91125, USA}

\author{N.\ Paul M.~Kuin}
\affiliation{Mullard Space Science Laboratory,
University College London,
Holmbury St. Mary, Dorking,
Surrey RH5 6NT, UK}

\author{S.\ R.~Kulkarni}
\affiliation{Cahill Center for Astrophysics, 
California Institute of Technology, MC 249-17, 
1200 E California Boulevard, Pasadena, CA, 91125, USA}

\author{Thomas Kupfer}
\affiliation{Kavli Institute for Theoretical Physics, University of California, Santa Barbara, CA 93106, USA
}

\author[0000-0001-9454-4639]{Ragnhild Lunnan}
\affiliation{The Oskar Klein Centre \& Department of Astronomy, Stockholm University, AlbaNova, SE-106 91 Stockholm, Sweden}

\author{Frank J. Masci}
\affiliation{IPAC, California Institute of Technology, 1200 E. California Blvd, Pasadena, CA 91125, USA}

\author[0000-0001-8771-7554]{Chow-Choong Ngeow}
\affiliation{Graduate Institute of Astronomy, National Central University, 32001, Taiwan}

\author{Peter E.~Nugent}
\affiliation{Lawrence Berkeley National Laboratory, 1 Cyclotron Road, Berkeley, CA, 94720, USA}

\author{Eran O.~Ofek}
\affiliation{Department of Particle Physics and Astrophysics, Weizmann Institute of Science, 234 Herzl St, 76100 Rehovot, Israel}

\author[0000-0002-4753-3387]{Maria T. Patterson}
\affiliation{DIRAC Institute, Department of Astronomy, University of Washington, 3910 15th Avenue NE, Seattle, WA 98195, USA}

\author{Glen Petitpas}
\affiliation{Harvard-Smithsonian Center for Astrophysics,
60 Garden Street, Cambridge, MA 02138, USA}

\author{Ben Rusholme}
\affiliation{IPAC, California Institute of Technology, 1200 E. California Blvd, Pasadena, CA 91125, USA}

\author{Hanna Sai}
\affiliation{Physics Department/Tsinghua Center for Astrophysics, Tsinghua University; Beijing, 100084, China}

\author{Itai Sfaradi}
\affiliation{Racah Institute of Physics, Hebrew University, Jerusalem 91904, Israel}

\author{David L. Shupe}
\affiliation{IPAC, California Institute of Technology, 1200 E. California Blvd, Pasadena, CA 91125, USA}

\author{Jesper Sollerman}
\affiliation{The Oskar Klein Centre \& Department of Astronomy, Stockholm University, AlbaNova, SE-106 91 Stockholm, Sweden}

\author[0000-0001-6753-1488]{Maayane T.~Soumagnac}
\affiliation{Department of Particle Physics and Astrophysics, Weizmann Institute of Science, 234 Herzl St, 76100 Rehovot, Israel}

\author{Yutaro Tachibana}
\affiliation{Department of Physics, Tokyo Institute of Technology, 2-12-1 Ookayama, Meguro-ku, Tokyo 152-8551, Japan}

\author{Francesco Taddia}
\affiliation{The Oskar Klein Centre \& Department of Astronomy, Stockholm University, AlbaNova, SE-106 91 Stockholm, Sweden}

\author{Richard Walters}
\affiliation{Cahill Center for Astrophysics, 
California Institute of Technology, MC 249-17, 
1200 E California Boulevard, Pasadena, CA, 91125, USA}

\author{Xiaofeng Wang}
\affiliation{Physics Department/Tsinghua Center for Astrophysics, Tsinghua University; Beijing, 100084, China}

\author[0000-0001-6747-8509]{Yuhan Yao}
\affiliation{Cahill Center for Astrophysics, 
California Institute of Technology, MC 249-17, 
1200 E California Boulevard, Pasadena, CA, 91125, USA}

\author{Xinhan Zhang}
\affiliation{Physics Department/Tsinghua Center for Astrophysics, Tsinghua University; Beijing, 100084, China}

\received{May 2019}
\revised{12 June 2019}
\submitjournal{ApJ}
\reportnum{astro-ph/1904.11009}

\begin{abstract}

We present detailed observations of ZTF18abukavn (\name),
discovered in high-cadence data from the Zwicky Transient Facility as a rapidly rising ($1.4\pm0.1$ mag/hr)
and luminous ($M_{g,\mathrm{peak}}=-20$ mag) transient.
It is spectroscopically classified as a
broad-lined stripped-envelope supernova (Ic-BL SN).
The high peak luminosity ($\lbol \gtrsim 3 \times 10^{44}\,\erg\,\psec$),
the short rise time ($\trise = 3\,\days$ in $g$-band),
and the blue colors at peak ($g-r\sim-0.4$) all
resemble the high-redshift Ic-BL iPTF16asu, as well as several other unclassified fast transients.
The early discovery of \name\ (within an hour of shock breakout) enabled an intensive spectroscopic campaign, including the highest-temperature ($\teff\gtrsim40,000\kelvin$) spectra of a stripped-envelope SN.
A retrospective search revealed luminous ($M_g \sim M_r \approx -14\,$mag) emission in the days to weeks before explosion, the first definitive detection of precursor emission for a Ic-BL.
We find a limit on the isotropic gamma-ray energy release $E_\mathrm{\gamma,iso}<4.9 \times 10^{48}\,\erg$,
a limit on X-ray emission $L_{\mathrm{X}} < 10^{40}\,\erg\,\psec$,
and a limit on radio emission 
$\nu L_\nu \lesssim 10^{37}\,\erg\,\psec$.
Taken together, we find that the early ($<10\,\days$) data are best explained by shock breakout in a massive shell of dense circumstellar material (0.02\,\msol) at large radii ($3 \times 10^{14}\,\cm$)
that was ejected in eruptive pre-explosion mass-loss episodes.
The late-time ($>10\,\days$) light curve requires an additional energy source, which could be the radioactive decay of Ni-56.
\end{abstract}

\section{Introduction}
\label{sec:introduction}

Recent discoveries by optical time-domain surveys challenge our understanding of how energy is deposited and transported in stellar explosions \citep{Kasen2017}.
For example, over 50 transients have been discovered with rise times and peak luminosities too rapid and too high, respectively, to be explained by radioactive decay \citep{Poznanski2010,Drout2014,Shivvers2016,Tanaka2016,Arcavi2016,Rest2018,Pursiainen2018}.
Possible powering mechanisms include interaction with extended circumstellar material (CSM; \citealt{ChevalierIrwin}), and energy injection from a long-lived central engine \citep{Kasen2010,Woosley2010,Kasen2016}.
These models have been difficult to test because
the majority of fast-luminous transients have been discovered \emph{post facto} and located at cosmological distances ($z\sim0.1$).

The discovery of iPTF16asu \citep{Whitesides2017,Wang2019} in the intermediate Palomar Transient Factory (iPTF; \citealt{Law2009}) 
showed that at least some of these fast-luminous transients
are energetic ($10^{52}\,\erg$)
high-velocity (``broad-lined''; $v \gtrsim20,000\,\km\,\psec$)
stripped-envelope (Ic) supernovae (Ic-BL SNe).
The light curve of iPTF16asu was unusual among Ic-BL SNe in being inconsistent with \nickel-decay \citep{Cano2013,Taddia2019}.
Suggested power sources include
energy injection by a magnetar,
ejecta-CSM interaction,
cooling-envelope emission, and an engine-driven explosion similar to low-luminosity gamma-ray bursts --- or some combination thereof.
Unfortunately,
the high redshift ($z=0.187$) precluded
a definitive conclusion.

Today, optical surveys such as ATLAS \citep{Tonry2018} and the Zwicky Transient Facility (ZTF; \citealt{Bellm2019a,Graham2019}) have the areal coverage to discover rare transients \emph{nearby},
as well as the cadence to discover transients when they are young ($<1\,\days$).
For example, the recent discovery of AT2018cow at 60\,\mpc\ \citep{Smartt2018,Prentice2018} represented an unprecedented opportunity to study a fast-luminous optical transient up close, in detail, and in real-time.
Despite an intense multiwavelength observing campaign, the nature of AT2018cow remains unknown -- possibilities include an engine-powered stellar explosion \citep{Prentice2018,Perley2019cow,Margutti2019,Ho2019}, the tidal disruption of a white dwarf by an intermediate-mass black hole \citep{Kuin2019,Perley2019cow}, and an electron capture SN \citep{Lyutikov2019}.
Regardless of the origin, it is clear that the explosion took place within dense material \citep{Perley2019cow,Margutti2019,Ho2019} confined to $\lesssim 10^{16}\,\cm$ \citep{Ho2019}. 

Here we present \name, discovered as a rapidly rising ($1.4\pm0.1$\,\magnitudes\,\phr) and luminous ($M_{g,\mathrm{peak}}=-20$) transient in high-cadence data from ZTF \citep{Ho2018a}.
The high inferred velocities ($>20,000\,\km\,\psec$), the spectroscopic evolution from a blue continuum to a Ic-BL SN \citep{Costantin2018},
the rapid rise ($\trise = 3\,\days$ in $g$-band) to high peak luminosity ($\lbol\gtrsim3\times10^{44}\,\erg\,\psec$)
all suggest that \name\ is a low-redshift ($z=0.03154$) analog to iPTF16asu.
The early discovery enabled an intensive follow-up campaign within the first day of the explosion,
including the highest-temperature ($\teff\gtrsim40,000\,\kelvin$) spectra of a stripped-envelope SN to-date.
A retrospective search in ZTF data revealed the first definitive detection of pre-explosion activity in a Ic-BL.

The structure of the paper is as follows.
We present our radio through X-ray data in Section \ref{sec:obs}.
In Section \ref{sec:basic-properties} we outline
basic properties of the explosion and its host galaxy.
In Section \ref{sec:interpretation} we attribute the power source for the light curve to shock breakout in extended CSM.
In Section \ref{sec:comparisons} we compare \name\ to unidentified fast-luminous transients at high redshift.
Finally, in Section \ref{sec:conclusions} we summarize our findings and look to the future.
Throughout the paper,
absolute times are reported in UTC and relative times are reported with respect to $t_0$, which is defined in Section \ref{sec:obs-discovery}.
We assume a standard $\Lambda$CDM cosmology \citep{Planck2016}.

\section{Observations}
\label{sec:obs}

\subsection{Zwicky Transient Facility Discovery}
\label{sec:obs-discovery}

ZTF observing time is divided between several different surveys, conducted using a custom mosaic camera \citep{Dekany2016} on the 48-inch Samuel Oschin Telescope (P48) at Palomar Observatory.
See \citet{Bellm2019a} for an overview of the observing system,
\citet{Bellm2019b} for a description of the surveys and scheduler, and
\citet{Masci2019} for details of the image processing system.

Every 5-$\sigma$ point-source detection is saved as an ``alert.'' Alerts are distributed in avro format \citep{Patterson2019}
and can be filtered based on a machine learning-based real-bogus metric \citep{Mahabal2019,Duev2019},
light-curve properties, and host characteristics (including a star-galaxy classifier; \citet{Tachibana2018}).
The ZTF collaboration uses a web-based system called the GROWTH marshal \citep{Kasliwal2019} to identify and keep track of transients of interest.

\begin{figure}[ht]
\centering
\includegraphics[scale=0.4]{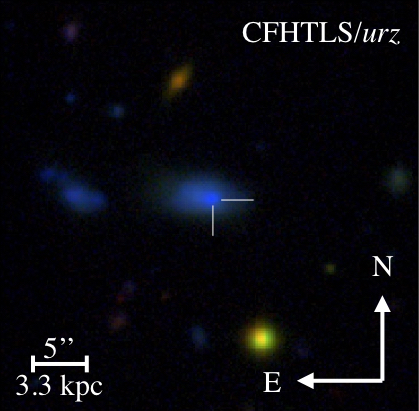}
\caption{
The position of \name\ (white crosshairs) in its host galaxy. Images from the Canada-France-Hawaii Telescope Legacy Survey (2004--2012), combined using the prescription in \citet{Lupton2004}.
}
\label{fig:hostim}
\end{figure}

ZTF18abukavn was discovered in an image obtained at 2018-09-09 03:55:18 (start of exposure) as part of
the ZTF extragalactic high-cadence partnership survey,
which covers 1725\,\degsq\ in six visits (3$g$, 3$r$) per night \citep{Bellm2019b}.
The discovery magnitude was $r=20.5 \pm 0.3 \,\magnitudes$,
and the source position was measured to be
$\alpha = 16^{\mathrm{h}}43^{\mathrm{m}}48.22^{\mathrm{s}}$, $\delta = +41^{\mathrm{d}}02^{\mathrm{m}}43.4^{\mathrm{s}}$ (J2000),
coincident with a compact galaxy (Figure \ref{fig:hostim})
at $z=\redshift$ or $d\approx\distance\,\mpc$.
As described in Section \ref{sec:obs-spectroscopy},
the redshift was unknown at the time of discovery;
we measured it from narrow galaxy emission lines in our follow-up spectra.
The host redshift along with key observational properties of the transient are listed in Table \ref{tab:obs-properties}. 

\begin{deluxetable*}{lrr}[!ht]
\tablecaption{Key observational properties of SN2018gep and its host galaxy \label{tab:obs-properties}} 
\tablewidth{0pt} 
\tablehead{ \colhead{Parameter} & \colhead{Value} & \colhead{Notes}} 
\tabletypesize{\normalsize} 
\startdata 
$z$ & $0.03154$ & From narrow host emission lines \\
$\lpeak$ & $\gtrsim 3 \times 10^{43}\,\erg$ & Peak UVOIR bolometric luminosity \\
$\trise$ & 0.5--3\,\days & Time from $t_0$ to \lpeak \\
$E_\mathrm{rad}$ & $10^{50}\,\erg$ & UVOIR output, $\Delta t=0.5$--40\,\days \\
$M_{r,\mathrm{prog}}$ & $-15$ & Peak luminosity of pre-explosion emission \\
$E_\mathrm{\gamma,iso}$ & $< 4.9 \times 10^{48}$ erg & Limit on prompt gamma-ray emission from \emph{Fermi}/GBM \\
$L_X$ & $< 2.5 \times 10^{41}\,\erg\,\psec$ & X-ray upper limit from \swift/XRT at $\Delta t=0.4$--14\,\days \\
& $< 10^{40}\,\erg\,\psec$ & X-ray upper limit from \chandra\ at $\Delta t=15$ and $\Delta t=70\,\days$ \\
$\nu L_\nu$ & $\approx 10^{37}\,\erg\,\psec$ & 9\,\ghz\ radio luminosity from VLA at $\Delta t=5$ and $\Delta t=16$ \\
$M_\mathrm{*,host}$ & $1.3 \times 10^{8}\,M_\odot$ & Host stellar mass \\
SFR$_\mathrm{host}$ & 0.12\,\msol\,\pyr & Host star-formation rate \\
Host metallicity & 1/5 solar & Oxygen abundance on O3N2 scale \\
\enddata 
\end{deluxetable*}

\begin{figure*}[ht]
\centering
\includegraphics[scale=0.7]{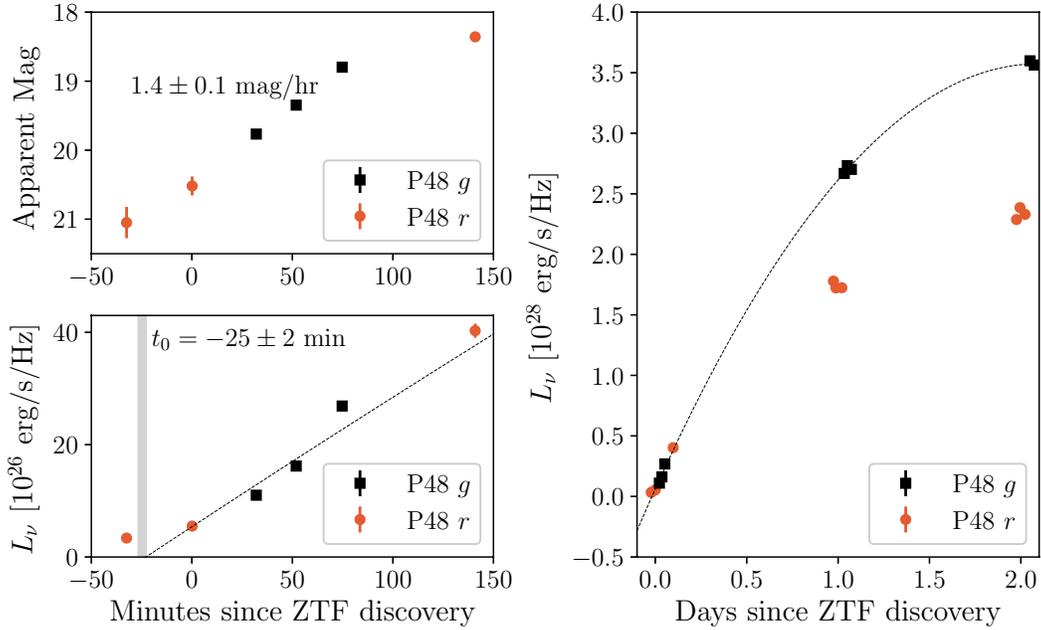}
\caption{
The rapid rise in the first few minutes and first few days after the ZTF discovery of \name. We also show an $r$-band point from prior to discovery that was found in retrospect by lowering the detection threshold from 5-$\sigma$ to 3-$\sigma$.
Top left: the rise in magnitudes gives an almost unprecedented rate of $1.4\pm0.1$\,\magnitudes\,\phr.
Bottom left: the rise in flux space together with the quadratic fit and definition of $t_0$.
Right: the rise in flux space showing the quadratic fit.
}
\label{fig:firstmins}
\end{figure*}

As shown in Figure \ref{fig:firstmins},
the source brightened by over two magnitudes
within the first three hours.
These early detections passed a filter written in the GROWTH marshal that was designed to find young SNe.
We announced the discovery and fast rise via the Astronomer's Telegram \citep{Ho2018a},
and reported the object to the IAU Transient Server (TNS\footnote{https://wis-tns.weizmann.ac.il}),
where it received the designation \name.

We triggered ultraviolet (UV) and optical observations with the UV/Optical Telescope (UVOT; \citealt{Roming2005}) aboard the \emph{Neil Gehrels Swift Observatory} \citep{Gehrels2004},
and observations began 10.2 hours after the ZTF discovery \citep{Schulze2018}.
A search of IceCube data found no temporally coincident high-energy neutrinos \citep{Blaufuss2018}.

Over the first two days, the source brightened by two additional magnitudes.
A linear fit to the early $g$-band photometry gives 
a rise of $1.4\pm0.1$\,\magnitudes\,\phr.
This rise rate is second only to the IIb SN 16gkg \citep{Bersten2018} but several orders of magnitude more luminous at discovery ($M_{g,\mathrm{disc}}\approx-17\,\magnitudes$).

To establish a reference epoch, we fit
a second-order polynomial to the first three days of the $g$-band light curve in flux space, and define $t_0$ as the time at which the flux is zero.
This gives $t_0$ as being $25 \pm 2$ minutes prior to the first detection, or $t_0 \approx$ UTC 2018-09-09 03:30.
The physical interpretation of $t_0$ is not straightforward, since the light curve flattens out at early times (see Figures \ref{fig:firstmins} and \ref{fig:lc}). We proceed using $t_0$ as a reference epoch
but caution against assigning it physical meaning.

\subsection{Photometry}
\label{sec:obs-photometry}

From $\Delta t\approx1\,\days$ to $\Delta t\approx60\,\days$,
we conducted a photometric follow-up campaign at UV and optical wavelengths
using \swift/UVOT,
the Spectral Energy Distribution Machine (SEDM; \citealt{Blagorodnova2018}) mounted on the automated 60-inch telescope at Palomar (P60; \citealt{Cenko2006}),
the optical imager (IO:O) on the Liverpool Telescope (LT; \citealt{Steele2004}), and the Lulin 1-m Telescope (LOT).

Basic reductions for the LT IO:O imaging were performed by the LT pipeline\footnote{\href{https://telescope.livjm.ac.uk/TelInst/Pipelines/\#ioo}{https://telescope.livjm.ac.uk/TelInst/Pipelines/\#ioo}}.  Digital image subtraction and photometry for the SEDM, LT and LOT imaging was performed using the Fremling Automated Pipeline (\texttt{FPipe}; \citealt{Fremling2016}).
\texttt{Fpipe} performs calibration and host subtraction against Sloan Digital Sky Survey reference images and catalogs (SDSS; \citealt{Ahn2014}).
SEDM spectra were reduced using \texttt{pysedm} \citep{Rigault2019}.

The UVOT data were retrieved from the NASA \swift\ Data Archive\footnote{\href{https://heasarc.gsfc.nasa.gov/cgi-bin/W3Browse/swift.pl}{https://heasarc.gsfc.nasa.gov/cgi-bin/W3Browse/swift.pl}} and reduced using standard software distributed with \package{HEAsoft} version 6.19\footnote{\href{https://heasarc.nasa.gov/lheasoft/}{https://heasarc.nasa.gov/lheasoft/}}.
Photometry was measured using \package{uvotmaghist} with a $3^{\prime\prime}$ circular aperture.
To remove the host contribution,
we obtained a final epoch in all broad-band filters on 18 October 2018 and built a host template using
\package{uvotimsum} and \package{uvotsource} with the same aperture used for the transient.

\begin{figure*}[ht]
\centering
\includegraphics[scale=0.8]{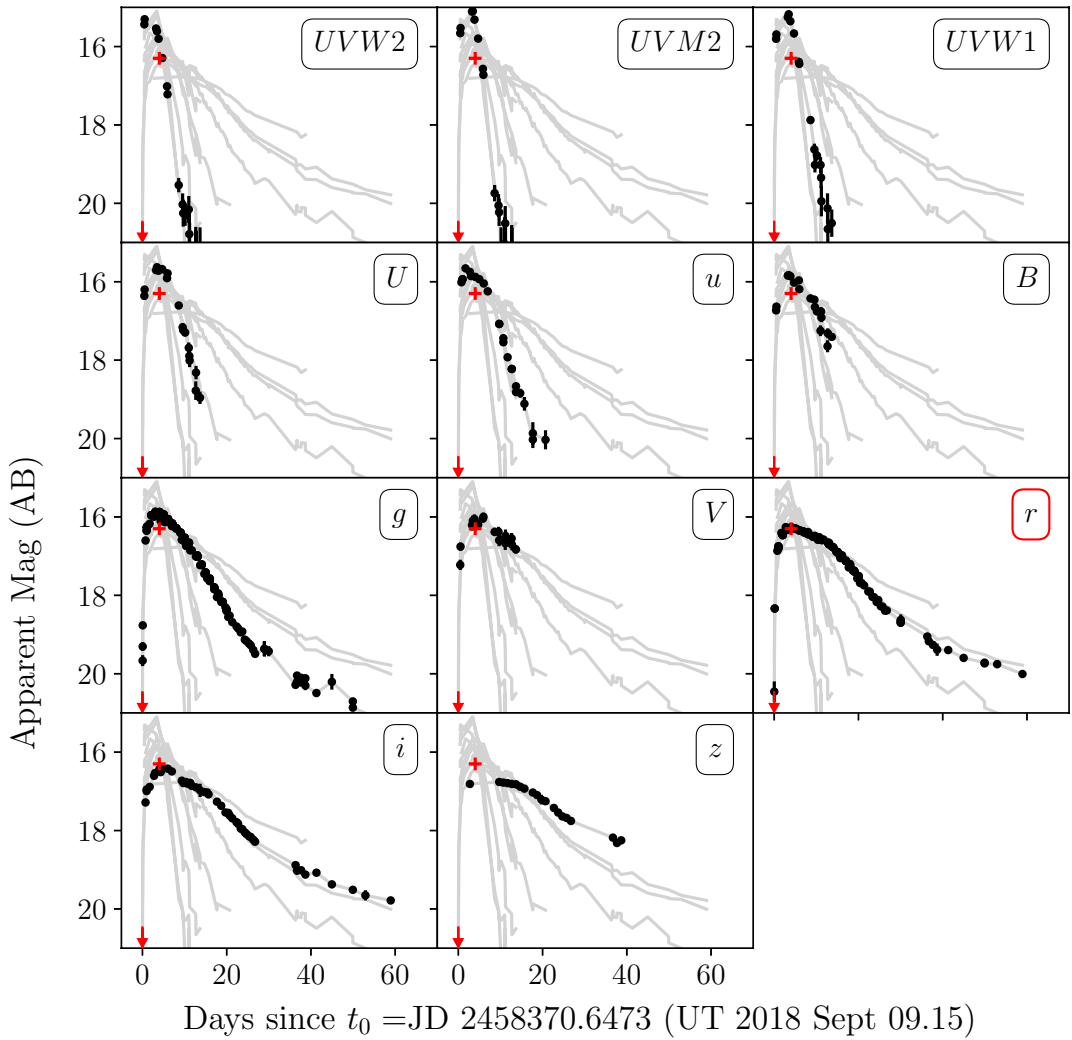}
\caption{UV and optical light curves from \swift\ and ground-based facilities.
The arrow marks the last non-detection, which was in $r$-band. The red cross marks the peak of the $r$-band light curve, which is 16.3\,\magnitudes\ at $\Delta t=4\,\days$.
The full set of light curves are shown as grey lines in the background, and each panel highlights an individual filter in black.
We correct for Galactic extinction using the attenuation curve from \citet{Fitzpatrick1999} and $E_{B-V}=A_V/R_V=0.01$ for $R_V=3.1$ and $A_V=0.029$ \citep{Schlafly2011}.
}
\label{fig:lc}
\end{figure*}

\begin{figure}[ht]
\centering
\includegraphics[width=\columnwidth]{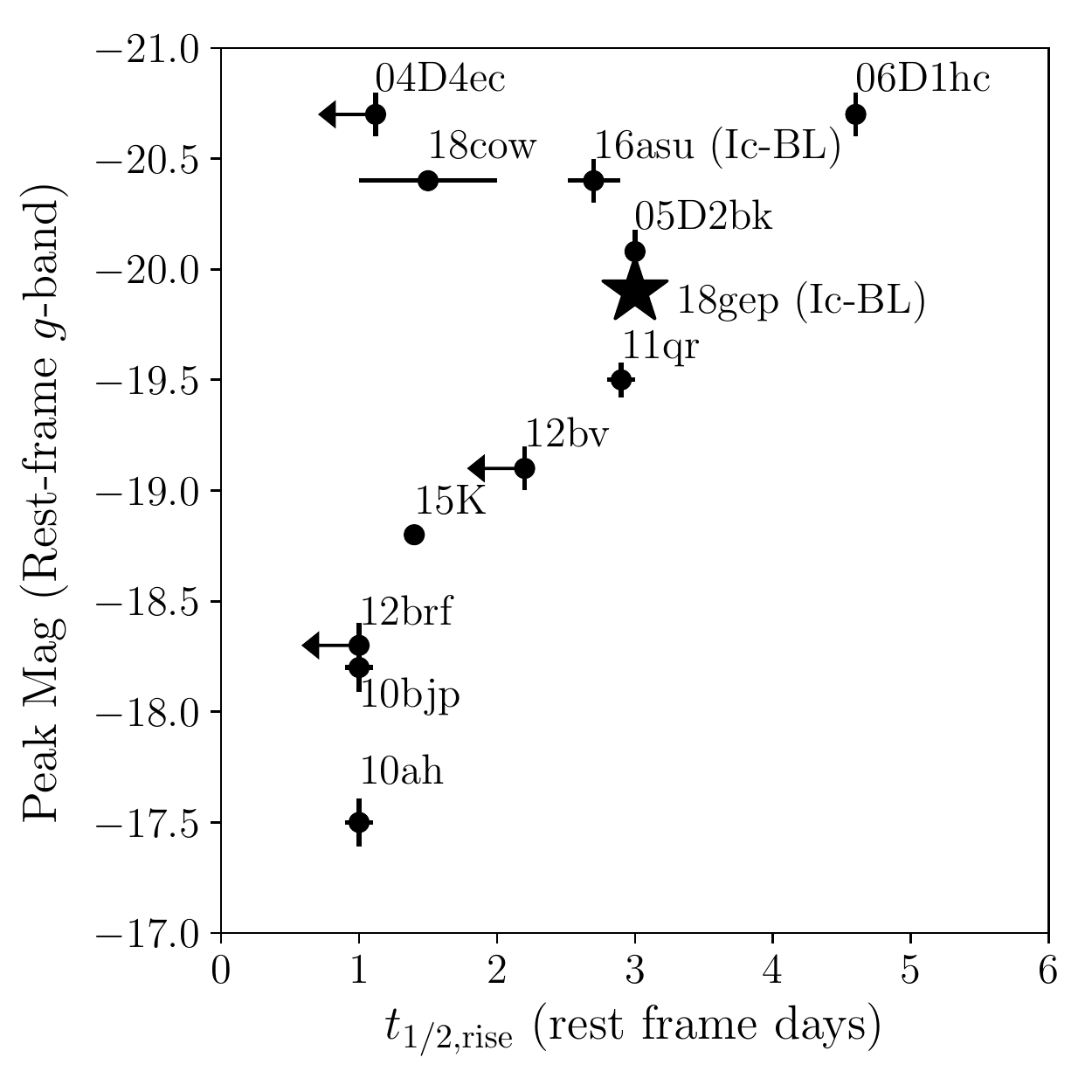}
\caption{
The rise time and peak absolute magnitude of \name,
iPTF16asu (a high-redshift analog),
and unclassified fast-luminous transients from \citet{Drout2014}, \citet{Arcavi2016}, \citet{Rest2018}, and \cite{Perley2019cow}.
When possible, 
we report measurements in rest-frame $g$-band,
and define ``rise time'' as time from half-max to max.
For iPTF16asu, we use the quadratic fit to the early $g$-band light curve from \citet{Whitesides2017} as well as their reported peak magnitude, but caution that this is rest-frame $r$-band.
For KSN2015K, there are only observations in the \emph{Kepler} white filter \citep{Rest2018}. 
}
\label{fig:mpeakrise}
\end{figure}

Figure \ref{fig:lc} shows the full set of light curves,
with a cross denoting the peak of the $r$-band light curve for reference.
The position of the cross is simply the
time and magnitude of our brightest $r$-band measurement, which is a good estimate given our cadence. The photometry is listed in Table \ref{tab:uvot-phot} in Appendix \ref{appendix:uv-opt-phot}. Note that despite the steep SED at early times, the K-correction is minimal. We estimate that the effect is roughly 0.03 mag, which is well within our uncertainties. In Figure \ref{fig:mpeakrise} we compare the rise time and peak absolute magnitude to other rapidly evolving transients from the literature.

\subsection{Spectroscopy}
\label{sec:obs-spectroscopy}

The first spectrum was taken 0.7\,\days\ after discovery by the Spectrograph for the Rapid Acquisition of Transients (SPRAT; \citealt{Piascik2014}) on the Liverpool Telescope (LT).
The spectrum showed a blue continuum with narrow galaxy emission lines, establishing this as a luminous transient ($M_\mathrm{g,peak} = -19.7$).
Twenty-three optical spectra were obtained from $\Delta t=0.7$--$61.1$\,\days,
using SPRAT,
the Andalusia Faint Object Spectrograph and Camera (ALFOSC) on the Nordic Optical Telescope (NOT),
the Double Spectrograph (DBSP; \citealt{Oke1982}) on the 200-inch Hale telescope at Palomar Observatory,
the Low Resolution Imaging Spectrometer (LRIS; \citealt{Oke1995}) on the Keck I 10-m telescope,
and the Xinglong 2.16-m telescope (XLT+BFOSC) of NAOC, China \citep{Wang2018}.
As discussed in Section \ref{sec:spec-evolution},
the early $\Delta t<5\,\days$ spectra show broad absorption features that evolve redward with time,
which we attribute to carbon and oxygen.
By $\Delta t\sim8\,\days$, the spectrum resembles a stripped-envelope SN, and the usual broad features of a Ic-BL emerge \citep{Costantin2018}.

We use the automated LT pipeline reduction and extraction for the LT spectra.
LRIS spectra were reduced and extracted using \texttt{Lpipe} \citep{Perley2019lpipe}.
The NOT spectrum was obtained at parallactic angle using a $1^{\prime\prime}$ slit, and was reduced in a standard way, including wavelength calibration against an arc lamp, and flux calibration using a spectrophotometric standard star.
The XLT+BFOSC spectra were reduced using the standard IRAF routines, including corrections for bias, flat field, and removal of cosmic rays. The Fe/Ar and Fe/Ne arc lamp spectra obtained during the observation night are used to calibrate the wavelength of the spectra, and the standard stars observed on the same night at similar airmasses as the supernova were used to calibrate the flux of spectra. The spectra were further corrected for continuum atmospheric extinction during flux calibration, using mean extinction curves obtained at Xinglong Observatory.
Furthermore, telluric lines were removed from the data.

\swift\ obtained three UV-grism spectra between 2018-09-15 3:29 and 6:58 UTC ($\Delta t\approx6.4\,\days$) for a 
total exposure time of 3918\,s. 
The data were processed using the calibration and 
software described by \citet{Kuin2015}.
During the observation, the source spectrum
was centered on the detector,
which is the default location for \swift/UVOT
observations.
Because of this, there is second-order contamination from a nearby star,
which was reduced by using
a narrow extraction width ($1.3^{\prime\prime}$ instead of $2.5^{\prime\prime}$).
The contamination renders the spectrum unreliable at wavelengths longer than 4100\,\AA,
but is negligible in the range 
2850--4100\,\AA
due to absorption from the ISM.
Below 2200\,\AA, the spectrum overlaps with the spectrum from another star in the field of view.

\begin{figure}[ht]
\centering
\includegraphics[scale=0.5]{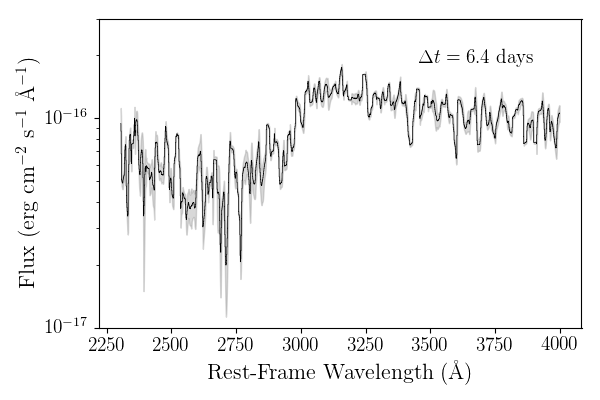}
\caption{\swift/UVOT grism spectrum shifted to the rest frame. Black line shows the data binned such that each bin size is 10\,\AA. Light grey represents 1-$\sigma$ uncertainties after binning. The spectrum has been scaled to match the UVOT $u$-band flux at this epoch (integrated from 3000\,\AA\ to 3900\,\AA), which was determined by interpolating the \swift\ $u$-band light curve.}
\label{fig:uvot-grism}
\end{figure}

The resulting spectrum (Figure \ref{fig:uvot-grism}) shows
a single broad feature between 2200\,\AA\ and 3000\,\AA\ (rest frame).
One possibility is that this is a blend of the UV features seen in SLSNe.
Line identifications for these features vary in the SLSN literature,
but are typically blends of \ion{Ti}{3}, \ion{Si}{3}, \ion{C}{2}, \ion{C}{3}, and \ion{Mg}{2} \citep{Quimby2011,Howell2013,Mazzali2016,Yan2017}.

The spectral log and a figure showing all the spectra are presented in Appendix \ref{appendix:uv-opt-spec}.
In Section \ref{sec:spec-evolution} we compare the early spectra to spectra at similar epochs in the literature.
We model one of the early spectra, which shows a ``W'' feature that has been seen in superluminous supernovae (SLSNe),
to measure the density, density profile, and element composition of the ejecta.
From the Ic-BL spectra, we measure the velocity evolution of the photosphere.

\subsection{Search for pre-discovery emission}
\label{sec:obs-progenitor}

The nominal ZTF pipeline only generates detections
above a 5-$\sigma$ threshold.
To extend the light curve further back in time,
we performed forced photometry at the position of \name\ on single-epoch difference images from the IPAC ZTF difference imaging pipeline. The ZTF forced photometry PSF-fitting code will be described in detail in a separate paper (Yao, Y. et al. in preparation).
As shown in Figure \ref{fig:firstmins}, forced photometry uncovered an earlier 3-$\sigma$ $r$-band detection.

Next, we searched for even fainter detections by constructing deeper reference images than those used by the nominal pipeline,
and subtracting them from 1-to-3 day stacks of ZTF science images.
The reference images were generated by performing an inverse-variance weighted coaddition of 298 $R$-band and 69 $g$-band images from PTF/iPTF taken between 2009 and 2016 using the \texttt{CLIPPED} combine strategy in \texttt{SWarp} \citep{Bertin2010,Gruen2014}.
PTF/iPTF images were used instead of ZTF images to build references as they were obtained years prior to the transient, and thus less likely to contain any transient flux.
No cross-instrument corrections were applied to the references prior to subtraction.
Pronounced regions of negative flux on the PTF/iPTF references caused by crosstalk from bright stars were masked out manually.

\begin{figure*}
    \centering
    \includegraphics[scale=0.7]{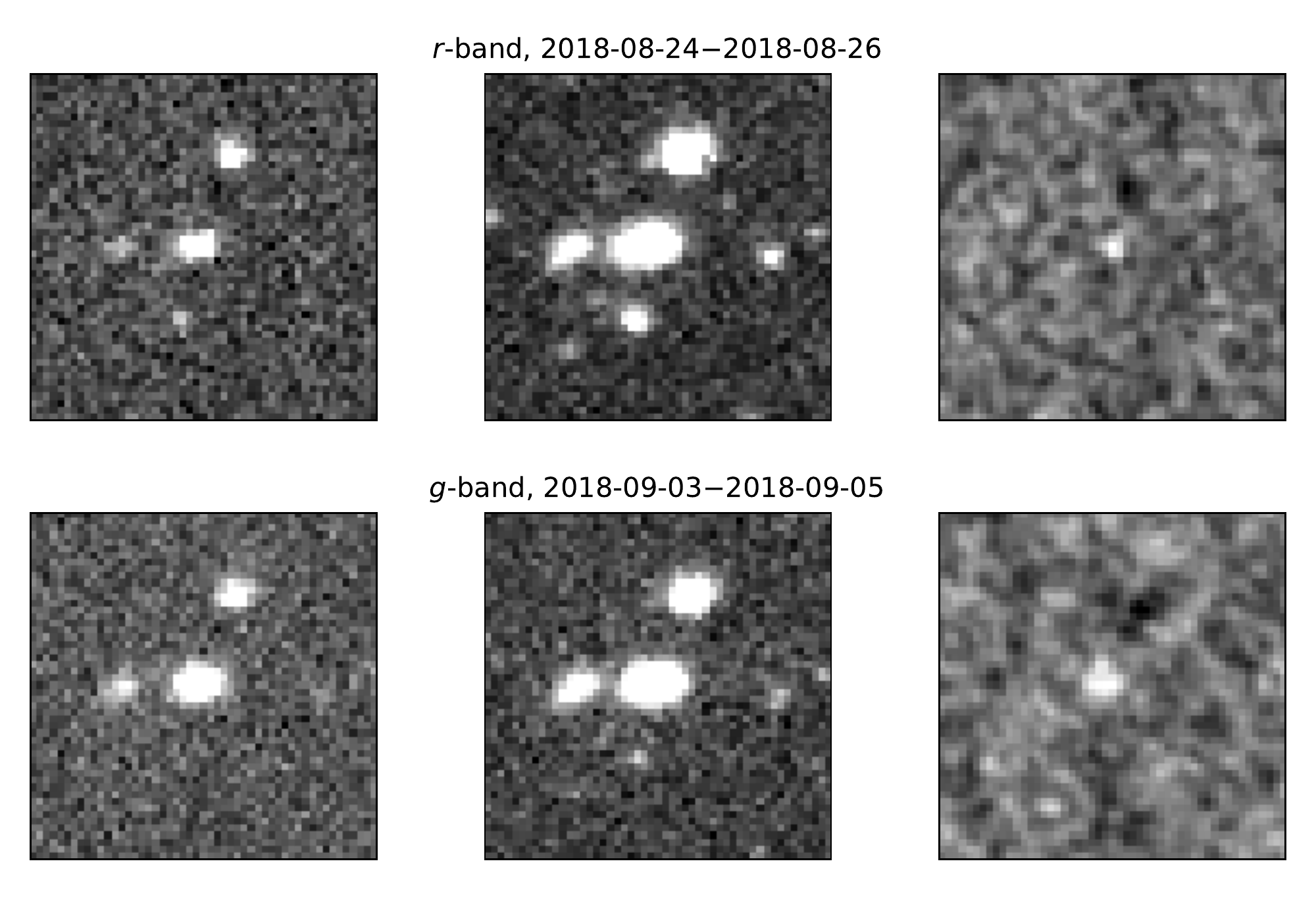}
    \caption{Sample pre-explosion subtractions of deep PTF/iPTF references from ZTF science images stacked in 3-day bins  (see Section \ref{sec:obs-progenitor}).
    Each cutout is centered on the location of \name. 
    The subtractions show clear emission at the location of the SN in both $g$ and $r$-bands days to weeks before the discovery of the SN in ZTF.}
    \label{fig:subtractions}
\end{figure*}

We stacked ZTF science images
obtained between 2018 Feb 22 and 2018 Aug 31 in a rolling window (segregated by filter) with a width of 3 days and a period of 1 day, also using the \texttt{CLIPPED} technique in \texttt{SWarp}.
Images taken between 2018 Sep 01 and $t_0$ were stacked in a window with a width of 1 day and a period of 1 day.
Subtractions were obtained using the \texttt{HOTPANTS} \citep{Becker2015} implementation of the \citet{Alard1998} PSF matching algorithm.
Many of the ZTF science images during this period were obtained under exceptional conditions, and the seeing on the ZTF science coadds was often significantly better than the seeing on the PTF/iPTF references. 
To correct for this effect, ZTF science coadds were convolved with their own point spread functions (PSFs),  extracted using \texttt{PSFEx}, prior to subtraction. 
During subtraction, PSF matching and convolution were performed on the template and the resulting subtractions were normalized to the photometric system of the science images. 
We show two example subtractions in
Figure \ref{fig:subtractions}.

Using these newly constructed deep subtractions,
PSF photometry was performed at the location of \name\ using the PSF of the science images.
To estimate the uncertainty on the flux measurements made on these subtractions, we employed a Monte Carlo technique, in which thousands of PSF fluxes were measured at random locations on the image, and the PSF-flux uncertainty was taken to be the 1$\sigma$ dispersion in  these measurements. 
We loaded this photometry into a local instance of SkyPortal \citep{skyportal}, an open-source web application that interactively displays astronomical datasets for annotation, analysis, and discovery.

We detected significant flux excesses at the location of \name\ in both $g$ and $r$ bands in the weeks preceding $t_0$ (i.e. its first detection in single-epoch ZTF subtractions).
The effective dates of these extended pre-discovery detections are determined by taking an inverse-flux variance weighted average of the input image dates.
The detections in the week leading up to explosion are
$m_g \sim m_r \approx 22$, which is approximately the magnitude limit of the coadd subtractions. 
However, in an $r$-band stack of images from August 24--26 (inclusive), we detect emission at $m_r \sim 21.5$ at $5\sigma$ above the background.

Assuming that the rapid rise we detected was close to the time of explosion,
this is the first definitive detection of pre-explosion emission in a Ic-BL SN.
There was a tentative detection in another source, PTF\,11qcj \citep{Corsi2014}, 1.5 and 2.5 years prior to the SN.
In Section \ref{sec:interpretation} we discuss possible mechanisms for this emission,
and conclude that it is likely related to a period of eruptive mass-loss immediately prior to the explosion.
We note that it is unlikely that this variability arises from AGN activity, due to the properties of the host galaxy (Section \ref{sec:host}).

\begin{figure*}[ht]
\centering
\includegraphics[scale=0.5]{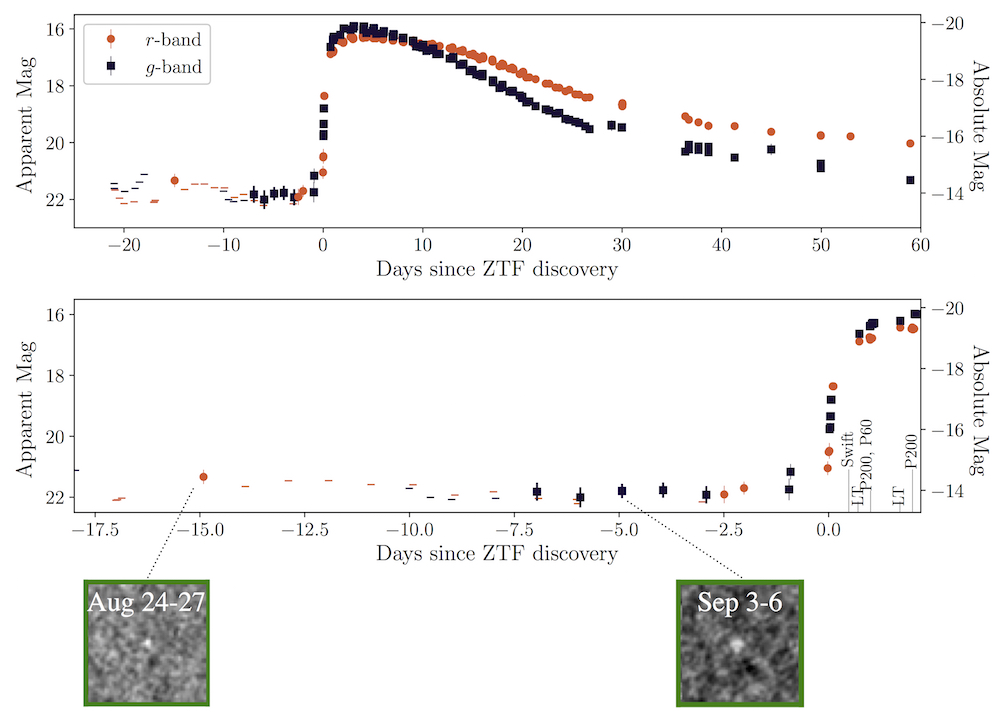}
\caption{
Full $r$ and $g$-band light curves of \name.
3-$\sigma$ upper limits are shown as horizontal lines.
Points at $t<0$ are from 3-day stacks of ZTF/P48 data as described in Section \ref{sec:obs-progenitor}.
Sample subtractions from two of these stacks are shown in the bottom row.
}
\label{fig:earlydata}
\end{figure*}

With forced photometry and faint detections from stacked images and deep references, we can construct a light curve that extends weeks prior to the rapid rise in the light curve,
shown in Figure \ref{fig:earlydata}.

\subsection{Radio follow-up}
\label{obs:radio}

We observed the field of \name\ with the Karl G. Jansky Very Large Array (VLA) on three epochs: on 2018 September 14 UT under the Program ID VLA/18A-242 (PI: D. Perley; \citealt{Ho2018b}), and on 2018 September 25 and 2018 November 23 UT under the Program ID VLA/18A-176 (PI: A. Corsi). We used 3C286 for flux calibration, and J1640+3946 for gain calibration. The observations were carried out in X- and Ku-band (nominal central frequencies of 9\,\ghz\ and 14\,\ghz, respectively) with a nominal bandwidth of 2\,\ghz. The data were calibrated using the automated VLA calibration pipeline available in the CASA package \citep{McMullin2007} then inspected for further flagging. The CLEAN procedure \citep{Hogbom1974} was used to form images in interactive mode. The image rms and the radio flux at the location of \name\ were measured using \texttt{imstat} in CASA. Specifically, we report the maximum flux within pixels contained in a circular region centered on the optical position of \name\ with radius comparable to the FWHM of the VLA synthesized beam at the appropriate frequency. The source was detected in the first two epochs, but not in the third (see Table \ref{tab:radio-flux}). As we discuss in Section \ref{sec:interpretation}, the first two epochs were conducted in a different array configuration than the third epoch,
and may have had a contribution from host galaxy light.

We also obtained three epochs of observations with the AMI large array (AMI-LA; \citealt{Zwart2008,Hickish2018}),
on UT 2018 Sept 12, 2018 Sept 23, and 2018 Oct 20.
AMI-LA is a radio interferometer comprised of eight, 12.8\,m diameter that extends from 18\,m up to 110\,m in length and operates with a 5\,GHz bandwidth around a central frequency of 15.5\,GHz.

We used a custom AMI data reduction software package \texttt{reduce\_dc} (e.g.\ \citealt{Perrott2013}) to perform initial data reduction, flagging, and calibration of phase and flux.
Phase calibration was conducted using short interleaved observations of J1646+4059, and for absolute flux calibration we used 3C286. Additional flagging and imaging were performed using CASA. 
All three observations resulted in null-detections with 3-$\sigma$ upper limits of $\approx 120\,\mu$Jy in the first two observations, and a 3-$\sigma$ upper limit of $\approx 120\,\mu$Jy in the last observation. 

Finally, we observed at higher frequencies using the Submillimeter Array (SMA; \citealt{Ho2004}) on UT 2018 Sep
15 under its target-of-opportunity program.
The project ID was 2018A-S068. Observations were performed in the sub-compact configuration using seven antennas. The observations were performed using RxA and RxB
receivers tuned to LO frequencies of 225.55 GHz and 233.55 GHz
respectively, providing 32 GHz of continuous bandwidth ranging from
213.55 GHz to 245.55 GHz with a spectral resolution of 140.0 kHz per
channel.
The
atmospheric opacity was around 0.16-0.19 with system temperatures
around 100-200\,\kelvin. 
The nearby quasars 1635+381 and 3C345 were used as the
primary phase and amplitude gain calibrators with absolute flux
calibration performed by comparison to Neptune. Passband calibration
was derived using 3C454.3. Data calibration was performed using the
MIR IDL package for the SMA, with subsequent analysis performed in
MIRIAD \citep{Sault1995}. For the flux measurements, all spectral channels were averaged together
into a single continuum channel and an rms of 0.6 mJy was achieved after 75 minutes on-source.

The full set of radio and sub-millimeter measurements are listed in Table \ref{tab:radio-flux}.

\startlongtable 
\begin{deluxetable*}{lrrrrrrr} 
\tablecaption{Radio flux density measurements for SN2018gep.\label{tab:radio-flux}} 
\tablewidth{0pt} 
\tablehead{ \colhead{Start Time} & \colhead{$\Delta t$} & \colhead{Instrument} & \colhead{$\nu$} & \colhead{$f_\nu$} & \colhead{$L_\nu$} & \colhead{$\theta_\mathrm{FWHM}$} & \colhead{Int. time} \\ \colhead{(UTC)} & \colhead{(days)} & \colhead{} & \colhead{(GHz)} & \colhead{($\mu$Jy)} & \colhead{(erg\,\psec\,\phz)} & \colhead{$^{\prime\prime}$} & \colhead{(hr)}} 
\tabletypesize{\scriptsize} 
\startdata 
2018-09-12 17:54 & 3.6 & AMI & 15 & $<120$ & $<2.9 \times 10^{27}$ & $43.53 \times 30.85$ & 4 \\ 
2018-09-23 15:35 & 14.5 & AMI & 15 & $<120$ & $<2.9 \times 10^{27}$ & $39.3 \times 29.29$ & 4 \\ 
2018-10-20 14:01 & 41.4 & AMI & 15 & $<120$ & $<2.9 \times 10^{27}$ & $43.53 \times 30.85$ & 4 \\ 
2018-09-15 02:33 & 6.0 & SMA & 230 & $<590$ & $<1.4 \times 10^{28}$ & $4.828 \times 3.920$ & 1.25 \\ 
2018-09-14 01:14 & 4.9 & VLA & 9.7 & $34 \pm 4$ & $8.3 \times 10^{26}$ & $7.06 \times 5.92$ & 0.5 \\ 
2018-09-25 00:40 & 15.9 & VLA & 9 & $24.4 \pm 6.8$ & $6.0 \times 10^{26}$ & $7.91 \times 6.89$ & 0.7 \\ 
2018-09-25 00:40 & 15.9 & VLA & 14 & $26.8 \pm 6.8$ & $6.6 \times 10^{26}$ & $4.73 \times 4.26$ & 0.5 \\ 
2018-11-23 13:30 & 75.4 & VLA & 9 & $<16$ & $<3.9 \times 10^{26}$ & $3.52 \times 2.08$ & 0.65 \\ 
2018-11-23 13:30 & 75.4 & VLA & 14 & $<17$ & $<4.2 \times 10^{26}$ & $2.77 \times 1.32$ & 0.65 \\ 
\enddata 
\tablecomments{For VLA measurements: The quoted errors are calculated as the quadrature sums of the image rms, plus a 5\% nominal absolute flux calibration uncertainty. When the peak flux density within the circular region is less than three times the RMS, we report an upper limit equal to three times the RMS of the image. For AMI measurements: non-detections are reported as 3-$\sigma$ upper limits. For SMA measurements: non-detections are reported as a 1-$\sigma$ upper limit.}\end{deluxetable*}

\subsection{X-ray follow-up}
\label{obs:xray}

We observed the position of \name\ with \swift/XRT
from $\Delta t\approx0.4$--14\,\days.
The source was not detected in any epoch.
To measure upper limits,
we used web-based tools developed by the \swift-XRT team \citep{Evans2009}.
For the first epoch, the 3-$\sigma$ upper limit was 0.003 ct/s.
To convert the upper limit from count rate to
flux, we assumed\footnote{\href{https://heasarc.gsfc.nasa.gov/cgi-bin/Tools/w3nh/w3nh.pl}{https://heasarc.gsfc.nasa.gov/cgi-bin/Tools/w3nh/w3nh.pl}} a Galactic neutral hydrogen column density of $1.3 \times 10^{20}\,\pcmsq$, and a power-law spectrum with photon index $\Gamma = 2$.
This gives\footnote{\href{https://heasarc.gsfc.nasa.gov/cgi-bin/Tools/w3pimms/w3pimms.pl}{https://heasarc.gsfc.nasa.gov/cgi-bin/Tools/w3pimms/w3pimms.pl}} an unabsorbed 0.3--10\,\kev\ flux of $<9.9 \times 10^{-14}\,\erg\,\pcmsq\,\psec$,
and $L_X<2.5 \times 10^{41}\,\erg\,\psec$.

We obtained two epochs of observations with the Advanced 
CCD Imaging Spectrometer (ACIS; \citealt{Garmire2003}) on the \chandra\ X-ray 
Observatory via our approved program (Proposal No. 19500451; PI: Corsi).
The first epoch began at 9:25 UTC on 10 October 2018
($\Delta t \approx 15\,\days$) under ObsId 20319 (integration time 12.2 ks), and the second began at
21:31 UTC on 4 December 2018 ($\Delta t \approx 70\,\days$) under ObsId 20320 (integration time 12.1 ks).
No X-ray emission is detected at the location of \name\
in either epoch, with 90\% upper limits on the 0.5--7.0\,keV
count rate of $\approx 2.7 \times 10^{-4}\,\mathrm{ct}\,\psec$.
Using the same values of hydrogen column density and power-law photon index as in our XRT measurements,
we find upper limits on the unabsorbed 0.5--7\,\kev\ X-ray flux
of $< 3.2 \times 10^{-15}$\,erg\,cm$^{-2}$\,s$^{-1}$,
or (for a direct comparison to the XRT band)
a 0.3--10\,\kev\ X-ray flux
of $< 4.2 \times 10^{-15}$\,erg\,cm$^{-2}$\,s$^{-1}$.
This corresponds to a 0.3--10\,\kev\ luminosity
upper limit of
$L_X < 1.0 \times 10^{40}\,\erg\,\psec$.

\subsection{Search for prompt gamma-ray emission}
\label{sec:obs-grbsearch}

We created a tool to search for prompt gamma-ray emission (GRBs) from \fermi-GBM \citep{Gruber2014,vonKienlin2014,Bhat2016},
the \swift\ Burst Alert Telescope (BAT; \citealt{Barthelmy2005}), and the IPN, which we have made available online\footnote{https://github.com/annayqho/HE\_Burst\_Search}. We did not find any GRB consistent with the position and $t_0$ of \name.

Our tool also determines whether a given position was visible to BAT and GBM at a given time, using the spacecraft pointing history. We use existing code\footnote{ \href{https://github.com/lanl/swiftbat\_python}{https://github.com/lanl/swiftbat\_python}} to determine the BAT history.
We find that the position of \name\ 
was in the BAT field-of-view from UTC 03:13:40 to 03:30:38, before \swift\ slewed to another location.

\begin{figure}[ht]
\centering
\includegraphics[scale=0.14]{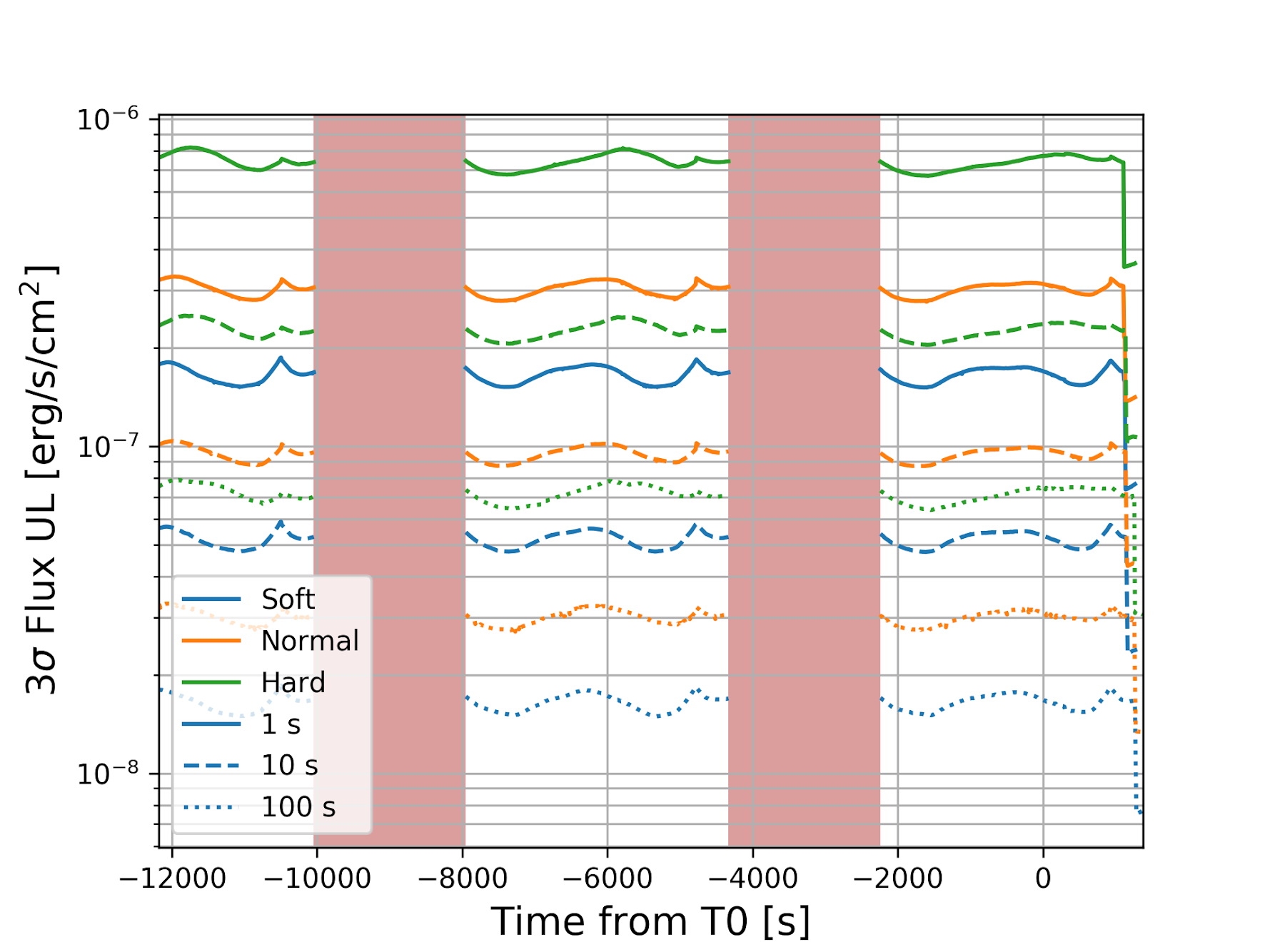}
\caption{3-$\sigma$ upper limits from GBM GRB search, which we performed for three hours prior to $t_0$. The red vertical bars indicate epochs when GBM was not taking data due to passing through the South Atlantic Anomaly.
The time of $t_0$ was estimated from a fit to the early data (Figure \ref{fig:earlydata}),
and is $26 \pm 5$ minutes prior to the first detection.
}
\label{fig:gbm-search}
\end{figure}

We also find that at $t_0$ \name\ was visible 
to the \fermi\ Gamma-Ray Burst Monitor (GBM; \citealt{Meegan2009}).
We ran a targeted GRB search in 10--1000\,\kev\ \fermi/GBM data from three hours prior to $t_0$ to half an hour after $t_0$.
We use the soft template,
which is a smoothly broken power law with low-energy index $-1.9$ and high-energy index $-2.7$, and an SED peak at 70\,\kev.
The search methodology (and parameters of the other templates) are described in \citet{Blackburn2015} and \citet{Goldstein2016}.
No signals with a consistent location were found.
For the 100\,s integration time, the fluence upper limit is $2 \times 10^{-6}\,\erg\,\pcmsq$.
This limit corresponds to a 10--1000\,\kev\ isotropic energy release
of $E_\mathrm{\gamma,iso} < 4.9 \times 10^{48}\,\erg$.
Limits for different spectral templates and integration times are shown in Figure \ref{fig:gbm-search}.

\subsection{Host galaxy data}
\label{obs:host}

We measure line fluxes using the Keck optical spectrum
obtained at $\Delta t\approx61\,\days$ (Figure \ref{fig:host_spectrum}).
We model the local continuum with a low-order polynomial and each emission line by a Gaussian profile of FHWM $\sim5.3$~\AA.
This is appropriate if Balmer absorption is negligible, which is generally the case for starburst galaxies.
For the host of \name, the Balmer decrement between H$\beta$, H$\gamma$ and H$\delta$ does not show any excess with respect to the expected values in \citet{Osterbrock2006}.
The resulting line fluxes are listed in Table \ref{tab:eml_host}.

We retrieved archival images of the host galaxy from \textit{Galaxy Evolution Explorer} (\galex) Data Release (DR) 8/9 \citep{Martin2005}, Sloan Digital Sky Survey (SDSS) DR9 \citep{Ahn2012a}, Panoramic Survey Telescope And Rapid Response System (PanSTARRS, PS1) DR1 \citep{Chambers2016a}, Two-Micron All Sky Survey \citep[2MASS;][]{Skrutskie2006a}, and \textit{Wide-Field Infrared Survey Explorer} \citep[\wise;][]{Wright2010}.
We also used UVOT photometry from \swift,
and NIR photometry from the Canada-France-Hawaii Telescope Legacy Survey \citep[CFHTLS;][]{Hudelot2012a}.

The images are characterized by different pixel scales (e.g., SDSS $0\farcs40$/px, GALEX $1\farcs$/px) and different point spread functions (e.g., SDSS/PS1 1--$2^{\prime\prime}$, \wise/W2 $6\farcs5$). To obtain accurate photometry, we use the matched-aperture photometry software package \package{Lambda Adaptive Multi-Band Deblending Algorithm in R} \citep[\package{LAMBDAR};][]{Wright2016} that is based a photometry software package developed by \citet{Bourne2012a}. To measure the total flux of the host galaxy, we defined an elliptical aperture that encircles the entire galaxy in the SDSS/$r'$-band image. This aperture was then convolved in \package{LAMBDAR} with the point-spread function of a given image that we specified directly (\galex\ and \wise\ data) or that we approximated by a two-dimensional Gaussian (2MASS, SDSS and PS1 images). After instrumental magnitudes were measured, we calibrated the photometry against instrument-specific zeropoints (\galex, SDSS and PS1 data), or as in the case of 2MASS and \wise\ images against a local sequence of stars from the 2MASS Point Source Catalogue and the AllWISE catalogue. The photometry from the UVOT images were extracted with the command \package{uvotsource} in \package{HEAsoft} and a circular aperture with a radius of $8^{\prime\prime}$. The photometry of the CFHT/WIRCAM data was done performed the software tool presented in \citet{Schulze2018b}\footnote{\href{https://github.com/steveschulze/aperture\_photometry}{https://github.com/steveschulze/aperture\_photometry}}. To convert the 2MASS, UVOT, WIRCAM and \wise\ photometry to the AB system, we applied the offsets reported in \citet{Blanton2007a}, \citet{Breeveld2011} and \citet{Cutri2013a}.
The resulting photometry is summarized in Table \ref{tab:host_phot}.

\section{Basic properties of the explosion and its host galaxy}
\label{sec:basic-properties}

The observations we presented in Section \ref{sec:obs} constitute some of the most detailed early-time observations of a stripped-envelope SN to date.
In this section we use this data to derive basic properties of the explosion: the evolution of bolometric luminosity, radius, and effective temperature over time (Section \ref{sec:phys-evol}), the velocity evolution of the photosphere and the density and composition of the ejecta as measured from the spectra (Section \ref{sec:spec-evolution}), and the mass, metallicity, and SFR of the host galaxy (Section \ref{sec:host}).
These properties are summarized in Table \ref{tab:obs-properties}.

\subsection{Physical evolution from blackbody fits}
\label{sec:phys-evol}

\begin{figure}[ht]
\centering
\includegraphics[scale=0.5]{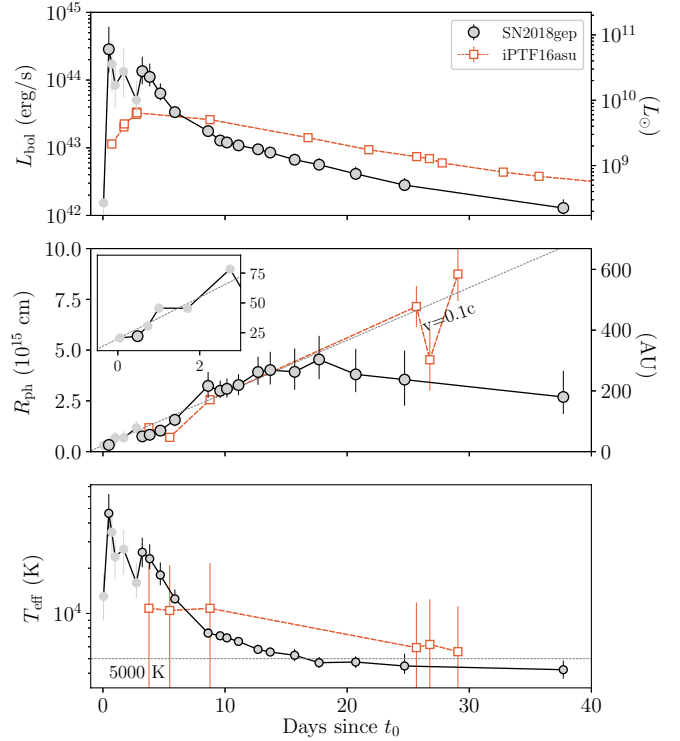}
\caption{Evolution of blackbody properties (luminosity, radius, temperature) over time compared to the Ic-BL SN iPTF16asu and the luminous fast-rising optical transient AT2018cow.
The light gray circles are derived from optical data only.
The outlined circles are derived from UV and optical data.
Middle panel: dotted line shows $v=0.1c$.
Note that $R\neq0$ at $t_0$, and instead $R(t=0)=3 \times 10^{14}\,\cm$.
Due to the scaling of our plot we do not show the radius evolution of AT2018cow,
which drops from $8 \times 10^{14}\,\cm$ to $10^{14}\,\cm$ on this timescale.
Bottom panel: dotted horizontal line shows 5000\,\kelvin, the recombination temperature for carbon and oxygen.
Once this temperature is reached, the photosphere flattens out (and potentially begins to recede).}
\label{fig:physevol}
\end{figure}

By interpolating the UVOT and ground-based photometry, we construct multi-band SEDs and fit a Planck function on each epoch, to measure the evolution of luminosity, radius, and effective temperature.
To estimate the uncertainties, we perform a Monte Carlo simulation with 600 trials, each time adding noise corresponding to a 15\% systematic uncertainty on each data point,
motivated by the need to obtain a combined $\chi^2$/dof $\sim$ 1 across all epochs.
The uncertainties for each parameter are taken as the 16-to-84 percentile range from this simulation.
The SED fits are shown in Appendix \ref{appendix:uv-opt-phot},
and the resulting evolution in bolometric luminosity, photospheric radius, and effective temperature is listed in Table \ref{tab:physevol}.
We plot the physical evolution in
Figure \ref{fig:physevol}, with a comparison to iPTF16asu and AT2018cow.

\begin{figure}[ht]
\centering
\includegraphics[scale=0.1]{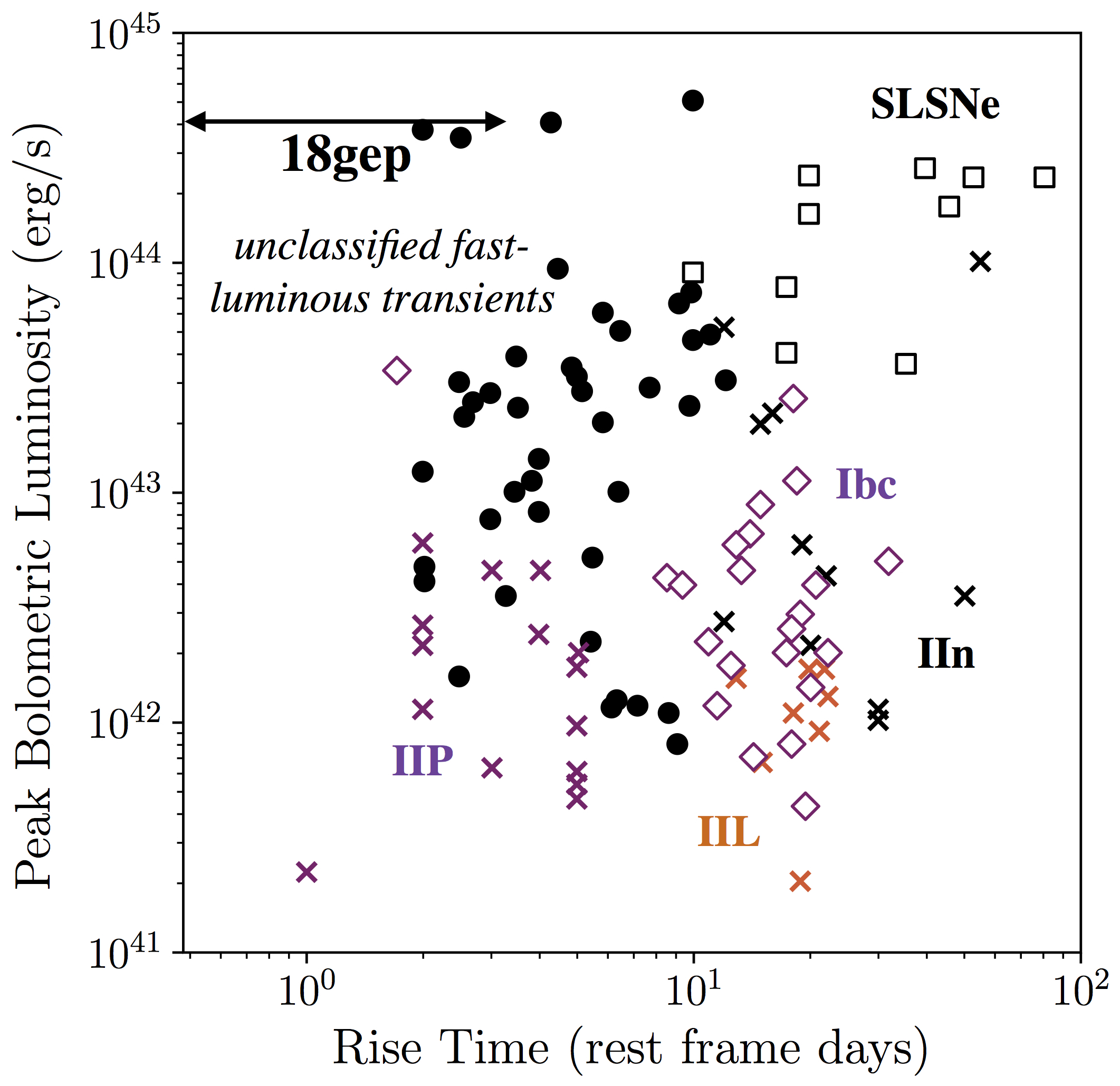}
\caption{Rise to peak bolometric luminosity compared to other classes of transients. Modified from Figure 1 in \citet{Margutti2019}.
}
\label{fig:lbolrise}
\end{figure}

The bolometric luminosity peaks between $\Delta t=0.5\,\days$ and $\Delta t=3\,\days$, at $>3 \times 10^{44}\,\erg\,\psec$.
In Figure \ref{fig:lbolrise} we compare this peak luminosity and time to peak luminosity with several classes of stellar explosions.
As in iPTF16asu, the bolometric luminosity falls as an exponential at late times ($t > 10\,\days$).
The total integrated UV and optical ($\approx 2000$--$9000$\AA) blackbody energy output from $\Delta t=0.5$--40\,\days\ is $\sim 10^{50}\,\erg$,
similar to that of iPTF16asu.

The earliest photospheric radius we measure is $\sim20\,$AU, at $\Delta t=0.05\,\days$.
Until $\Delta t \approx 17\,\days$
the radius expands over time with a very large inferred velocity of $v\approx0.1c$.
After that, it remains flat, and even appears to recede.
This possible recession corresponds to a flattening in the temperature at $\sim 5000\,\kelvin$,
which is the recombination temperature of carbon and oxygen.
This effect was not seen in iPTF16asu, which remained hotter (and more luminous) for longer.
Finally, the effective temperature rises before falling as $\sim t^{-1}$.
We interpret these properties in the context of shock-cooling emission in Section \ref{sec:interpretation}.

\begin{table}[!ht] 
\centering 
\caption{Physical evolution of AT2018gep from blackbody fits.} 
\begin{tabular}{lrrr} 
\hline 
$\Delta t$ & $L (10^{10} L_\odot)$ & $R$ (AU) & $T$ (kK) \\ 
\hline$0.05$ & $0.04^{+0.04}_{-0.02}$ & $21^{+14}_{-6}$ & $13^{+5}_{-4}$ \\ 
$0.48$ & $7.4^{+8.6}_{-4.1}$ & $22^{+7}_{-5}$ & $46^{+16}_{-13}$ \\ 
$0.73$ & $4.5^{+5.5}_{-2.8}$ & $31^{+11}_{-6}$ & $35^{+12}_{-11}$ \\ 
$1.0$ & $2.2^{+2.1}_{-1.2}$ & $46^{+18}_{-9}$ & $24^{+6}_{-6}$ \\ 
$1.7$ & $3.5^{+4.2}_{-2.1}$ & $46^{+22}_{-10}$ & $27^{+9}_{-8}$ \\ 
$2.7$ & $1.3^{+1.2}_{-0.4}$ & $78^{+22}_{-20}$ & $16^{+5}_{-3}$ \\ 
$3.2$ & $3.5^{+2.2}_{-1.3}$ & $50^{+14}_{-8}$ & $26^{+6}_{-5}$ \\ 
$3.8$ & $2.9^{+1.7}_{-0.8}$ & $56^{+11}_{-11}$ & $23^{+5}_{-3}$ \\ 
$4.7$ & $1.7^{+0.7}_{-0.3}$ & $69^{+16}_{-14}$ & $18^{+3}_{-2}$ \\ 
$5.9$ & $0.88^{+0.17}_{-0.08}$ & $100^{+14}_{-21}$ & $13^{+1}_{-0}$ \\ 
$8.6$ & $0.46^{+0.08}_{-0.06}$ & $220^{+46}_{-39}$ & $7.4^{+0.6}_{-0.5}$ \\ 
$9.6$ & $0.33^{+0.04}_{-0.03}$ & $200^{+33}_{-24}$ & $7.1^{+0.4}_{-0.4}$ \\ 
$10.0$ & $0.31^{+0.04}_{-0.03}$ & $210^{+34}_{-28}$ & $6.9^{+0.4}_{-0.4}$ \\ 
$11.0$ & $0.28^{+0.04}_{-0.03}$ & $220^{+35}_{-33}$ & $6.5^{+0.4}_{-0.3}$ \\ 
$13.0$ & $0.25^{+0.04}_{-0.03}$ & $260^{+50}_{-42}$ & $5.8^{+0.3}_{-0.3}$ \\ 
$14.0$ & $0.22^{+0.04}_{-0.03}$ & $270^{+60}_{-47}$ & $5.5^{+0.4}_{-0.3}$ \\ 
$16.0$ & $0.17^{+0.04}_{-0.03}$ & $260^{+76}_{-58}$ & $5.3^{+0.5}_{-0.5}$ \\ 
$18.0$ & $0.15^{+0.04}_{-0.02}$ & $300^{+77}_{-64}$ & $4.7^{+0.4}_{-0.4}$ \\ 
$21.0$ & $0.11^{+0.03}_{-0.02}$ & $250^{+83}_{-58}$ & $4.7^{+0.4}_{-0.4}$ \\ 
$25.0$ & $0.073^{+0.02}_{-0.013}$ & $240^{+95}_{-85}$ & $4.5^{+0.9}_{-0.5}$ \\ 
$38.0$ & $0.034^{+0.012}_{-0.007}$ & $180^{+86}_{-55}$ & $4.2^{+0.6}_{-0.5}$ \\ 
\hline 
\end{tabular} 
\label{tab:physevol} 
\end{table} 

\subsection{Spectral evolution and velocity measurements}
\label{sec:spec-evolution}

\subsubsection{Comparisons to early spectra in the literature}

\begin{figure*}[ht]
\centering
\includegraphics[scale=0.12]{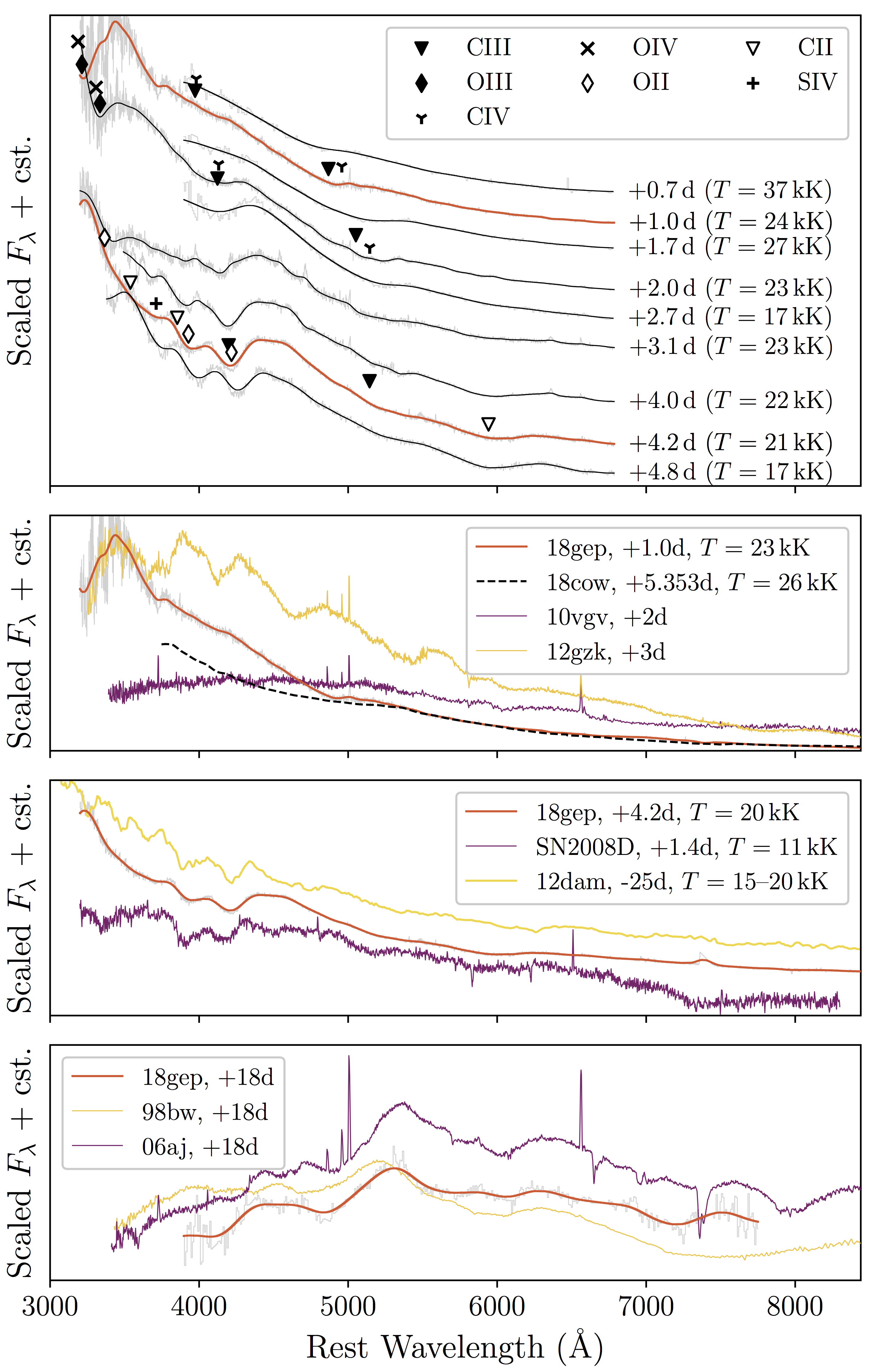}
\caption{Top panel: spectra of \name\ taken in the first five days. Broad absorption features are consistent with ionized carbon and oxygen, which evolve redward with time.
Second-from-top panel: an early spectrum of 18gep compared to spectra from other stellar explosions at a comparable phase.
Second-from-bottom panel: The spectrum at $\Delta t=4.2\,\days$ shows a ``W'' feature, which we compare to similar ``W'' features seen in an early spectrum of SN2008D from \cite{Modjaz2009}, and a typical pre-max spectrum of a SLSN-I (PTF12dam, from \citealt{Nicholl2013}).
We boost the SLSN spectrum by an additional expansion velocity of $\sim15000$\,km\,s$^{-1}$,
and apply reddening of $E(B-V)=0.63$ to SN\,2008D.
Weak features in the red are also similar to what are seen in PTF12dam, and are consistent with arising from CII and CIII lines, following the analysis of \cite{AGY2018}. The lack of narrow carbon features as well as the smooth spectrum below 3700\,\AA\ suggest a large velocity dispersion leading to significant line broadening, compared to the intrinsically narrow features observed in SLSNe-I \citep{AGY2018,Quimby2018}.
Bottom panel: a spectrum of 18gep when it resembled an ordinary Ic-BL SN, compared to spectra at similar phases of Ic-BL SNe accompanying GRBs.
}
\label{fig:earlyspec}
\end{figure*}

We obtained nine spectra of \name\ in the first five days after discovery.
These early spectra are shown in
Figure \ref{fig:earlyspec}, when the effective temperature declined from 50,000\,\kelvin\ to 20,000\,\kelvin.
To our knowledge, our early spectra have no analogs in the literature,
in that there has never been a spectrum of a stripped-envelope SN at such a high temperature (excluding spectra during the afterglow phase of GRBs).\footnote{There is however a spectrum of a Type II SN at a comparable temperature:
iPTF13dqy was $\sim 50,000\,\kelvin$ at the time of the first spectrum \citep{Yaron2017}.}
Two of the earliest spectra in the literature, one at $\Delta t=2\,\days$ for Type Ic SN PTF10vgv \citep{Corsi2012} and one at $\Delta t=3\,\days$ for Type Ic SN PTF12gzk \citep{BenAmi2012}
are redder and exhibit more features than the spectrum of \name. We show the comparison in Figure \ref{fig:earlyspec}.

At $\Delta t\approx4\,\days$,
a ``W'' feature emerges in the rest-frame wavelength range 3800--4350\,\AA.
In the second-from-bottom panel of Figure \ref{fig:earlyspec}
we make a comparison to ``W'' features seen
in SN\,2008D (e.g. \citealt{Modjaz2009}),
which was a Type Ib SN associated with an X-ray flash \citep{Mazzali2008},
and in a typical pre-max
stripped-envelope superluminous supernova (Type I SLSN; \citealt{Moriya2018,GalYam2018}).
The absorption lines are broadened much more than in PTF12dam \citep{Nicholl2013} and probably more than in SN2008D as well.
Finally, \name\ cooled more slowly than SN 2008D:
only after 4.25 days did it reach the temperature that SN\,2008D reached after $<2$\,days.

\subsubsection{Origin of the ``W'' feature}

The lack of comparison data at such early epochs (high temperatures) motivated us to model one of the early spectra, in order to determine the composition and density profile of the ejecta.
We used the spectral synthesis code JEKYLL \citep{Ergon2018}, configured to run in steady-state using a full NLTE-solution. An inner blackbody boundary was placed at an high continuum optical depth ($\sim$50), and the temperature at this boundary was iteratively determined to reproduce the observed luminosity. The atomic data used is based on what was specified in \citet{Ergon2018}, but has been extended as described in Appendix \ref{appendix:atomic-data}.
We explored models with C/O (mass fractions: 0.23/0.65) and O/Ne/Mg (mass fractions: 0.68/0.22/0.07) compositions taken from a model by \citet{Woosley2007}\footnote{ The model was divided into compositional zones by \citet{Jer2015} and a detailed specification of the C/O and O/Ne/Mg zones is given in Table D.2 therein.} and a power-law density profile, where the density at the inner border was adjusted to fit the observed line velocities. Except for the density at the inner border, various power-law indices where also explored, but in the end an index of -9 worked out best.

\begin{figure}[ht]
\centering
\includegraphics[scale=0.4]{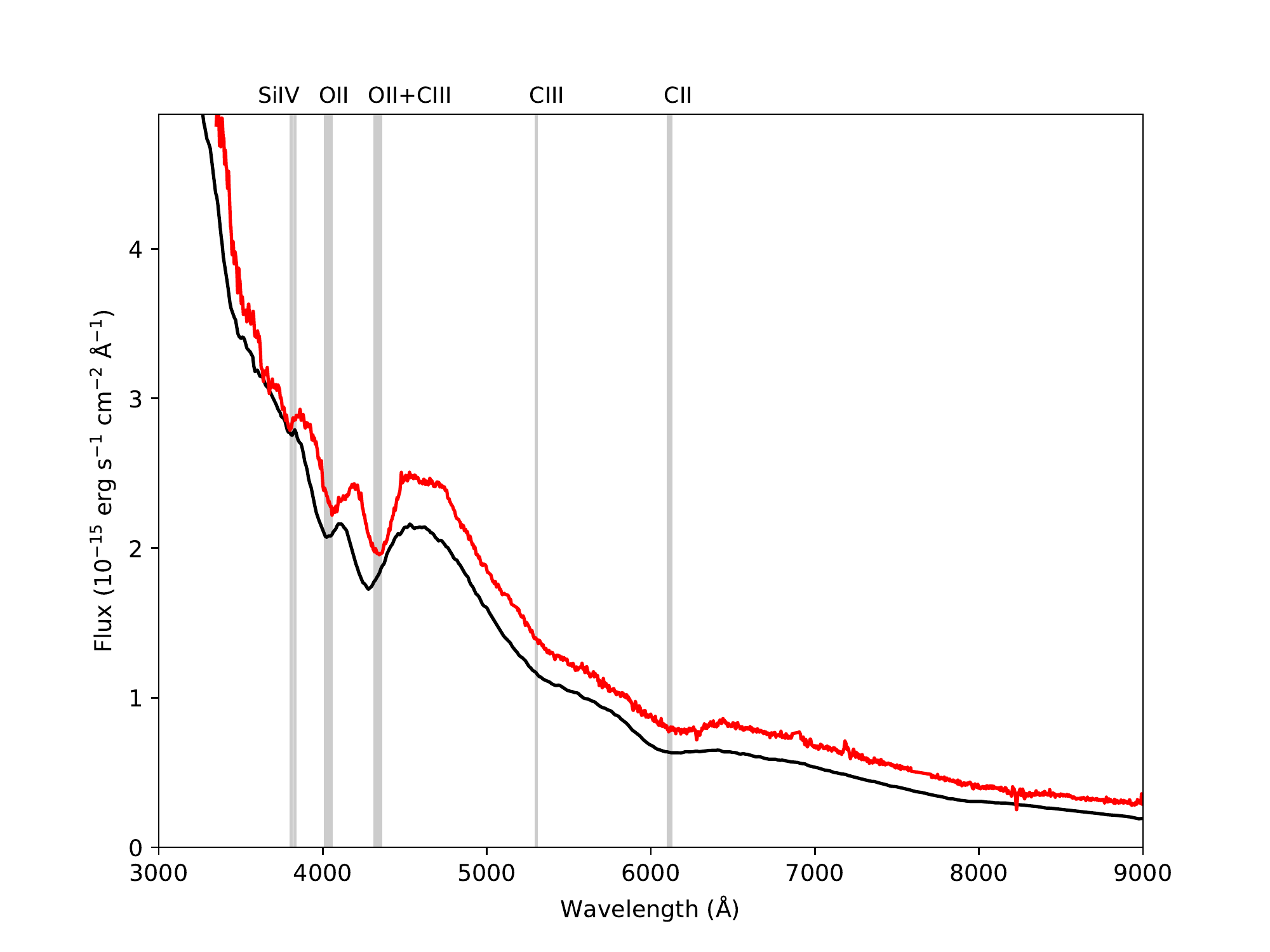}
\caption{
Observed spectrum (red) at 4.2\,\days,
compared to our model spectrum (black)
from the spectral synthesis code JEKYLL configured to run in steady-state using a full NLTE solution.
The model has a C/O composition, an inner border at 22,000\,\km\,\psec, a density of $4 \times 10^{-12}\,\gm\,\pcmcub$, and a density profile with a power-law index of $-9$.
The absolute (but not relative) flux of the spectrum was calibrated using the interpolated P48 g and r magnitudes. We also show the \ion{O}{2}, \ion{C}{2}, \ion{C}{3} and \ion{Si}{4} lines discussed in the text shifted to the velocity of the model photosphere.}
\label{fig:spec_model_comp}
\end{figure}

\begin{figure}[ht]
\centering
\includegraphics[scale=0.4]{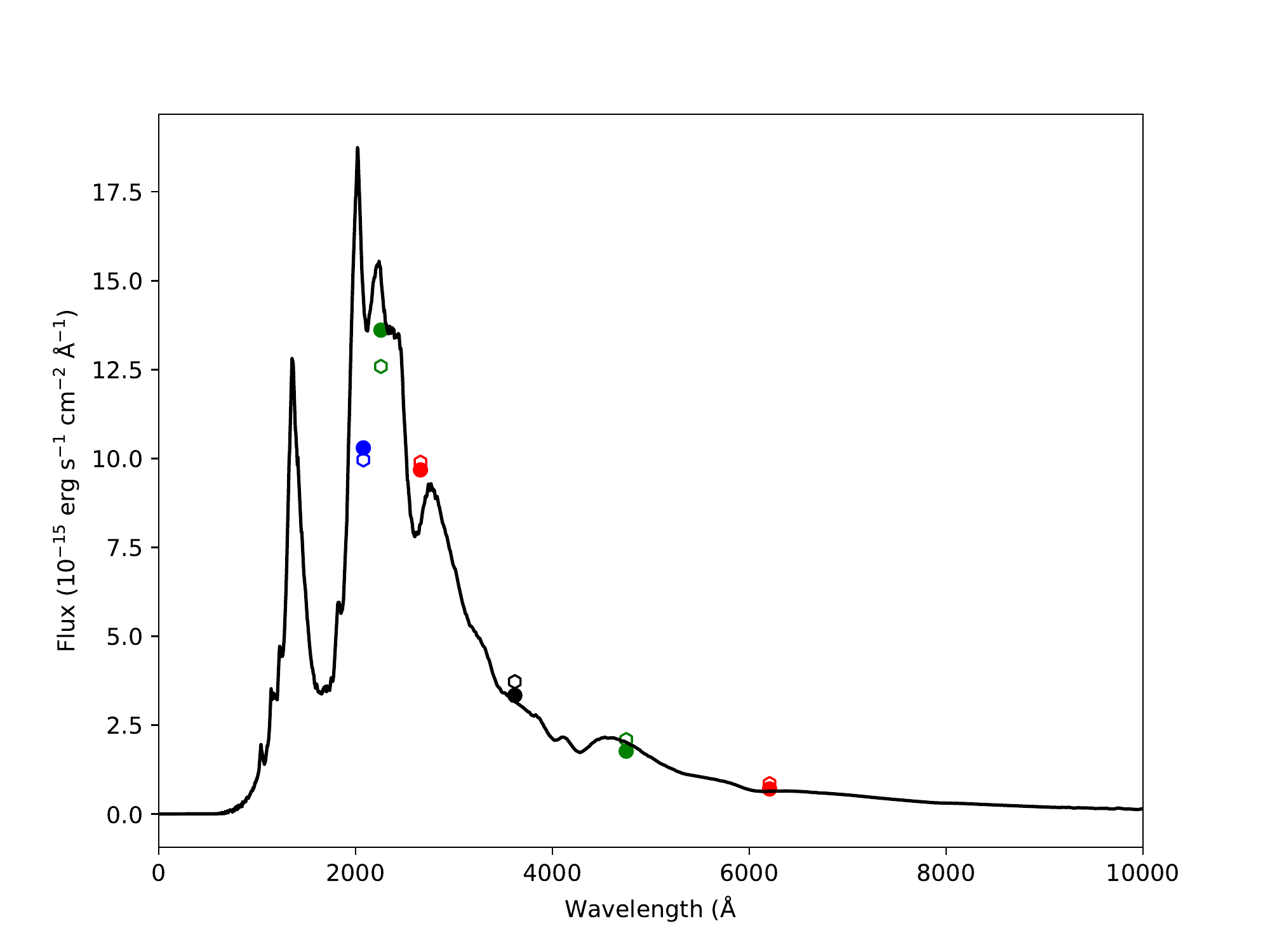}
\caption{Comparison of model (filled circles) and observed (unfilled circles) mean fluxes through the \swift\ UVW1 (blue), UVM2 (green), UVW2 (red), and the SDSS u (black), g (green) and r (red) filters. We also show the model spectrum in black.}
\label{fig:sed_model_comp}
\end{figure}

Figures \ref{fig:spec_model_comp} and \ref{fig:sed_model_comp} show the model with the best overall agreement with the spectra and the SED (as listed in Table \ref{tab:opt-spec} the spectrum was obtained at high airmass, making it difficult to correct for telluric features). The model has a C/O composition, an inner border at 22,000\,\km\,\psec\ (corresponding to an optical depth of $\sim$50), a density of 4$\times$10$^{-12}$ g cm$^{-3}$ at this border and a density profile with a power-law index of $-9$. In Figure \ref{fig:spec_model_comp} we show that the model does a good job of reproducing both the spectrum and the SED of \name. In particular, it is interesting to note that the ``W" feature seem to arise naturally in C/O material at the observed conditions. 
A similar conclusion was reached by \citet{Dessart2019}, whose magnetar-powered SLSN-I models, calculated using the NLTE code CMFGEN, show the ``W" feature even when non-thermal processes where not included in the calculation (as in our case).

In the model, the ``W" feature mainly arises from the \ion{O}{2} 2p$^2$(3P)3s 4P $\leftrightarrow$ 2p$^2$(3P)3p 4D$^\circ$ (4639--4676 \AA), \ion{O}{2} 2p$^2$(3P)3s 4P $\leftrightarrow$ 2p$^2$(3P)3p 2D (4649 \AA) and \ion{O}{2} 2p$^2$(3P)3s 4P $\leftrightarrow$ 2p$^2$(3P)3p 4P$^\circ$ (4317--4357\,\AA) transitions.
The departure from LTE is modest in the line-forming region,
and the departure coefficients for the \ion{O}{2} states are small. The spectrum redward of the ``W" feature is shaped by carbon lines, and the features near 5700 and 6500\,\AA~arise from the \ion{C}{2} 3s 2S $\leftrightarrow$ 3p 2P$^\circ$ (6578,6583\,\AA) and \ion{C}{3} 2s3p 1P$^\circ$ $\leftrightarrow$ 2s3d 1D (5696\,\AA) transitions, respectively. In the model, the \ion{C}{2} feature is too weak, suggesting that the ionization level is too high in the model. There is also a contribution from the \ion{C}{3} 2s3s 3S $\leftrightarrow$ 2s3p 3P$^\circ$ (4647--4651\,\AA) transition to the red part of the ``W" feature, which could potentially be what is seen in the spectra from earlier epochs. In addition, there is a contribution from \ion{Si}{4} 4s 2S $\leftrightarrow$ 4p 2P$^\circ$ (4090, 4117\,\AA) near the blue side of the ``W" feature, which produce a distinct feature in models with lower velocities and which could explain the observed feature on the blue side of the ``W" feature.

In spite of the overall good agreement, there are also some differences between the model and the observations. In particular the model spectrum is bluer and the velocities are higher. These two quantities are in tension and a better fit to one of them would result in a worse fit to the other. As mentioned above, the ionization level might be too high in the model, which suggests that the temperature might be too high as well. It should be noted that adding host extinction (which is assumed to be zero) or reducing the distance (within the error bars) would help in making the model redder (in the observer frame), and the latter would also help in reducing the temperature. The (modest) differences between the model and the observations could also be related to physics not included in the model, like a non-homologous velocity field, departures from spherical asymmetry, and clumping.

The total luminosity of the model is 6.2$\times$10$^{43}$ erg s$^{-1}$, the photosphere is located at $\sim$33,000\,\km\,\psec\ and the temperature at the photosphere is $\sim$17,500\,\kelvin, which is consistent with the values estimated from the blackbody fits (although the blackbody radius and temperature fits refer to the thermalization layer). As mentioned, we have also tried models with a O/Ne/Mg composition.
However, these models failed to reproduce the carbon lines redwards of the ``W" feature. We therefore conclude that the (outer) ejecta probably has a C/O-like composition, and that this composition in combination with a standard power-law density profile reproduce the spectrum of \name\ at the observed conditions (luminosity and velocity) 4.2 days after explosion. 

In our model, the broad feature seen in our \swift\ UVOT grism spectrum is dominated by the strong \ion{Mg}{2} (2796,2803\,\AA) resonance line.
However, a direct comparison is not reliable because the ionization is probably lower at this epoch than what we consider for our model.

\subsubsection{Photospheric velocity from Ic-BL spectra}

At $\Delta t\gtrsim7.8\,\days$, the spectra of \name\ qualitatively resemble those of a stripped-envelope SN. We measure velocities using the method in \citet{Modjaz2016},
which accommodates blending
of the \ion{Fe}{2}$\lambda$5169 line 
(which has been shown to be a good tracer of photospheric velocity; \citealt{Branch2002}) with 
the nearby \ion{Fe}{2}$\lambda\lambda$4924,5018 lines.

\begin{figure}[ht]
\centering
\includegraphics[scale=0.55]{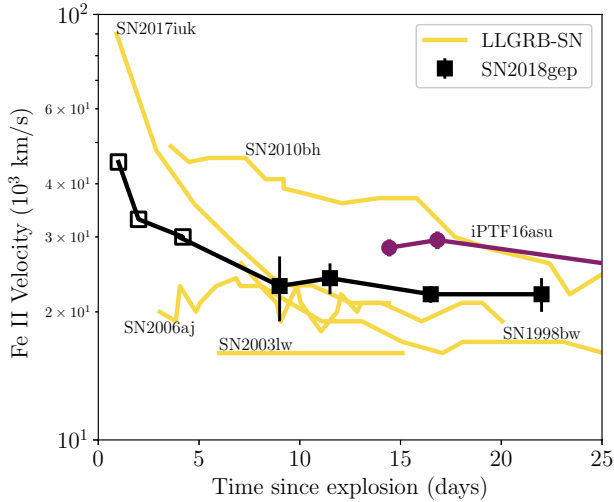}
\caption{Velocity evolution over time as measured from spectral absorption features. Open symbols for \name\ come from C/O velocities measured from line minima.
Closed symbols come from the \ion{Fe}{2} feature in the Ic-BL spectra.
The velocities are comparable to those measured for Ic-BL SNe associated with low-luminosity GRBs (LLGRBs).
The velocity evolution for SN2017iuk is taken from \citet{Izzo2019}.
Velocities for iPTF16asu are taken from \citet{Whitesides2017}.
Velocities for the other Ic-BL SNe are taken from \citet{Modjaz2016}
and shifted from V-band max using data from \citet{Galama1998}, \citet{Campana2006}, \citet{Malesani2004}, and \citet{Bufano2012}.
}
\label{fig:feiivel}
\end{figure}

At earlier times, when the spectra do not resemble typical Ic-BL SNe, we use our line identifications of ionized C and O to measure velocities.
As shown in Figure \ref{fig:feiivel},
the velocity evolution we measure is comparable to that seen in Ic-BL SNe associated with GRBs (more precisely, low-luminosity GRBs; LLGRBs)
which are systematically higher than those of Ic-BL SNe lacking GRBs \citep{Modjaz2016}.
However, as discussed in Section 
\ref{sec:obs-grbsearch}, no GRB was detected.

\subsection{Properties of the host galaxy}
\label{sec:host}

We infer a star-formation rate of $0.09\pm0.01~M_\odot\,{\rm yr}^{-1}$ from the H$\alpha$ emission line using the \citet{Kennicutt1998a} relation converted to use a Chabrier initial mass function \citep{Chabrier2003a, Madau2014}. We note that this is a lower limit as the slit of the Keck observation did not enclose the entire galaxy.
We estimate a correction factor of 2--3:
the slit diameter in the Keck spectra was 1.0'', and the extraction radius was $\sim1.75''$ in the February observation and $\sim1.21''$ in the March observation.
The host diameter is roughly 4''.

We derive an electron temperature of $13,100^{+900}_{-1000}\,\kelvin$ from the flux ratio between {[\ion{O}{3}]}$\lambda$4641 and {[\ion{O}{3}]}$\lambda$5007, using the software package \package{PyNeb} version 1.1.7 \citep{Luridiana2015}. In combination with the flux measurements of {[\ion{O}{2}]}$\lambda\lambda$3226,3729, {[\ion{O}{3}]}$\lambda$4364, {[\ion{O}{3}]}$\lambda$4960, {[\ion{O}{3}]}$\lambda$5008, and H$\beta$, we infer a total oxygen abundance of $8.01^{+0.10}_{-0.09}$ (statistical error; using Eqs. 3 and 5 in \citealt{Izotov2006}). Assuming a solar abundance of 8.69 \citep{Asplund2009}, the metallicity of the host is $\sim20\%$ solar.

We also compute the oxygen abundance using the strong-line metallicity indicator O3N2 \citep{Pettini2004} with the updated calibration reported in \citet{Marino2013}. The oxygen abundance in the O3N2 scale is $8.05\pm0.01\,({\rm stat})\pm0.10\,({\rm sys})$.\footnote{Note, the oxygen abundance  of {\name}'s host lies outside of the domain calibrated by \citet{Marino2013}. However, we will use the measurement from the O3N2 indicator only to put the host in context of other galaxy samples that are on average more metal-enriched.}

We also estimate mass and star-formation rate
by modeling the host SED;
see Appendix \ref{appendix:host-data} for a table of measurements, and details on where we obtained them.
We use the software package \package{LePhare} version 2.2 \citep{Arnouts1999a,Ilbert2006a}. We generated $3.9\times10^6$ templates based on the \citet{Bruzual2003a} stellar population-synthesis models with the Chabrier initial mass function \citep[IMF;][]{Chabrier2003a}. The star formation history (SFH) was approximated by a declining exponential function of the form $\exp \left(−t/\tau\right)$, where $t$ is the age of the stellar population and $\tau$ the e-folding time-scale of the SFH (varied in nine steps between 0.1 and 30 Gyr). These templates were attenuated with the Calzetti attenuation curve \citep{Calzetti2000} varied in 22 steps from $E(B-V)=0$ to 1 mag . 

\begin{figure}[!ht]
    \centering
    \includegraphics[width=1\columnwidth]{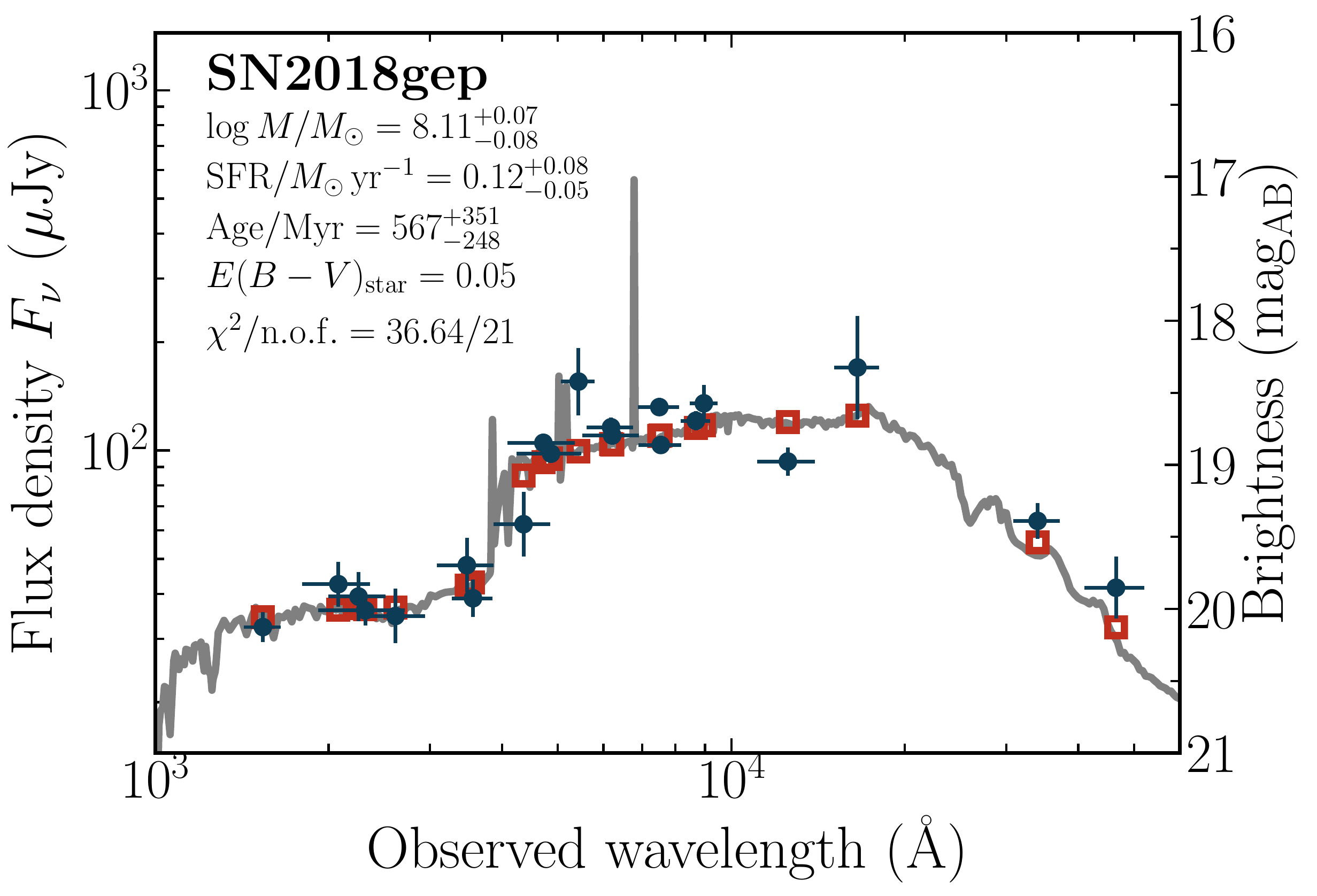}
    \caption{The spectral energy distribution of the host galaxy of \name\ from 1,000 to 60,000~\AA\ and the best fit (solid line) in the observer frame. Filled data points represent photometric measurements. The error bars in the `x' direction indicate the full-width half maximum of each filter response function. The open data points signify the model-predicted magnitudes. The quoted values of the host properties represent the median values and the corresponding 1-$\sigma$ errors.
    }
    \label{fig:host_sed}
    \end{figure}

As shown in Figure \ref{fig:host_sed}, the SED is well characterized by a galaxy mass of $\log M/M_\odot = 8.11^{+0.07}_{-0.08}$ and an attenuation-corrected star-formation rate of $0.12^{+0.08}_{-0.05}\,M_\odot\,\pyr$. The derived star-formation rate is comparable to measurement inferred from H$\alpha$. The attenuation of the SED is marginal, with $E(B-V)_{\rm star}=0.05$, and consistent with the negligible Balmer decrement \ref{obs:host}.

\begin{figure}[]
    \centering
    \includegraphics[width=0.9\columnwidth]{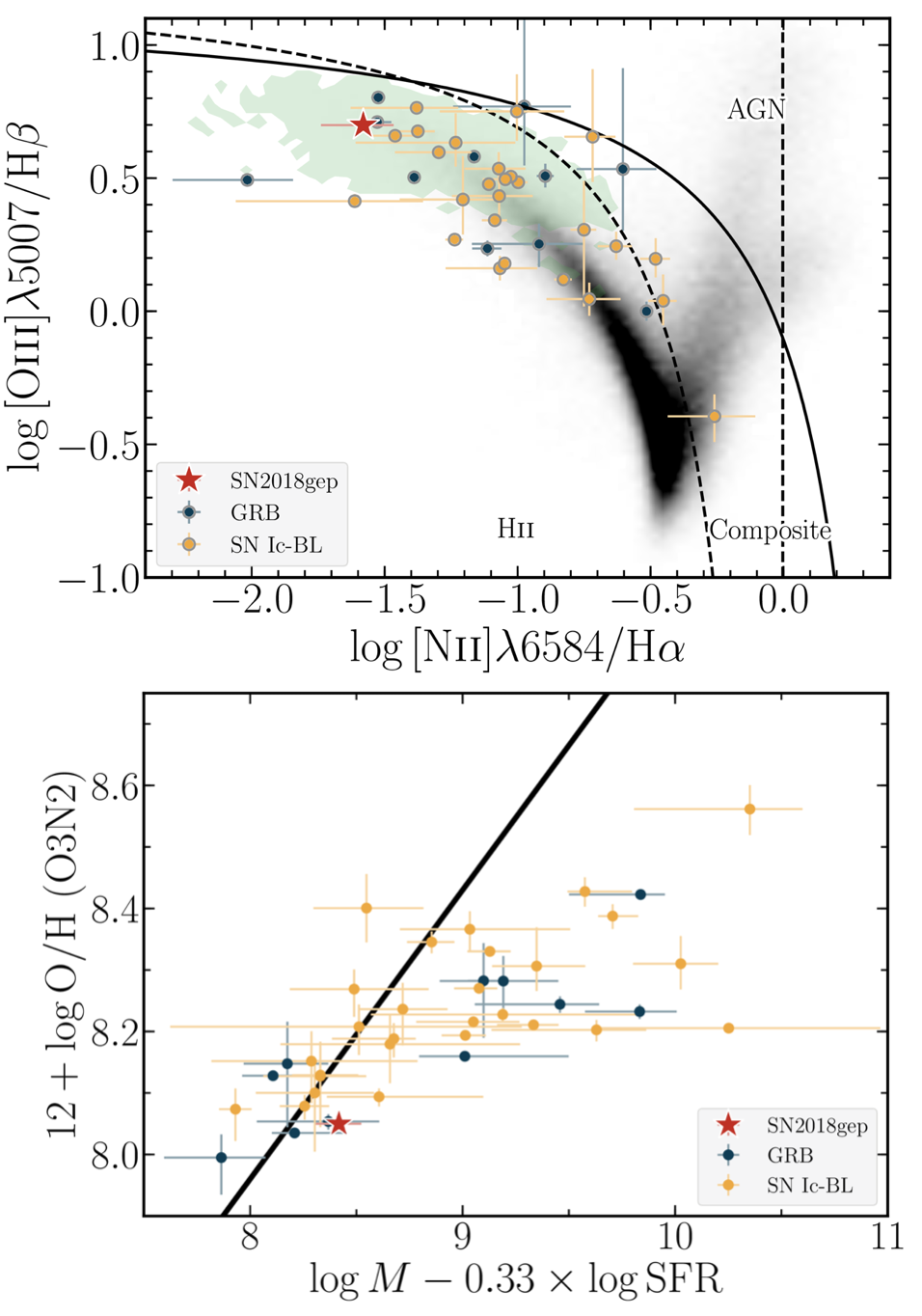}
    \caption{
    Top: BPT diagram. The host of \name\ is a low-metallicity galaxy with an intense ionizing radiation field (green shaded region indicates extreme emission line galaxies).
    The majority of Ic-BL SNe and long-duration GRBs are found in more metal enriched galaxies (parameterized by [\ion{N}{2}]/H$\alpha$), and galaxies with less intense radiation fields (parameterized by [\ion{O}{3}]/H$\alpha$). Field galaxies from SDSS DR15 are shown as a background density distribution.
    The thick solid line separates star formation- and AGN-dominated galaxies \citep{Kewley2001a}.
    The thick dashed lines encircle the region of composite galaxies \citep{Kauffmann2003a}.
    Bottom: The mass-metallicity-star-formation-rate plane. The bulk of the the SN-Ic-BL and GRB host populations are found in hosts that are more metal enriched. For reference, the host of AT2018cow had $\log{M}-0.33 \times \log{\mathrm{SFR}}\approx9.4$.
    The black line is the fundamental metallicity relation in \citet{Mannucci2010}.
    }
    \label{fig:host_comparison}
\end{figure}

Figure \ref{fig:host_comparison} shows that
the host galaxy of \name\ is even more low-mass and metal-poor than the typical host galaxies of Ic-BL SNe,
which are low-mass and metal-poor compared to the overall core collapse SN population to begin with.
The figure uses data for 28 Ic-BL SNe from PTF and iPTF \citep{Modjaz2019, Taddia2019}
and a sample of 11 long-duration GRBs (including LLGRBs, all at $z<0.3$).
We measured the emission lines from the spectra presented in \citet{Taddia2019} and used line measurements reported in \citet{Modjaz2019} for objects with missing line fluxes. The photometry was taken from Schulze, S. et al. (in preparation). Photometry and spectroscopy were taken from a variety of sources\footnote{\citet{Gorosabel2005}, \citet{Bersier2006}, \citet{Margutti2007}, \citet{Ovaldsen2007} \citet{Kocevski2007}, \citet{Thoene2008}, \citet{Michalowski2009}, \citet{Han2010}, \citet{Levesque2010}, \citet{Starling2011}, \citet{Hjorth2012}, \citet{Thoene2014}, \citet{Schulze2014}, \citet{Kruehler2015}, \citet{Stanway2015}, \citet{Toy2016}, \citet{Izzo2017}, and \citet{Cano2017b}}. The oxygen abundances were measured in the O3N2 scale like for \name\ and their SEDs were modelled with the same set of galaxy templates.
For reference, the mass and SFR of the host of AT2018cow was $1.4 \times 10^{9}\,\msol$ and $0.22\,\msol\,\pyr$, respectively \citep{Perley2019cow}.
The mass and SFR of the host of iPTF16asu was $4.6^{+6.5}_{-2.3} \times 10^{8}\,\msol$ and $0.7\,\msol\,\pyr$, respectively \citep{Whitesides2017}.

\section{Interpretation}
\label{sec:interpretation}

In Sections \ref{sec:obs} and \ref{sec:basic-properties},
we presented our observations and basic inferred properties of \name\ and its host galaxy.
Now we consider what we can learn about the progenitor,
beginning with the power source for the light curve.

\subsection{Radioactive decay}

The majority of stripped-envelope SNe have light curves powered by the radioactive decay of \nickel.
As discussed in \citet{Kasen2017}, this mechanism can be ruled out for light curves that rise rapidly to a high peak luminosity,
because this would require the unphysical condition of a nickel mass that exceeds the total ejecta mass.
With a peak luminosity exceeding $10^{44}\,\erg\,\psec$ and a rise to peak of a few days, \name\ clearly falls into the disallowed region (see Figure 1 in \citealt{Kasen2017}).
Thus, we rule out radioactive decay as the mechanism powering the peak of the light curve.

\begin{figure}[ht]
\centering
\includegraphics[width=1.0\columnwidth]{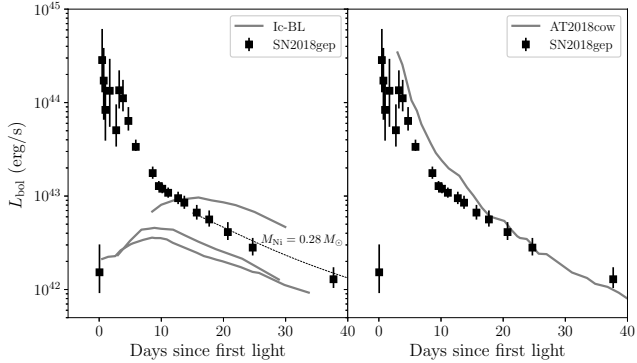}
\caption{
The bolometric light curve of \name\ compared to (left) other Ic-BL SNe from the literature \citep{Cano2013} and (right) to AT2018cow \citep{Perley2019cow}.
The dotted line shows the expected contribution from the radioactive decay of \nickel, for a gamma-ray escape time of 30\,\days\ and \mni=0.28\,\msol.
In order of decreasing \lbol,
the three Ic-BL SNe are SN1998bw,
SN2010bh, and SN2006aj.
}
\label{fig:bol-lc-comparison}
\end{figure}

We now consider whether radioactive decay could dominate the light curve at late times ($t\gg\tpeak$).
The left panel of
Figure \ref{fig:bol-lc-comparison}
shows the bolometric light curve of \name\ compared to several other Ic-BL SNe from the literature \citep{Cano2013}, whose light curves are thought to be dominated by the radioactive decay of \nickel\ (although see \citet{Moriya2017} for another possible interpretation).
The luminosity of \name\ at $t\sim20\,\days$ is about half that of SN1998bw, and double that of SN2010bh and SN2006aj.
By modeling the light curves of the three Ic-BL SNe shown, \citet{Cano2013} infers nickel masses of 0.42\,\msol, 0.12\,\msol, and 0.21\,\msol, respectively.
On this scale, \name\ has $\mni\sim0.1$--$0.2\,\msol$.

The right panel of Figure \ref{fig:bol-lc-comparison} shows the light curve of \name\ compared to that of
AT2018cow \citep{Perley2019cow}.
To estimate the nickel mass of AT2018cow,
\citet{Perley2019cow} compared the bolometric luminosity at $t\sim20\,\days$ to that of SN2002ap (whose nickel mass was derived via late-time nebular spectroscopy; \citealt{Foley2003})
and found $\mni<0.05\,\msol$.
On this scale, we would expect $\mni\lesssim0.05\,\msol$ for \name\ as well.

Finally, \citet{Katz2013} and \citet{Wygoda2019}
present an analytical technique for testing whether a light curve is powered by radioactive decay.
At late times, the bolometric luminosity is equal to the rate of energy deposition by radioactive decay $Q(t)$, because the diffusion time is much shorter than the dynamical time: $\lbol(t) = Q(t)$.
At any given time, the energy deposition rate $Q(t)$ is 

\begin{equation}
\label{eq:qdep}
    Q(t) = Q_\gamma(t) \left( 1-e^{-(t_0/t)^2} \right)
    + Q_\mathrm{pos}(t)
\end{equation}

\noindent where $Q_\gamma(t)$ is the energy release rate of gamma-rays and $t_0$ is the time at which the ejecta becomes optically thin to gamma rays.
The expression for $Q_\gamma(t)$ is

\begin{equation}
    \frac{Q_\gamma(t) }{10^{43}\,\erg\,\psec}
    = \frac{\mni}{\msol} 
    \left( 6.45 e^{-t/8.76\,\mathrm{d}} + 1.38 e^{-t/111.4\,\mathrm{d}} \right).
\end{equation}

$Q_\mathrm{pos}(t)$ is the energy deposition rate of positron kinetic energy, and the expression is

\begin{equation}
    \frac{Q_\mathrm{pos}(t)}{10^{41}\,\erg\,\psec} = 4.64 \frac{\mni}{\msol} 
    \left(-e^{-t/8.76\,\mathrm{d}} + e^{-t/111.4\,\mathrm{d}} \right).
\end{equation}

The dotted line in Figure \ref{fig:bol-lc-comparison} shows a model track with $\mni=0.28\,M_\odot$ and $t_0=30\,\days$.
Lower nickel masses produce tracks that are too low to reproduce the data, and larger values of $t_0$ produce tracks that drop off too rapidly.
Thus on this scale it seems that $\mni\sim0.3\,\msol$,
similar to other Ic-BL SNe \citep{Lyman2016}.
Because the data have not yet converged to model tracks,
we cannot solve directly for $t_0$ and \mni\ using the technique for Ia SNe in \citet{Wygoda2019}.

We can also try to solve directly for $t_0$ and \mni\ using the technique for Ia SNe in \citet{Wygoda2019}. The first step is to solve for $t_0$ using Equation \ref{eq:qdep} and a second equation resulting from the fact that the expansion is adiabatic,

\begin{equation}
\label{eq:integral}
    \int_0^t Q(t') \, t' \, dt' 
    = \int_0^t L_\mathrm{bol}(t') \, t'\, dt'.
\end{equation}

\begin{figure}[ht]
\centering
\includegraphics[width=1.0\columnwidth]{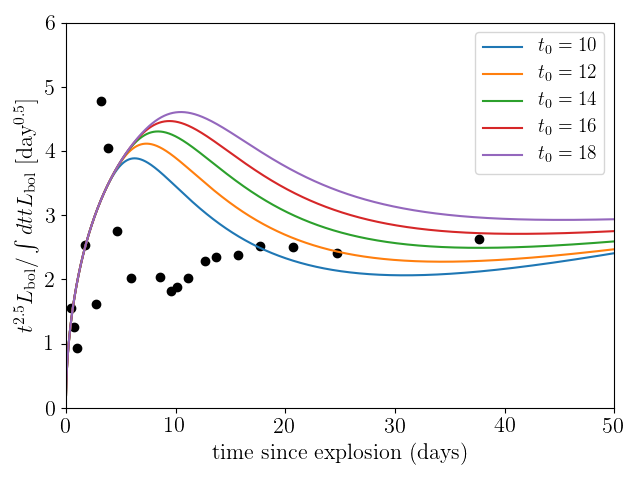}
\caption{
To test whether a light curve is powered by radioactive decay, the ratio of the bolometric luminosity to the time-weighted integrated bolometric luminosity should converge to model tracks, as described in \citet{Katz2013} and \citet{Wygoda2019}.
This enables a direct measurement of the gamma-ray escape time $t_0$ and the nickel mass \mni.
However, our data have not converged to these tracks, suggesting that either radioactive decay is not dominant, or that we are not yet in a phase where we can perform this measurement.
}
\label{fig:wygoda}
\end{figure}

The ratio of Equation \ref{eq:qdep} to Equation \ref{eq:integral} removes the dependence on \mni,
and enables $t_0$ to be measured.
However, as shown in Figure \ref{fig:wygoda},
the data have not yet converged to model tracks.

\subsection{Interaction with extended material}
\label{sec:interpretation-shock-cooling}

One way to power a rapid and luminous light curve
is to deposit energy into circumstellar material (CSM) at large radii \citep{Nakar2010,Nakar2014,Piro2015}.
Since this is a Ic-BL SN,
we expect the progenitor to be stripped of its envelope and therefore compact ($R\sim0.5\,R_\odot\sim10^{10}\,\cm$; \citealt{Groh2013}),
although there have never been any direct progenitor detections for a Ic-BL SN.

With this expectation,
extended material at larger radii would have to arise from mass-loss.
This would not be surprising, as massive stars are known to shed a significant fraction of their mass in winds and eruptive episodes; see \citet{Smith2014} for a review.

First we perform an order-of-magnitude calculation
to see whether the rise time and peak luminosity could be explained by a model in which shock interaction powers the light curve (``wind shock breakout'').
Assuming that the progenitor ejected material with a velocity $v_w$ at a time $t$ prior to explosion, the radius of this material at any given time is

\begin{equation}
\begin{split}
    R_\mathrm{sh}  
    & = R_* + v_\mathrm{w}t \\
    & \approx 
    (8.64 \times 10^{12}\,\cm)
    \left( \frac{v_w}{1000\,\km\,\psec} \right)
    \left( \frac{t}{\days} \right).
\end{split}
\end{equation}

For material ejected 15 days prior to explosion,
traveling at 1000\,\km\,\psec, the radius would be $R_\mathrm{CSM}\sim10^{14}\,\cm$ at the time of explosion.
The shock crossing timescale is $t_\mathrm{cross}$:

\begin{equation}
    t_\mathrm{cross} \sim R_\mathrm{CSM}/v_s \approx (0.4\,\days) \left( \frac{R}{10^{14}\,\cm} \right) 
    \left( \frac{v_s}{0.1c} \right)^{-1}
\end{equation}

\noindent where $v_s$ is the velocity of the shock.
The shock heats the CSM with an energy density that is roughly half of the kinetic energy of the shock, so $e_s \sim (1/2)(\rho v_s^2/2)$.
The luminosity is the total energy deposited divided by $t_\mathrm{cross}$,

\begin{equation}
\begin{split}
    L_\mathrm{BO} & \sim \frac{E_\mathrm{BO}}{t_\mathrm{cross}} \sim \frac{v_s^3}{4} \frac{dM}{dR} \\
    & = (8 \times 10^{44}\,\erg\,\psec)
    \left( \frac{v_s}{0.1c} \right)^3
    \left( \frac{dM}{\msol} \right)
    \left( \frac{dR}{10^{14}\,\cm} \right)^{-1}
\end{split}
\end{equation}

\noindent assuming a constant density.
Thus, for shock velocities on the order of the observed photospheric radius expansion ($0.1c$),
and a CSM radius on the order of the first photospheric radius that we measure ($3 \times 10^{14}\,\cm$),
it is easy to explain the rise time and peak luminosity that we observe.

To test whether shock breakout (and subsequent post-shock cooling) can explain the evolution of the physical properties we measured in Section \ref{sec:basic-properties},
we ran one-dimensional numerical radiation hydrodynamics simulations of a SN running into a circumstellar shell with CASTRO \citep{Almgren2010,Zhang2011}. We assume spherical symmetry and solve the coupled equations of radiation hydrodynamics using a grey flux-limited non-equilibrium diffusion approximation. The setup is similar to the models presented in \citet{Rest2018} but with parameters modified to fit \name.

The ejecta is assumed to be homologously expanding, characterized by a broken power-law density profile, an ejecta mass $M_{\rm ej}$, and energy $E_{\rm ej}$.
The ejecta density profile has an inner power-law index of $n=0$ (that is, $\rho(r)\propto r^{-n}$) then steepens to an index $n=10$, as is appropriate for core-collapse SN explosions \citep{Matzner1999}. The circumstellar shell is assumed to be uniform in density with radius $R_{\rm CSM}$ and mass $M_{\rm CSM}$. We adopt a uniform opacity of $\kappa=0.2$ cm$^2$ g$^{-1}$, which is characteristic of hydrogen-poor electron scattering.

\begin{figure}[ht]
\centering
\includegraphics[width=1.0\columnwidth]{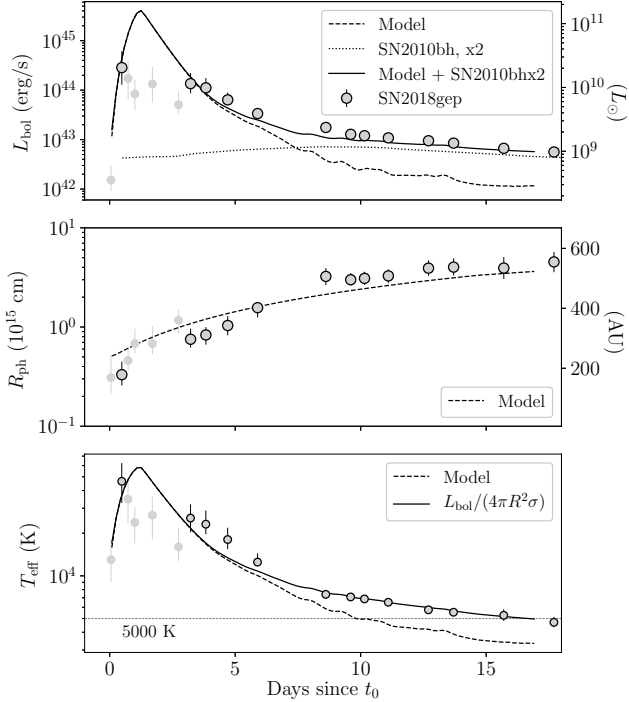}
\caption{
Best-fit CSM interaction model with the light curve of the Ic-BL SN\,2010bh \citep{Cano2013} scaled up by a factor of two.
The model parameters are $M_\mathrm{ej}=8\,\msol$, $E_\mathrm{ej} = 2 \times 10^{52}\,\erg$, $M_\mathrm{CSM} = 0.02\,\msol$,  and $R_\mathrm{CSM} = 3 \times 10^{14}\,\cm$.
As in Figure \ref{fig:physevol},
the outlined circles are derived from UV and optical data, while the light grey circles are derived from
optical data only.
}
\label{fig:shock-cooling-fit}
\end{figure}

The best-fit model, shown in Figure
\ref{fig:shock-cooling-fit},
used the following parameters:
$M_\mathrm{ej}=8\,\msol$, $E_\mathrm{ej} = 2 \times 10^{52}\,\erg$, $M_\mathrm{CSM} = 0.02\,\msol$,  and $R_\mathrm{CSM} = 3 \times 10^{14}\,\cm$.
The inferred kinetic energy is consistent with typical values measured for Ic-BL SNe (e.g. \citealt{Cano2017a,Taddia2019}),
and $R_\mathrm{CSM}$ is similar in value to the first photospheric radius we measure (at $\Delta t=0.05\,\days$; see Figure \ref{fig:physevol}).

The inferred values presented here are likely uncertain to within a factor of a few, given the degeneracies of the rise time and peak luminosity with the CSM mass and radius. Qualitatively, a larger CSM radius will result in a higher peak luminosity and longer rise time. The peak luminosity is relatively independent of the CSM mass, which instead affects the photospheric velocity and temperature (i.e. a larger CSM mass slows down the post-interaction velocity to a greater extent and increases the shock-heated temperature).   A full discussion of the dependencies of the light curve and photospheric properties on the CSM parameters will be presented in an upcoming work (Khatami, D. et al., in preparation).

In this framework, the shockwave sweeps through the CSM prior to peak luminosity, so that at maximum luminosity the outer parts of the CSM have been swept into a dense shell moving at SN-like velocities ($v_\mathrm{post-shock}\approx3v_s/4$).
This scenario was laid out in \citet{ChevalierIrwin} and discussed in \citet{Kasen2017}.
This explains the high velocities we measure at early times and the absence of narrow emission features in our spectra.
For another discussion of the absence of narrow emission lines due to an abrupt cutoff in CSM density, see \citet{Moriya2012}.
Following \citet{ChevalierIrwin},
the rapid rise corresponds to shock breakout from the CSM, and begins at a time $R_\mathrm{CSM}/v_\mathrm{sh}$ after the explosion,
where $v_\mathrm{sh}$ is the velocity of the shock.
The time to peak luminosity (1.2\,\days) is longer than this delay time by a factor ($R_w/R_d$).
Given the best-fit $R_w=3 \times 10^{14}\,\cm$,
and assuming $R_d \sim R_w$, we find $v_\mathrm{sh}=0.1c$,
and an explosion time $\sim 1\,\days$ prior to $t_0$.
This model also predicts an increasing temperature while the shock breaks out (i.e. during the rise to peak bolometric luminosity).

Other Ic SNe have shown early evidence for interaction
in their light curves, but in other cases the emission has been attributed to post-shock cooling in expanding material rather than shock breakout itself.
For example, the first peak observed in iPTF14gqr \citep{De2018} was short-lived ($\lesssim2\,\days$) and attributed to shock-cooling emission from material stripped by a compact companion.
iPTF14gqr is different in a number of ways from \name:
the spectra showed high-ionization emission lines, including \ion{He}{2},
and the explosion had a much smaller kinetic energy ($E_K\approx10^{50}\,\erg$) and smaller velocities (10,000\,\km\,\psec).
The main peak in iPTF16asu was also modeled as shock-cooling emission rather than shock breakout \citep{Whitesides2017}.

\begin{figure}[ht]
\centering
\includegraphics[width=1.0\columnwidth]{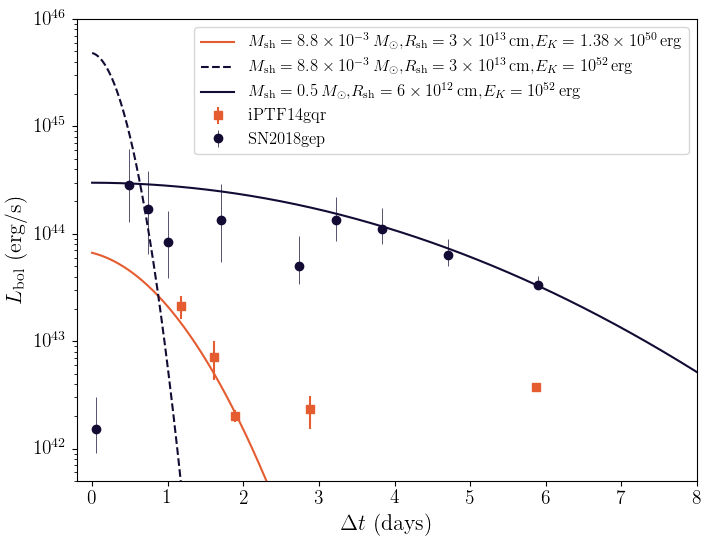}
\caption{
Estimated CSM and explosion properties using models from \citet{Piro2015}.
The shell mass is much larger than the one in iPTF14gqr, which is the reason for the more extended shock-cooling peak.}
\label{fig:piro-fit}
\end{figure}

Under the assumption that the light curve represented
post-shock cooling emission, \citet{De2018} and \citet{Whitesides2017}
both used one-zone analytic models from \citet{Piro2015} to estimate the properties of the explosion and the CSM.
This approximation assumes that the emitting region is a uniformly heated expanding sphere.
In iPTF14gqr the inferred properties of the extended material were $M_e\sim8\times10^{-3}\,M_\odot$ at $R_e\sim3\times10^{13}\,\cm$.
In iPTF16asu the inferred properties of the extended material
were $M_e\sim0.45\,M_\odot$ at $R_e\sim1.7\times10^{12}\,\cm$.
The fit also required a more energetic explosion than iPTF14gqr ($4 \times 10^{51}\,\erg$).
By applying the same framework to the decline of the bolometric light curve of \name, we arrive at similar values to those inferred for iPTF16asu, as shown in Figure \ref{fig:piro-fit}.

We model the main peak of \name\ as shock breakout rather than post-shock cooling emission.
Our motivation for this choice is that the timescale over which we detect the precursor emission is more consistent with a large radius and lower shell mass.
From the shell mass and radius,
we can also estimate the mass-loss rate immediately prior to explosion,

\begin{equation}
    \frac{\dot{M}}{\msol\,\pyr} \approx 
    32
    \left( \frac{M_\mathrm{sh}}{\msol} \right)
    \left( \frac{v_w}{1000 \km\,\psec} \right)
    \left( \frac{R_\mathrm{sh}}{10^{14} \cm} \right)^{-1}.
\end{equation}

\noindent For our best-fit parameters $M_\mathrm{sh}=0.02\,M_\odot$ and $R_\mathrm{sh}=3\times10^{14}\,\cm$,
and taking $v_w=1000\,\km\,\psec$, we find $\dot{M}\approx0.6\,\msol\,\pyr$,
4--6 orders of magnitude higher than what is typically
expected for Ic-BL SNe \citep{Smith2014}.

In the shock breakout model, the shock sweeps through confined CSM and passes into lower-density material.
Thus, it is not surprising that we do not observe the X-ray or radio emission that would indicate interaction with high-density material.
From our VLA observations of \name,
the radio flux marginally decreased from $\Delta t=5\,\days$ to $\Delta t=75\,\days$.
This could be astrophysical, but could also be instrumental (change in beamsize due to change in VLA configuration).
Using the relation of \citet{Murphy2011},
the estimated contribution from the host galaxy (for a 
SFR of $0.12^{+0.08}_{-0.05}\,\msol\,\pyr$; see Section \ref{sec:host})
is 

\begin{equation}
\begin{split}
    \left( \frac{L_\mathrm{1.4\,\mathrm{GHz}}}{\erg\,\psec\,\phz} \right) 
    & \approx 1.57 \times 10^{28}
    \left( \frac{\mathrm{SFR}_\mathrm{radio}}{\msol\,\pyr} \right) \\
    & \approx 1.9 \times 10^{27}\,\erg\,\psec\,\phz.
\end{split}
\end{equation}

\noindent Taking a spectral index of $-0.7$ (a synchrotron spectrum), the expected 9\,\ghz\ luminosity would be between
$3.0 \times 10^{26}\,\erg\,\psec\,\phz$ and 
$8.6 \times 10^{26}\,\erg\,\psec\,\phz$.
From Table \ref{tab:radio-flux},
the measured spectral luminosity is
$8.3 \times 10^{26}\,\erg\,\psec\,\phz$ (at 10\,\ghz)
in the first epoch,
and
$6 \times 10^{26}\,\erg\,\psec\,\phz$ (at 9\,\ghz)
in the second epoch.
The slit covering fraction of our LRIS observations is again relevant here;
as discussed in Section \ref{sec:host},
the true SFR is likely a factor of a few higher than what we inferred from modeling the galaxy SED.
So, it is plausible that the first two radio detections are entirely due to the host galaxy. 

\begin{figure}[ht]
\centering
\includegraphics[width=1.0\columnwidth]{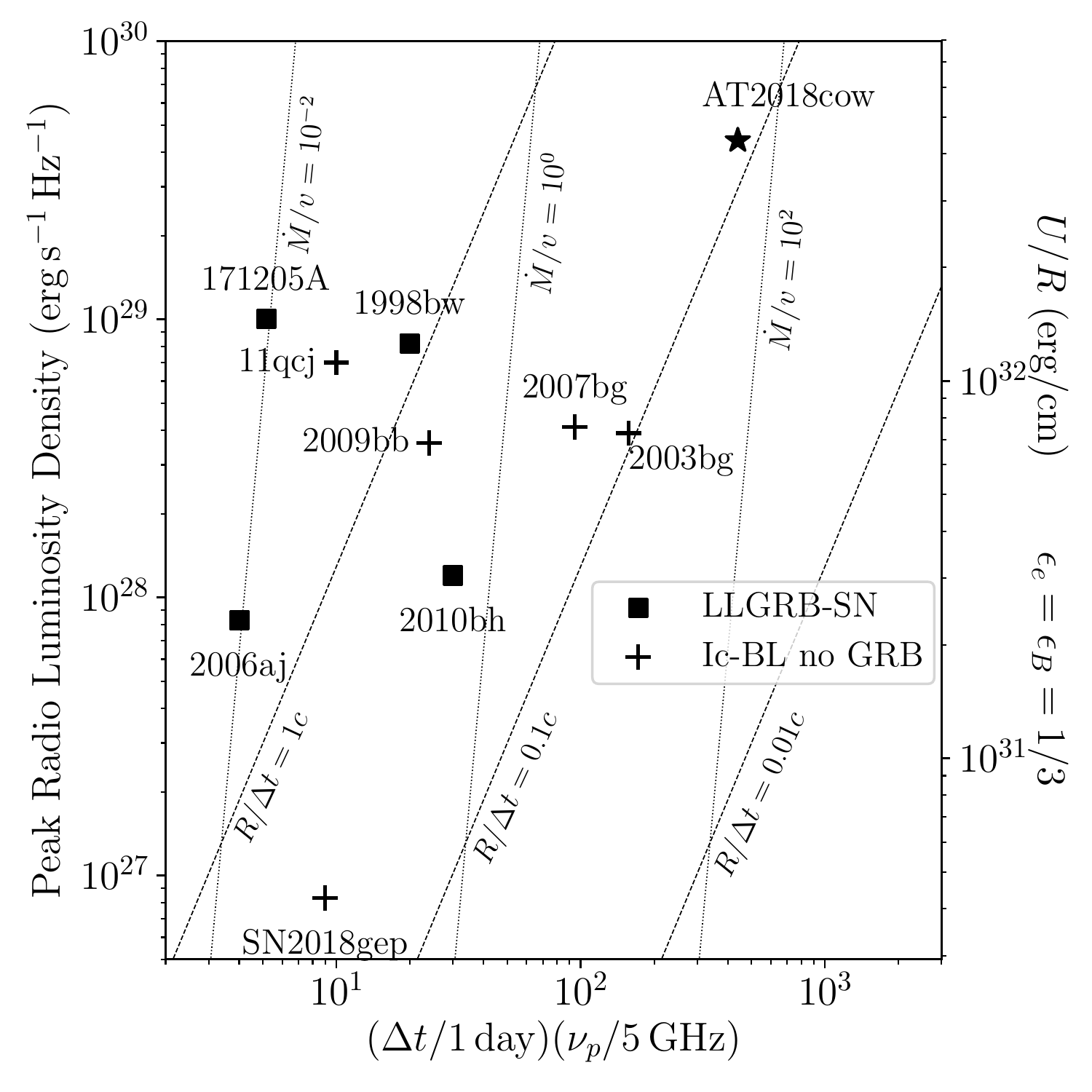}
\caption{The radio luminosity of SN2018gep compared to AT2018cow and radio-loud Ic-BL SNe (assuming $\epsilon_e=\epsilon_B=1/3$, cf. \citealt{Chevalier1998,Soderberg2010,Ho2019}).
Lines of constant mass-loss rate (scaled to wind velocity) are shown in units of $10^{-4}\,M_\odot\,\pyr/1000\,\km\,\psec$.
The radio luminosity for GRB\,171205A was taken from VLA observations reported by \citet{Laskar2017}, but we note that this is a lower limit in luminosity and in peak frequency because the source was heavily self-absorbed at this epoch.
}
\label{fig:lum-tnu}
\end{figure}

In the third epoch, the luminosity is (at 9\,\ghz)
is $< 3.9 \times 10^{26}\,\erg\,\psec\,\phz$,
although the difference from the first two epochs
may be due to the different array configuration.
Taking the peak of the 9--10\,\ghz\ light curve to be
$8.3 \times 10^{26}\,\erg\,\psec\,\phz$ at $\Delta t\approx5\,\days$,
Figure \ref{fig:lum-tnu} shows that \name\ would be an order of magnitude less luminous in radio emission
than any other Ic-BL SN.
If the luminosity truly decreased, then the implied mass-loss rate is $\dot{M} \sim 3 \times 10^{-6}$,
consistent with the idea that the shock has passed from confined CSM into much lower-density material.

If the emission is constant and due entirely to the host galaxy,
the point shown in Figure \ref{fig:lum-tnu} is an upper limit in luminosity.
Assuming that the peak of the SED of any radio emission from the SN is not substantially different from the frequencies we measure (i.e.\ that the spectrum is not self-absorbed at these frequencies),
we have a limit on the 9\,\ghz\ radio luminosity of $L_p \lesssim 10^{27}\,\erg\,\psec\,\phz$ at $\Delta t\approx 5$--15\,\days.

The shell mass and radius also give an estimate of the optical depth: $\tau \approx \kappa M / r^2 \approx 100 >> 1 $, which means that the shell would be optically thick. The lack of detected X-ray emission is consistent with the expectation that any X-ray photons produced in the collision would be thermalized by the shell and reradiated as blackbody emission.

Finally, assuming that the rapid rise to peak is indeed caused by shock breakout,
we examine whether our model is consistent with our detections in the weeks prior to explosion.
Material ejected 10 days prior to the explosion at the escape velocity of a Wolf-Rayet star ($v_\mathrm{esc}\sim1000\,\km\,\psec$) would lie at $R \sim 10^{14}\,\cm$, which is consistent with our model.
Assuming that the emission mechanism is internal shocks between shells of ejected material traveling at different velocities, we can estimate the amount of mass required:

\begin{equation}
    \frac{1}{2} \epsilon M v^2 = L\tau
\end{equation}

\noindent where $v \approx 1000\,\km\,\psec$, $\epsilon\approx0.5$ is the efficiency of thermalizing the kinetic energy of the shells, $M$ is the shell mass, $L\approx 10^{39}\,\erg\,\psec$ is the luminosity we observe, and $\tau\approx10\,\days$ is the timescale over which we observe the emission.
We find $M\approx0.02\,\msol$, again consistent with our model.

We conclude that the data are consistent with a scenario in which a compact Ic-BL progenitor underwent a period of eruptive mass-loss shortly prior to explosion.
In the terminal explosion, the light curve was initially dominated by shock breakout through (and post-shock cooling of) this recently-ejected material.

Finally, we return to the question of the emission detected in the first few minutes, which showed an inflection point prior to the rapid rise to peak 
(Figure \ref{fig:firstmins}).
Given the pre-explosion activity and inference of CSM interaction, it is not surprising that the rise is not well-modeled by a simple quadratic function.
One possibility is that we are seeing ejecta already heated from earlier precursor activity.
Another possibility is that we are seeing the effects of a finite light travel time.
For a sphere of $R\sim 3 \times 10^{14}\,\cm$,
the light crossing time is $\sim20$ minutes.
The slower rising phase could represent the time for photons to reach us across the extent of the emitting sphere.

In Table \ref{tab:model-properties},
we summarize the key properties inferred from
Section \ref{sec:interpretation}.

\begin{deluxetable}{lrr}[ht]
\tablecaption{Key model properties of SN2018gep \label{tab:model-properties}} 
\tablewidth{0pt} 
\tablehead{ \colhead{Parameter} & \colhead{Value} & \colhead{Notes}} 
\tabletypesize{\scriptsize} 
\startdata 
\trise & 1.2\,\days \\
$E_\mathrm{SN}$ & $2 \times 10^{52}\,\erg$ \\
$\mej$ & 8\,\msol \\
$M_\mathrm{CSM}$ & 0.02\,\msol \\
$R_\mathrm{CSM}$ & $3 \times 10^{14}\,\cm$ \\
$\dot{M}$ & $0.6\,\msol\,\pyr$ & Assuming $v_w=1000\,\km\,\psec$\\
$\mni$ & $< 0.2$--0.3\,\msol \\
\enddata 
\end{deluxetable}

\section{Comparison to unclassified rapidly evolving transients at high redshift}
\label{sec:comparisons}

\begin{figure}[ht]
\centering
\includegraphics[width=1.0\columnwidth]{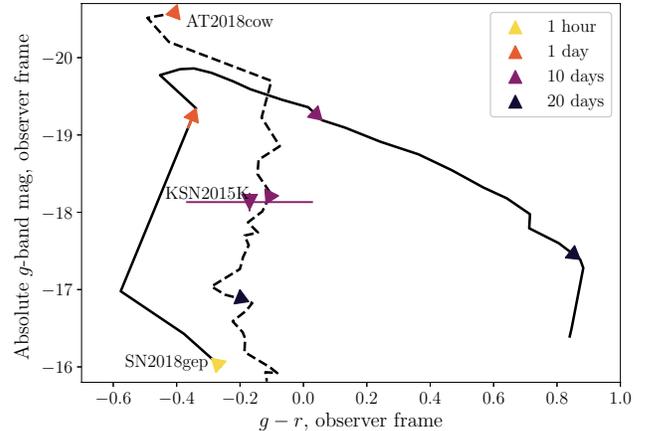}
\caption{A ``color-magnitude'' diagram of AT2018cow and \name, showing the evolution of color with time from first light ($t_0$). Like AT2018cow, the fast transient KSN2015K stayed persistently blue even after it had faded to half-maximum. \name\ has more typical SN evolution, reddening with time (cooling in temperature).
}
\label{fig:colmag}
\end{figure}

In terms of the timescale of its light curve evolution,
\name\ is similar to AT2018cow in fulfilling
the criteria that optical surveys use to identify rapidly evolving transients (e.g. \citealt{Drout2014,Tanaka2016,Pursiainen2018}).
However, there are a number of ways in which \name\
is more of a ``typical'' member of these populations
than AT2018cow.
In particular, \name\ has an expanding photospheric radius and declining effective temperature.
By contrast, one of the challenges in explaining AT2018cow as a stellar explosion was its nearly constant temperature (persistent blue color) and \emph{declining} photospheric radius.
In Figure \ref{fig:colmag} we show these two different kinds of evolution as very different tracks in color-magnitude space.
We also show a late-time point for KSN2015K \citep{Rest2018},
which shows blue colors even after the transient had faded to half-max.
The mass-loss rate inferred for \citet{Rest2018}
was $2 \times 10^{-3}\,\msol\,\pyr$.

Of the PS-1 events,
most appear to expand, cool, and redden with time \citep{Drout2014}.
That said, there are few co-eval data points in multiple filters,
even in the gold sample transients.
The transients are also faint; all but one lie at $z>0.1$.
Of the DES sample,
most also show evidence for declining temperatures and increasing radii, although three show evidence of a constant temperature and decreasing radius: 15X3mxf, 16X1eho, and 15C3opk.
The peak bolometric luminosities for these three transients are reported as $3 \times 10^{43}\,\erg\,\psec$,
$9 \times 10^{43}\,\erg\,\psec$,
and $5 \times 10^{43}\,\erg\,\psec$, respectively \citep{Pursiainen2018}.

To estimate a rate of Ic-BL SNe that have a light curve powered by shock breakout, we used the sample of 25 nearby ($z<0.1$) Ic-BL SNe from PTF \citep{Taddia2019},
because these were found in an untargeted survey.
Of these, we could not draw a conclusion about eight (either because the peak was not resolved or there was no multi-color photometry available around peak, or both).
The remaining clearly lacked the rise time or blue colors of SN2018gep.
Furthermore, SN2018gep is unique among the sample of 12 nearby ($z<0.1$) Ic-BL SNe from ZTF discovered so far, which will be presented in a separate publication.
From this, we estimate that the rate of Ic-BL SNe with a main peak dominated by shock breakout is no more than 10\% of the rate of Ic-BL SNe.

\section{Summary and Future Work}
\label{sec:conclusions}

In this paper, we presented 
an unprecedented dataset that connects late-stage eruptive mass loss in a stripped massive star to its subsequent explosion as a rapidly rising luminous transient.
Here we summarize our key findings:

\begin{enumerate}
    \item High-cadence dual-band observations with ZTF (six observations in 3 hours) captured a rapid rise ($1.4\pm0.1$ mag/hr) to peak luminosity, and a corresponding increase in temperature.
    This rise rate is second only to that of SN\,2016gkg \citep{Bersten2018}, which was attributed to shock breakout in extended material surrounding a Type IIb progenitor.
    However, the signal in \name\ is two magnitudes more luminous.
    \item A retrospective search in ZTF data revealed clear detections of precursor emission in the days and months leading up to the terminal explosion. The luminosity of these detections ($M=-14$) and evidence for variability suggests that they arise from eruptive mass-loss, rather than the luminosity of a quiescent progenitor. This is the first definitive pre-explosion detection of a Ic-BL SN to date.
    \item The bolometric light curve peaks after a few days at $>3 \times 10^{44}\,\erg\,\psec$.
    At late times, a power-law and an exponential decay are both acceptable fits to the data.
    \item The temperature rises to $50,000\,\kelvin$ in the first day,
    then declines as $t^{-1}$ then flattens at 5000\,K, which we attribute to recombination of carbon and oxygen.
    \item The photosphere expands at $v=0.1c$, and flattens once recombination sets in.
    \item We obtained nine spectra in the first five days of the explosion, as the effective temperature declined from 50,000\,K to 20,000\,K. To our knowledge, these represent the earliest-ever spectra of a stripped-envelope SN, in terms of temperature evolution.
    \item The early spectra exhibit a ``W'' feature similar to what has been seen in stripped-envelope superluminous SNe. From a NLTE spectral synthesis model, we find that this can be reproduced with a carbon and oxygen composition.
    \item The velocities inferred from the spectra are among the highest observed for stripped-envelope SNe, and are most similar to the velocities of Ic-BL SNe accompanied by GRBs.
    \item The host galaxy has a star-formation rate of 0.12\,\msol\,\pyr, and a lower mass and lower metallicity than galaxies hosting GRB-SNe, which are low-mass and low-metallicity compared to the overall core collapse SN population.
    \item The early light curve is best-described by shock breakout in extended but confined CSM, with $M=0.02\,\msol$ at $R=3 \times 10^{14}\,\cm$. The implied mass-loss rate is $0.6\,\msol\,\pyr$ in the days leading up to the explosion, consistent with our detections of precursor emission. After the initial breakout, the shock runs through CSM of much lower density, hence the lack of narrow emission features and lack of strong radio and X-ray emission.
    \item Although \name\ is similar to AT2018cow in terms of its bolometric light curve, it has a very different color evolution. In this sense, the ``rapidly evolving transients'' in the PS-1 and DES samples are more similar to \name\ than to AT2018cow.
    \item The late-time light curve seems to require an energy deposition mechanism distinct from shock-interaction. Radioactive decay is one possibility, but further monitoring is needed to test this.
\end{enumerate}

The code used to produce the results described in this paper was written in Python
and is available online in an open-source repository\footnote{https://github.com/annayqho/SN2018gep}.
When the paper has been accepted for publication, the data will be made publicly available via WISeREP, an interactive repository of supernova data \citep{Yaron2012}.

\acknowledgements

The authors would like to thank the anonymous referee whose comments improved the flow, precision, and clarify of the paper.
It is a pleasure to thank Tony Piro, Dan Kasen, E. Sterl Phinney, Eliot Quataert, Maryam Modjaz, Jim Fuller, Lars Bildsten, Udi Nakar, Paul Duffell, and Luc Dessart for helpful discussions.
A.Y.Q.H. is particularly grateful to Tony Piro and the community at Carnegie Observatories for their hospitality on Tuesdays during the period in which this work was performed.
Thank you to the staff at the SMA, AMI, the VLA, \swift, and Chandra for rapidly scheduling and executing the observations.
Thank you to David Palmer (LANL) for his assistance in searching the pointing data for \swift/BAT.
Thank you to Michael J. Koss (Eureka Scientific Int),
Andrew Drake (Caltech), Scott Adams (Caltech),
Matt Hankins (Caltech), Kevin Burdge (Caltech), and Kirsty Taggart (LJMU) for assisting with optical spectroscopic observations.
Thank you to Erik Petigura and David Hogg for their advice on figure aesthetics.
Thank you to David Alexander Kann (IAA-CSIC) and Nhan Nguyen (Bonn) for pointing out errors in early versions of the paper posted to the arXiv.
D.A.G. thanks St\'efan van der Walt and Ari Crellin-Quick for assistance with skyportal, which enabled the search for pre-explosion emission.

A.Y.Q.H. is supported by a National Science Foundation Graduate Research Fellowship under Grant No.\,DGE‐1144469.
This work was supported by the GROWTH project funded by the National Science Foundation under PIRE Grant No.\,1545949.
A.G.-Y. is supported by the EU via ERC grant No. 725161, the ISF, the BSF Transformative program and by a Kimmel award.
Y.T. studied as a GROWTH intern at Caltech during the summer and fall of 2017.
C.C.N. thanks the funding from MOST grant 104-2923-M-008 -004-MY5.
R.L. is supported by a Marie Sk\l{}odowska-Curie Individual Fellowship within the Horizon 2020 European Union (EU) Framework Programme for Research and Innovation (H2020-MSCA-IF-2017-794467).
A.H. acknowledges support by the I-Core Program of the Planning and Budgeting Committee
and the Israel Science Foundation, and support by the ISF grant 647/18. This research was supported by a Grant from the GIF, the German-Israeli Foundation for Scientific Research and Development.
This research was funded in part by the Gordon and Betty Moore Foundation through Grant GBMF5076, and a grant from the Heising-Simons Foundation.
A.C. acknowledges support from the NSF CAREER award N. 1455090 and from the NASA/Chandra GI award N. GO8-19055A.
Research support to I.A. is provided by the GROWTH project, funded by the National Science Foundation under Grant No 1545949.

Based on observations obtained with the Samuel Oschin Telescope 48-inch and the 60-inch Telescope at the Palomar Observatory as part of the Zwicky Transient Facility project. Major funding has been provided by the U.S National Science Foundation under Grant No.\,AST-1440341 and by the ZTF partner institutions: the California Institute of Technology, the Oskar Klein Centre, the Weizmann Institute of Science, the University of Maryland, the University of Washington, Deutsches Elektronen-Synchrotron, the University of Wisconsin-Milwaukee, and the TANGO Program of the University System of Taiwan. Partially based on observations made with the Nordic Optical Telescope, operated by the Nordic Optical Telescope Scientific Association at the Observatorio del Roque de los Muchachos, La Palma, Spain, of the Instituto de Astrofisica de Canarias. The Liverpool Telescope is operated by Liverpool John Moores University with financial support from the UK Science and Technology Facilities Council. LT is located  on the island of La Palma, in the Spanish Observatorio del Roque de los Muchachos of the Instituto de Astrofisica de Canarias.
The scientific results reported in this article are based in part on observations made by the Chandra X-ray Observatory.
The data presented here were obtained in part with ALFOSC, which is provided by the Instituto de Astrofisica de Andalucia (IAA) under a joint agreement with the University of Copenhagen and NOTSA.
The Submillimeter Array is a joint project between the Smithsonian
Astrophysical Observatory and the Academia Sinica Institute of
Astronomy and Astrophysics and is funded by the Smithsonian
Institution and the Academia Sinica.
We acknowledge the support of the staff of the Xinglong 2.16-m telescope. This work is supported by the National Natural Science Foundation of China (NSFC grants 11325313 and 11633002), and the National Program on Key Research and Development Project (grant no.\,2016YFA0400803). 
SED Machine is based upon work supported by the National Science Foundation under Grant No.\,1106171.
This publication has made use of data collected at Lulin Observatory, partly supported by MoST grant 105-2112-M-008-024-MY3.
The JEKYLL simulations were performed on resources provided by the Swedish National Infrastructure for Computing (SNIC) at Parallelldatorcentrum (PDC).

\software{
\code{Astropy} \citep{Astropy2013, Astropy2018},
\code{IPython} \citep{ipython},
\code{matplotlib} \citep{matplotlib},
\code{numpy} \citep{numpy},
\code{scipy} \citep{Virtanen2019},
\code{extinction} \citep{Barbary2016}
\code{SkyPortal} \citep{skyportal}}

\facilities{CFHT, Keck:I (LRIS), Hale (DBSP), AMI, Liverpool:2m (IO:O, SPRAT), DCT, Swift (UVOT, XRT), Beijing:2.16m, EVLA, SMA, LO:1m, NOT (ALFOSC)}

\appendix

Here we provide supplementary figures and tables.
In Appendix \ref{appendix:uv-opt-phot} we provide the full set of optical and UV photometry and the blackbody fits to this photometry.
In Appendix \ref{appendix:uv-opt-spec} we provide the log of optical and UV spectroscopic observations,
as well as a figure showing all of our optical spectra.
In Appendix \ref{appendix:atomic-data} we include more details about the atomic data used for our spectral modeling.
In Appendix \ref{appendix:host-data} we show the spectrum, line-flux measurements, and photometry that was used to derive properties of the host galaxy.

\section{UV and Optical Photometry}
\label{appendix:uv-opt-phot}

\begin{figure*}[!ht]
\centering
\includegraphics[]{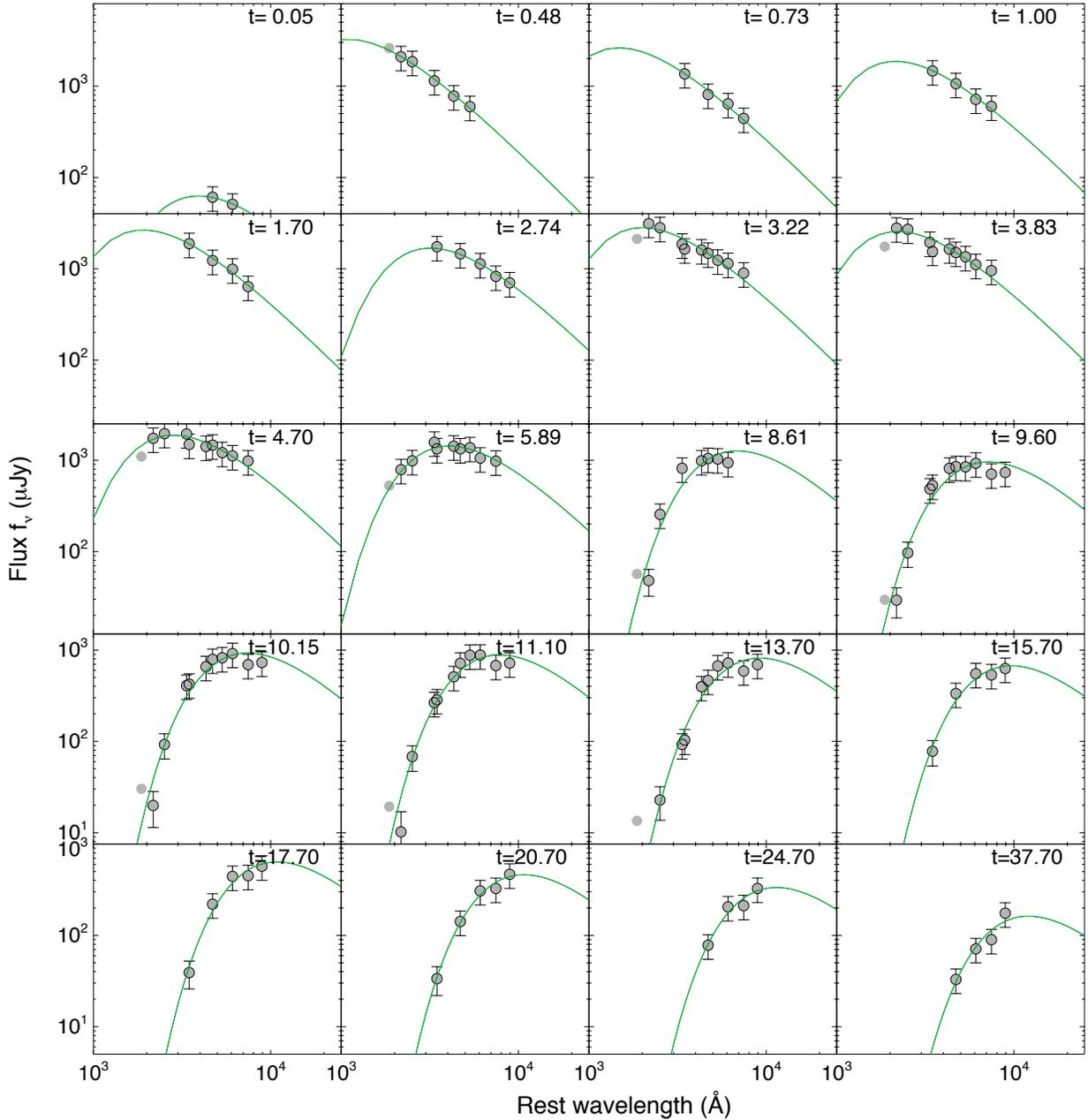}
\caption{Blackbody fits to \swift/UVOT and optical photometry for \name. Since the UVOT and ground-based observations were taken at slightly different epochs, we interpolated the data in time using UVOT epochs at early times and LT epochs at later times.
}
\label{fig:bbfits}
\end{figure*}

Here we provide our optical and UV photometry (Table \ref{tab:uvot-phot}) and the blackbody fits to this photometry used to derive the photospheric evolution (Figure \ref{fig:bbfits}).

\startlongtable 
\begin{deluxetable}{lrrrrr} 
\tablecaption{Optical and ultraviolet photometry for SN2018gep\label{tab:uvot-phot}} 
\tablewidth{0pt} 
\tablehead{ \colhead{Date (JD)} & \colhead{$\Delta t$} & \colhead{Instrument} & \colhead{Filter} & \colhead{AB Mag} & \colhead{Error in AB Mag} } 
\tabletypesize{\scriptsize} 
\startdata 
2458370.6634 & 0.02 & P48+ZTF & r & 20.48 & 0.26 \\ 
2458370.6856 & 0.04 & P48+ZTF & g & 19.70 & 0.14 \\ 
2458370.6994 & 0.05 & P48+ZTF & g & 19.34 & 0.11 \\ 
2458370.7153 & 0.07 & P48+ZTF & g & 18.80 & 0.08 \\ 
2458370.7612 & 0.11 & P48+ZTF & r & 18.36 & 0.08 \\ 
2458370.7612 & 0.11 & P48+ZTF & r & 18.36 & 0.08 \\ 
2458371.6295 & 0.98 & P60+SEDM & r & 16.78 & 0.01 \\ 
2458371.6323 & 0.99 & P60+SEDM & g & 16.39 & 0.02 \\ 
2458371.6351 & 0.99 & P60+SEDM & i & 17.01 & 0.01 \\ 
2458371.6369 & 0.99 & P48+ZTF & r & 16.83 & 0.03 \\ 
2458371.6378 & 0.99 & P48+ZTF & r & 16.81 & 0.04 \\ 
2458371.6378 & 0.99 & P48+ZTF & r & 16.81 & 0.04 \\ 
2458371.6392 & 0.99 & P60+SEDM & u & 15.98 & 0.02 \\ 
2458371.642 & 0.99 & P60+SEDM & r & 16.77 & 0.01 \\ 
2458371.6448 & 1.0 & P60+SEDM & g & 16.37 & 0.02 \\ 
2458371.6476 & 1.0 & P60+SEDM & i & 16.97 & 0.01 \\ 
2458371.6514 & 1.0 & P48+ZTF & r & 16.80 & 0.03 \\ 
2458371.6517 & 1.0 & P60+SEDM & u & 15.98 & 0.02 \\ 
2458371.6838 & 1.04 & P48+ZTF & r & 16.78 & 0.04 \\ 
2458371.6959 & 1.05 & P48+ZTF & g & 16.31 & 0.02 \\ 
2458371.6968 & 1.05 & P48+ZTF & g & 16.29 & 0.03 \\ 
2458371.6968 & 1.05 & P48+ZTF & g & 16.29 & 0.03 \\ 
2458371.7138 & 1.07 & P48+ZTF & g & 16.29 & 0.03 \\ 
2458371.7138 & 1.07 & P48+ZTF & g & 16.29 & 0.03 \\ 
2458371.7359 & 1.09 & P48+ZTF & g & 16.28 & 0.03 \\ 
2458372.6396 & 1.99 & P48+ZTF & r & 16.48 & 0.04 \\ 
2458372.6396 & 1.99 & P48+ZTF & r & 16.48 & 0.04 \\ 
2458372.6586 & 2.01 & P48+ZTF & r & 16.49 & 0.06 \\ 
2458372.6586 & 2.01 & P48+ZTF & r & 16.49 & 0.06 \\ 
2458372.6861 & 2.04 & P48+ZTF & r & 16.47 & 0.04 \\ 
2458372.6861 & 2.04 & P48+ZTF & r & 16.47 & 0.04 \\ 
2458372.7134 & 2.07 & P48+ZTF & g & 15.99 & 0.03 \\ 
2458372.7371 & 2.09 & P48+ZTF & g & 15.99 & 0.02 \\ 
2458372.7371 & 2.09 & P48+ZTF & g & 15.99 & 0.02 \\ 
2458373.6276 & 2.98 & P48+ZTF & r & 16.36 & 0.03 \\ 
2458373.6447 & 3.0 & P60+SEDM & r & 16.32 & 0.01 \\ 
2458373.6464 & 3.0 & P60+SEDM & g & 15.99 & 0.01 \\ 
2458373.6481 & 3.0 & P60+SEDM & i & 16.55 & 0.01 \\ 
2458373.6498 & 3.0 & P60+SEDM & u & 15.90 & 0.02 \\ 
2458373.6627 & 3.02 & P48+ZTF & r & 16.35 & 0.03 \\ 
2458373.6627 & 3.02 & P48+ZTF & r & 16.35 & 0.03 \\ 
2458373.685 & 3.04 & P48+ZTF & r & 16.34 & 0.03 \\ 
2458373.685 & 3.04 & P48+ZTF & r & 16.34 & 0.03 \\ 
2458373.6984 & 3.05 & P48+ZTF & g & 15.91 & 0.02 \\ 
2458373.7189 & 3.07 & P48+ZTF & g & 15.90 & 0.03 \\ 
2458373.736 & 3.09 & P48+ZTF & g & 15.91 & 0.02 \\ 
2458374.6316 & 3.98 & P48+ZTF & r & 16.30 & 0.04 \\ 
2458374.6316 & 3.98 & P48+ZTF & r & 16.30 & 0.04 \\ 
2458374.6429 & 4.0 & P48+ZTF & r & 16.32 & 0.03 \\ 
2458374.6495 & 4.0 & P48+ZTF & r & 16.30 & 0.03 \\ 
2458374.6551 & 4.01 & P60+SEDM & r & 16.29 & 0.01 \\ 
2458374.6569 & 4.01 & P60+SEDM & g & 16.03 & 0.01 \\ 
2458374.6586 & 4.01 & P60+SEDM & i & 16.45 & 0.01 \\ 
2458374.6603 & 4.01 & P60+SEDM & u & 15.92 & 0.03 \\ 
2458374.6845 & 4.04 & P48+ZTF & r & 16.29 & 0.04 \\ 
2458374.6845 & 4.04 & P48+ZTF & r & 16.29 & 0.04 \\ 
2458374.6994 & 4.05 & P48+ZTF & g & 15.91 & 0.03 \\ 
2458374.6994 & 4.05 & P48+ZTF & g & 15.91 & 0.03 \\ 
2458374.7041 & 4.06 & P48+ZTF & g & 15.93 & 0.02 \\ 
2458374.7264 & 4.08 & P48+ZTF & g & 15.92 & 0.03 \\ 
2458374.7428 & 4.1 & P48+ZTF & g & 15.91 & 0.03 \\ 
2458374.7428 & 4.1 & P48+ZTF & g & 15.91 & 0.03 \\ 
2458375.6247 & 4.98 & P60+SEDM & r & 16.31 & 0.01 \\ 
2458375.6265 & 4.98 & P60+SEDM & g & 16.07 & 0.01 \\ 
2458375.6282 & 4.98 & P60+SEDM & i & 16.43 & 0.01 \\ 
2458375.6299 & 4.98 & P60+SEDM & u & 15.98 & 0.03 \\ 
2458375.6757 & 5.03 & P48+ZTF & r & 16.33 & 0.04 \\ 
2458375.6757 & 5.03 & P48+ZTF & r & 16.33 & 0.04 \\ 
2458375.7144 & 5.07 & P48+ZTF & g & 15.97 & 0.04 \\ 
2458375.7144 & 5.07 & P48+ZTF & g & 15.97 & 0.04 \\ 
2458375.7381 & 5.09 & P48+ZTF & g & 15.99 & 0.03 \\ 
2458376.62 & 5.97 & P48+ZTF & r & 16.37 & 0.04 \\ 
2458376.6623 & 6.02 & P60+SEDM & r & 16.37 & 0.01 \\ 
2458376.6626 & 6.02 & P48+ZTF & r & 16.37 & 0.04 \\ 
2458376.664 & 6.02 & P60+SEDM & g & 16.16 & 0.02 \\ 
2458376.6657 & 6.02 & P60+SEDM & i & 16.44 & 0.01 \\ 
2458376.6674 & 6.02 & P60+SEDM & u & 16.09 & 0.03 \\ 
2458376.6739 & 6.03 & P48+ZTF & r & 16.36 & 0.04 \\ 
2458376.7272 & 6.08 & P48+ZTF & g & 16.10 & 0.03 \\ 
2458376.7272 & 6.08 & P48+ZTF & g & 16.10 & 0.03 \\ 
2458376.7423 & 6.1 & P48+ZTF & g & 16.09 & 0.03 \\ 
2458376.7423 & 6.1 & P48+ZTF & g & 16.09 & 0.03 \\ 
2458377.6186 & 6.97 & P60+SEDM & r & 16.40 & 0.01 \\ 
2458377.6204 & 6.97 & P60+SEDM & g & 16.27 & 0.02 \\ 
2458377.6221 & 6.97 & P60+SEDM & i & 16.51 & 0.01 \\ 
2458377.6238 & 6.98 & P60+SEDM & u & 16.29 & 0.01 \\ 
2458377.6301 & 6.98 & P48+ZTF & r & 16.41 & 0.03 \\ 
2458377.6301 & 6.98 & P48+ZTF & r & 16.41 & 0.03 \\ 
2458377.6513 & 7.0 & P48+ZTF & r & 16.41 & 0.03 \\ 
2458377.6639 & 7.02 & P48+ZTF & r & 16.40 & 0.03 \\ 
2458377.6639 & 7.02 & P48+ZTF & r & 16.40 & 0.03 \\ 
2458377.6761 & 7.03 & P48+ZTF & r & 16.41 & 0.03 \\ 
2458377.6761 & 7.03 & P48+ZTF & r & 16.41 & 0.03 \\ 
2458377.6935 & 7.05 & P48+ZTF & g & 16.21 & 0.03 \\ 
2458377.7038 & 7.06 & P48+ZTF & g & 16.20 & 0.03 \\ 
2458377.7165 & 7.07 & P48+ZTF & g & 16.20 & 0.04 \\ 
2458377.7165 & 7.07 & P48+ZTF & g & 16.20 & 0.04 \\ 
2458377.7458 & 7.1 & P48+ZTF & g & 16.22 & 0.03 \\ 
2458377.7458 & 7.1 & P48+ZTF & g & 16.22 & 0.03 \\ 
2458378.6164 & 7.97 & P48+ZTF & r & 16.45 & 0.04 \\ 
2458378.6437 & 8.0 & P48+ZTF & r & 16.46 & 0.05 \\ 
2458378.665 & 8.02 & P48+ZTF & g & 16.33 & 0.03 \\ 
2458378.665 & 8.02 & P48+ZTF & g & 16.33 & 0.03 \\ 
2458378.6844 & 8.04 & P48+ZTF & g & 16.32 & 0.03 \\ 
2458378.693 & 8.05 & P60+SEDM & r & 16.41 & 0.01 \\ 
2458378.7039 & 8.06 & P48+ZTF & g & 16.32 & 0.03 \\ 
2458378.7158 & 8.07 & P48+ZTF & r & 16.47 & 0.03 \\ 
2458379.6623 & 9.02 & P48+ZTF & g & 16.44 & 0.04 \\ 
2458379.6823 & 9.04 & P48+ZTF & g & 16.43 & 0.03 \\ 
2458379.6823 & 9.04 & P48+ZTF & g & 16.43 & 0.03 \\ 
2458379.6977 & 9.05 & P48+ZTF & g & 16.44 & 0.03 \\ 
2458379.7176 & 9.07 & P48+ZTF & r & 16.51 & 0.04 \\ 
2458379.7409 & 9.09 & P48+ZTF & r & 16.52 & 0.04 \\ 
2458379.7577 & 9.11 & P48+ZTF & r & 16.52 & 0.03 \\ 
2458380.6214 & 9.97 & P48+ZTF & g & 16.63 & 0.04 \\ 
2458380.6251 & 9.98 & P48+ZTF & g & 16.66 & 0.04 \\ 
2458380.6778 & 10.03 & P48+ZTF & g & 16.56 & 0.03 \\ 
2458380.6778 & 10.03 & P48+ZTF & g & 16.56 & 0.03 \\ 
2458381.6238 & 10.98 & P48+ZTF & r & 16.58 & 0.04 \\ 
2458381.6289 & 10.98 & P48+ZTF & r & 16.59 & 0.04 \\ 
2458381.659 & 11.01 & P48+ZTF & r & 16.59 & 0.04 \\ 
2458381.6837 & 11.04 & P48+ZTF & g & 16.69 & 0.04 \\ 
2458381.7053 & 11.06 & P48+ZTF & g & 16.69 & 0.05 \\ 
2458381.7122 & 11.06 & P48+ZTF & g & 16.71 & 0.04 \\ 
2458383.6141 & 12.97 & P48+ZTF & r & 16.72 & 0.05 \\ 
2458383.6141 & 12.97 & P48+ZTF & r & 16.72 & 0.05 \\ 
2458383.6342 & 12.99 & P48+ZTF & r & 16.68 & 0.03 \\ 
2458383.6555 & 13.01 & P48+ZTF & r & 16.70 & 0.04 \\ 
2458383.6829 & 13.04 & P48+ZTF & g & 17.05 & 0.06 \\ 
2458383.6829 & 13.04 & P48+ZTF & g & 17.05 & 0.06 \\ 
2458383.6838 & 13.04 & P48+ZTF & g & 17.05 & 0.04 \\ 
2458383.6838 & 13.04 & P48+ZTF & g & 17.05 & 0.04 \\ 
2458383.705 & 13.06 & P48+ZTF & g & 17.01 & 0.05 \\ 
2458383.7143 & 13.07 & P48+ZTF & g & 17.05 & 0.05 \\ 
2458383.7143 & 13.07 & P48+ZTF & g & 17.05 & 0.05 \\ 
2458384.6451 & 14.0 & P48+ZTF & r & 16.80 & 0.05 \\ 
2458384.6525 & 14.01 & P48+ZTF & r & 16.80 & 0.05 \\ 
2458384.6741 & 14.03 & P48+ZTF & r & 16.80 & 0.04 \\ 
2458384.717 & 14.07 & P48+ZTF & g & 17.26 & 0.06 \\ 
2458384.717 & 14.07 & P48+ZTF & g & 17.26 & 0.06 \\ 
2458384.7384 & 14.09 & P48+ZTF & g & 17.24 & 0.05 \\ 
2458385.6151 & 14.97 & P48+ZTF & g & 17.45 & 0.05 \\ 
2458385.633 & 14.99 & P48+ZTF & g & 17.45 & 0.04 \\ 
2458385.633 & 14.99 & P48+ZTF & g & 17.45 & 0.04 \\ 
2458385.6622 & 15.01 & P48+ZTF & g & 17.46 & 0.05 \\ 
2458385.6622 & 15.01 & P48+ZTF & g & 17.46 & 0.05 \\ 
2458385.6844 & 15.04 & P48+ZTF & r & 16.92 & 0.04 \\ 
2458385.6844 & 15.04 & P48+ZTF & r & 16.92 & 0.04 \\ 
2458385.6919 & 15.04 & P48+ZTF & r & 16.92 & 0.04 \\ 
2458385.6919 & 15.04 & P48+ZTF & r & 16.92 & 0.04 \\ 
2458385.7117 & 15.06 & P48+ZTF & r & 16.93 & 0.04 \\ 
2458385.7117 & 15.06 & P48+ZTF & r & 16.93 & 0.04 \\ 
2458386.6167 & 15.97 & P48+ZTF & g & 17.62 & 0.07 \\ 
2458386.6242 & 15.98 & P48+ZTF & g & 17.67 & 0.06 \\ 
2458386.6242 & 15.98 & P48+ZTF & g & 17.67 & 0.06 \\ 
2458386.6404 & 15.99 & P48+ZTF & g & 17.65 & 0.06 \\ 
2458386.6546 & 16.01 & P48+ZTF & g & 17.60 & 0.06 \\ 
2458386.6994 & 16.05 & P48+ZTF & r & 17.02 & 0.05 \\ 
2458386.6994 & 16.05 & P48+ZTF & r & 17.02 & 0.04 \\ 
2458386.7013 & 16.05 & P48+ZTF & r & 17.04 & 0.05 \\ 
2458386.7158 & 16.07 & P48+ZTF & r & 17.04 & 0.04 \\ 
2458386.7377 & 16.09 & P48+ZTF & r & 17.00 & 0.05 \\ 
2458387.6227 & 16.98 & P48+ZTF & r & 17.14 & 0.04 \\ 
2458387.6227 & 16.98 & P48+ZTF & r & 17.14 & 0.04 \\ 
2458387.6399 & 16.99 & P48+ZTF & r & 17.14 & 0.05 \\ 
2458387.6541 & 17.01 & P48+ZTF & r & 17.15 & 0.04 \\ 
2458387.6541 & 17.01 & P48+ZTF & r & 17.15 & 0.04 \\ 
2458387.6822 & 17.03 & P48+ZTF & g & 17.85 & 0.08 \\ 
2458387.6822 & 17.03 & P48+ZTF & g & 17.85 & 0.08 \\ 
2458387.7041 & 17.06 & P48+ZTF & g & 17.83 & 0.10 \\ 
2458387.7041 & 17.06 & P48+ZTF & g & 17.83 & 0.10 \\ 
2458387.7232 & 17.08 & P48+ZTF & g & 17.88 & 0.09 \\ 
2458387.7232 & 17.08 & P48+ZTF & g & 17.88 & 0.09 \\ 
2458388.6124 & 17.97 & P60+SEDM & r & 17.22 & 0.02 \\ 
2458388.6154 & 17.97 & P48+ZTF & g & 18.04 & 0.06 \\ 
2458388.6154 & 17.97 & P48+ZTF & g & 18.04 & 0.06 \\ 
2458388.6396 & 17.99 & P48+ZTF & g & 17.99 & 0.07 \\ 
2458388.6396 & 17.99 & P48+ZTF & g & 17.99 & 0.07 \\ 
2458388.6542 & 18.01 & P48+ZTF & g & 18.04 & 0.06 \\ 
2458388.6542 & 18.01 & P48+ZTF & g & 18.04 & 0.06 \\ 
2458388.6834 & 18.04 & P48+ZTF & r & 17.30 & 0.06 \\ 
2458388.6936 & 18.05 & P48+ZTF & r & 17.30 & 0.05 \\ 
2458388.7203 & 18.07 & P48+ZTF & r & 17.25 & 0.06 \\ 
2458389.6156 & 18.97 & P48+ZTF & r & 17.39 & 0.06 \\ 
2458389.6227 & 18.98 & P48+ZTF & r & 17.40 & 0.05 \\ 
2458389.6317 & 18.98 & P48+ZTF & g & 18.20 & 0.06 \\ 
2458389.6317 & 18.98 & P48+ZTF & g & 18.20 & 0.06 \\ 
2458389.6416 & 18.99 & P48+ZTF & g & 18.20 & 0.06 \\ 
2458389.6416 & 18.99 & P48+ZTF & g & 18.20 & 0.06 \\ 
2458389.6804 & 19.03 & P48+ZTF & g & 18.21 & 0.08 \\ 
2458389.6804 & 19.03 & P48+ZTF & g & 18.21 & 0.08 \\ 
2458389.6947 & 19.05 & P48+ZTF & g & 18.21 & 0.09 \\ 
2458389.6947 & 19.05 & P48+ZTF & g & 18.21 & 0.09 \\ 
2458389.7166 & 19.07 & P48+ZTF & r & 17.41 & 0.05 \\ 
2458389.7476 & 19.1 & P48+ZTF & r & 17.43 & 0.04 \\ 
2458390.6228 & 19.98 & P48+ZTF & g & 18.43 & 0.06 \\ 
2458390.6228 & 19.98 & P48+ZTF & g & 18.43 & 0.06 \\ 
2458390.6326 & 19.99 & P48+ZTF & g & 18.40 & 0.06 \\ 
2458390.6326 & 19.99 & P48+ZTF & g & 18.40 & 0.06 \\ 
2458390.6797 & 20.03 & P48+ZTF & r & 17.55 & 0.04 \\ 
2458390.7209 & 20.07 & P48+ZTF & r & 17.56 & 0.07 \\ 
2458390.7347 & 20.09 & P48+ZTF & r & 17.53 & 0.05 \\ 
2458399.5989 & 28.95 & P48+ZTF & g & 19.40 & 0.19 \\ 
2458399.5989 & 28.95 & P48+ZTF & g & 19.40 & 0.19 \\ 
2458400.6307 & 29.98 & P48+ZTF & g & 19.46 & 0.13 \\ 
2458400.6638 & 30.02 & P48+ZTF & r & 18.68 & 0.11 \\ 
2458400.6756 & 30.03 & P48+ZTF & r & 18.72 & 0.10 \\ 
2458400.6756 & 30.03 & P48+ZTF & r & 18.72 & 0.10 \\ 
2458400.6987 & 30.05 & P48+ZTF & r & 18.65 & 0.14 \\ 
2458415.6169 & 44.97 & P60+SEDM & r & 19.62 & 0.10 \\ 
2458415.6196 & 44.97 & P60+SEDM & g & 20.24 & 0.20 \\ 
2458415.6223 & 44.98 & P60+SEDM & i & 19.39 & 0.06 \\ 
2458420.593 & 49.95 & P60+SEDM & r & 19.74 & 0.03 \\ 
2458420.5958 & 49.95 & P60+SEDM & g & 20.74 & 0.06 \\ 
2458420.5984 & 49.95 & P60+SEDM & i & 19.53 & 0.03 \\ 
2458420.6011 & 49.95 & P60+SEDM & r & 19.76 & 0.04 \\ 
2458420.6038 & 49.96 & P60+SEDM & g & 20.90 & 0.08 \\ 
2458423.584 & 52.94 & P60+SEDM & r & 19.78 & 0.10 \\ 
2458423.5894 & 52.94 & P60+SEDM & i & 19.67 & 0.13 \\ 
2458429.5848 & 58.94 & P60+SEDM & r & 20.03 & 0.06 \\ 
2458429.5875 & 58.94 & P60+SEDM & g & 21.32 & 0.11 \\ 
2458429.5902 & 58.94 & P60+SEDM & i & 19.80 & 0.04 \\ 
2458371.3802 & 0.73 & LT & u & 16.06 & 0.01 \\ 
2458372.3561 & 1.71 & LT & u & 15.70 & 0.01 \\ 
2458373.3944 & 2.75 & LT & u & 15.79 & 0.01 \\ 
2458380.3607 & 9.71 & LT & u & 17.12 & 0.02 \\ 
2458380.3612 & 9.71 & LT & u & 17.12 & 0.02 \\ 
2458381.3403 & 10.69 & LT & u & 17.49 & 0.03 \\ 
2458381.3409 & 10.69 & LT & u & 17.59 & 0.04 \\ 
2458382.3451 & 11.7 & LT & u & 17.97 & 0.04 \\ 
2458383.3399 & 12.69 & LT & u & 18.27 & 0.06 \\ 
2458383.3404 & 12.69 & LT & u & 18.27 & 0.06 \\ 
2458384.34 & 13.69 & LT & u & 18.71 & 0.08 \\ 
2458384.3405 & 13.69 & LT & u & 18.86 & 0.10 \\ 
2458385.339 & 14.69 & LT & u & 18.89 & 0.11 \\ 
2458386.3369 & 15.69 & LT & u & 19.16 & 0.17 \\ 
2458388.3375 & 17.69 & LT & u & 20.07 & 0.22 \\ 
2458388.338 & 17.69 & LT & u & 19.91 & 0.28 \\ 
2458391.3458 & 20.7 & LT & u & 20.08 & 0.24 \\ 
2458371.3794 & 0.73 & LT & g & 16.64 & 0.01 \\ 
2458372.3554 & 1.71 & LT & g & 16.21 & 0.01 \\ 
2458373.3951 & 2.75 & LT & g & 16.03 & 0.01 \\ 
2458380.3599 & 9.71 & LT & g & 16.63 & 0.01 \\ 
2458381.3396 & 10.69 & LT & g & 16.74 & 0.01 \\ 
2458382.3438 & 11.7 & LT & g & 16.88 & 0.01 \\ 
2458383.3391 & 12.69 & LT & g & 17.04 & 0.01 \\ 
2458384.3392 & 13.69 & LT & g & 17.27 & 0.01 \\ 
2458385.3377 & 14.69 & LT & g & 17.49 & 0.01 \\ 
2458386.3362 & 15.69 & LT & g & 17.62 & 0.05 \\ 
2458388.3367 & 17.69 & LT & g & 18.08 & 0.01 \\ 
2458389.3394 & 18.69 & LT & g & 18.20 & 0.01 \\ 
2458390.367 & 19.72 & LT & g & 18.34 & 0.06 \\ 
2458391.3445 & 20.7 & LT & g & 18.55 & 0.01 \\ 
2458393.3452 & 22.7 & LT & g & 18.88 & 0.02 \\ 
2458394.3463 & 23.7 & LT & g & 18.96 & 0.01 \\ 
2458395.3462 & 24.7 & LT & g & 19.20 & 0.03 \\ 
2458396.3496 & 25.7 & LT & g & 19.31 & 0.03 \\ 
2458397.3884 & 26.74 & LT & g & 19.53 & 0.04 \\ 
2458407.3531 & 36.71 & LT & g & 20.08 & 0.06 \\ 
2458407.3537 & 36.71 & LT & g & 20.24 & 0.07 \\ 
2458408.3179 & 37.67 & LT & g & 20.26 & 0.10 \\ 
2458408.3186 & 37.67 & LT & g & 20.14 & 0.07 \\ 
2458409.3255 & 38.68 & LT & g & 20.15 & 0.09 \\ 
2458409.3262 & 38.68 & LT & g & 20.34 & 0.12 \\ 
2458371.3787 & 0.73 & LT & r & 16.89 & 0.01 \\ 
2458372.3546 & 1.71 & LT & r & 16.43 & 0.01 \\ 
2458373.3958 & 2.75 & LT & r & 16.28 & 0.01 \\ 
2458380.3592 & 9.71 & LT & r & 16.51 & 0.01 \\ 
2458381.3389 & 10.69 & LT & r & 16.54 & 0.01 \\ 
2458382.3431 & 11.7 & LT & r & 16.60 & 0.01 \\ 
2458383.3384 & 12.69 & LT & r & 16.68 & 0.01 \\ 
2458384.3385 & 13.69 & LT & r & 16.78 & 0.01 \\ 
2458385.337 & 14.69 & LT & r & 16.91 & 0.01 \\ 
2458386.3354 & 15.69 & LT & r & 17.07 & 0.01 \\ 
2458388.336 & 17.69 & LT & r & 17.31 & 0.01 \\ 
2458389.3387 & 18.69 & LT & r & 17.40 & 0.01 \\ 
2458390.3663 & 19.72 & LT & r & 17.59 & 0.04 \\ 
2458391.3438 & 20.7 & LT & r & 17.70 & 0.01 \\ 
2458393.3444 & 22.7 & LT & r & 17.93 & 0.01 \\ 
2458394.3456 & 23.7 & LT & r & 18.07 & 0.01 \\ 
2458395.3455 & 24.7 & LT & r & 18.14 & 0.02 \\ 
2458396.3489 & 25.7 & LT & r & 18.31 & 0.02 \\ 
2458397.3877 & 26.74 & LT & r & 18.41 & 0.02 \\ 
2458407.3524 & 36.71 & LT & r & 19.20 & 0.03 \\ 
2458408.317 & 37.67 & LT & r & 19.29 & 0.04 \\ 
2458409.3246 & 38.68 & LT & r & 19.41 & 0.16 \\ 
2458371.378 & 0.73 & LT & i & 17.30 & 0.01 \\ 
2458372.3539 & 1.71 & LT & i & 16.90 & 0.01 \\ 
2458373.3965 & 2.75 & LT & i & 16.62 & 0.01 \\ 
2458380.3585 & 9.71 & LT & i & 16.81 & 0.01 \\ 
2458381.3381 & 10.69 & LT & i & 16.82 & 0.01 \\ 
2458382.3424 & 11.7 & LT & i & 16.88 & 0.01 \\ 
2458383.3377 & 12.69 & LT & i & 16.92 & 0.01 \\ 
2458384.3378 & 13.69 & LT & i & 16.99 & 0.17 \\ 
2458385.3363 & 14.69 & LT & i & 17.04 & 0.01 \\ 
2458386.3347 & 15.69 & LT & i & 17.10 & 0.01 \\ 
2458388.3353 & 17.69 & LT & i & 17.28 & 0.01 \\ 
2458389.338 & 18.69 & LT & i & 17.39 & 0.01 \\ 
2458390.3656 & 19.72 & LT & i & 17.56 & 0.04 \\ 
2458391.3431 & 20.7 & LT & i & 17.63 & 0.01 \\ 
2458393.3437 & 22.7 & LT & i & 17.84 & 0.01 \\ 
2458394.3449 & 23.7 & LT & i & 17.98 & 0.01 \\ 
2458395.3448 & 24.7 & LT & i & 18.10 & 0.04 \\ 
2458396.3481 & 25.7 & LT & i & 18.18 & 0.01 \\ 
2458397.3869 & 26.74 & LT & i & 18.30 & 0.02 \\ 
2458407.3517 & 36.7 & LT & i & 19.04 & 0.03 \\ 
2458408.3162 & 37.67 & LT & i & 19.03 & 0.07 \\ 
2458409.3238 & 38.68 & LT & i & 19.14 & 0.07 \\ 
2458373.3972 & 2.75 & LT & z & 16.82 & 0.01 \\ 
2458380.3577 & 9.71 & LT & z & 16.77 & 0.01 \\ 
2458381.3374 & 10.69 & LT & z & 16.79 & 0.01 \\ 
2458382.3416 & 11.69 & LT & z & 16.80 & 0.01 \\ 
2458383.3369 & 12.69 & LT & z & 16.82 & 0.01 \\ 
2458384.337 & 13.69 & LT & z & 16.83 & 0.01 \\ 
2458385.3355 & 14.69 & LT & z & 16.89 & 0.01 \\ 
2458386.334 & 15.69 & LT & z & 16.94 & 0.01 \\ 
2458388.3345 & 17.69 & LT & z & 17.04 & 0.01 \\ 
2458389.3372 & 18.69 & LT & z & 17.11 & 0.02 \\ 
2458390.3648 & 19.72 & LT & z & 17.23 & 0.12 \\ 
2458391.3423 & 20.7 & LT & z & 17.26 & 0.01 \\ 
2458393.343 & 22.7 & LT & z & 17.44 & 0.01 \\ 
2458394.3441 & 23.7 & LT & z & 17.55 & 0.02 \\ 
2458395.344 & 24.7 & LT & z & 17.65 & 0.05 \\ 
2458396.3474 & 25.7 & LT & z & 17.69 & 0.02 \\ 
2458397.3862 & 26.74 & LT & z & 17.77 & 0.03 \\ 
2458407.3509 & 36.7 & LT & z & 18.19 & 0.03 \\ 
2458408.3155 & 37.67 & LT & z & 18.33 & 0.06 \\ 
2458409.3231 & 38.68 & LT & z & 18.26 & 0.07 \\ 
2458374.9769 & 4.33 & LOT & g & 16.14 & 0.01 \\ 
2458375.9702 & 5.32 & LOT & g & 16.17 & 0.01 \\ 
2458379.9736 & 9.33 & LOT & g & 16.62 & 0.01 \\ 
2458381.0023 & 10.36 & LOT & g & 16.77 & 0.01 \\ 
2458381.9909 & 11.34 & LOT & g & 16.89 & 0.01 \\ 
2458386.0102 & 15.36 & LOT & g & 17.60 & 0.01 \\ 
2458391.0243 & 20.38 & LOT & g & 18.59 & 0.01 \\ 
2458391.9648 & 21.32 & LOT & g & 18.72 & 0.01 \\ 
2458392.9823 & 22.34 & LOT & g & 18.83 & 0.01 \\ 
2458393.9679 & 23.32 & LOT & g & 18.98 & 0.01 \\ 
2458394.9508 & 24.3 & LOT & g & 19.16 & 0.02 \\ 
2458395.9525 & 25.31 & LOT & g & 19.26 & 0.01 \\ 
2458396.9584 & 26.31 & LOT & g & 19.43 & 0.02 \\ 
2458406.9893 & 36.34 & LOT & g & 20.32 & 0.07 \\ 
2458411.95 & 41.3 & LOT & g & 20.53 & 0.10 \\ 
2458374.9847 & 4.34 & LOT & i & 16.53 & 0.01 \\ 
2458379.9812 & 9.33 & LOT & i & 16.75 & 0.01 \\ 
2458381.01 & 10.36 & LOT & i & 16.78 & 0.01 \\ 
2458381.9986 & 11.35 & LOT & i & 16.80 & 0.01 \\ 
2458386.018 & 15.37 & LOT & i & 17.05 & 0.01 \\ 
2458391.0321 & 20.38 & LOT & i & 17.57 & 0.01 \\ 
2458391.9726 & 21.33 & LOT & i & 17.71 & 0.01 \\ 
2458392.9901 & 22.34 & LOT & i & 17.80 & 0.01 \\ 
2458393.9756 & 23.33 & LOT & i & 17.97 & 0.01 \\ 
2458394.9692 & 24.32 & LOT & i & 18.07 & 0.01 \\ 
2458395.9603 & 25.31 & LOT & i & 18.16 & 0.01 \\ 
2458396.978 & 26.33 & LOT & i & 18.25 & 0.01 \\ 
2458406.9971 & 36.35 & LOT & i & 18.90 & 0.03 \\ 
2458411.9578 & 41.31 & LOT & i & 19.09 & 0.04 \\ 
2458374.9807 & 4.33 & LOT & r & 16.30 & 0.01 \\ 
2458375.974 & 5.33 & LOT & r & 16.33 & 0.01 \\ 
2458379.9774 & 9.33 & LOT & r & 16.54 & 0.01 \\ 
2458381.0061 & 10.36 & LOT & r & 16.61 & 0.01 \\ 
2458381.9947 & 11.35 & LOT & r & 16.62 & 0.01 \\ 
2458386.014 & 15.37 & LOT & r & 16.97 & 0.01 \\ 
2458391.0282 & 20.38 & LOT & r & 17.71 & 0.01 \\ 
2458391.9686 & 21.32 & LOT & r & 17.77 & 0.01 \\ 
2458392.9862 & 22.34 & LOT & r & 17.93 & 0.01 \\ 
2458393.9717 & 23.32 & LOT & r & 18.06 & 0.01 \\ 
2458394.9653 & 24.32 & LOT & r & 18.19 & 0.01 \\ 
2458395.9564 & 25.31 & LOT & r & 18.30 & 0.01 \\ 
2458396.9623 & 26.31 & LOT & r & 18.42 & 0.01 \\ 
2458406.9932 & 36.35 & LOT & r & 19.07 & 0.02 \\ 
2458411.9538 & 41.31 & LOT & r & 19.42 & 0.03 \\ 
2458371.0917 & 0.44 & UVOT & B & 16.77 & 0.06 \\ 
2458371.1601 & 0.51 & UVOT & B & 16.67 & 0.06 \\ 
2458373.8837 & 3.24 & UVOT & B & 15.88 & 0.07 \\ 
2458374.0828 & 3.44 & UVOT & B & 15.87 & 0.08 \\ 
2458374.481 & 3.83 & UVOT & B & 15.90 & 0.07 \\ 
2458375.3416 & 4.69 & UVOT & B & 16.06 & 0.06 \\ 
2458376.48 & 5.83 & UVOT & B & 16.00 & 0.06 \\ 
2458376.599 & 5.95 & UVOT & B & 16.23 & 0.07 \\ 
2458379.2575 & 8.61 & UVOT & B & 16.46 & 0.07 \\ 
2458380.184 & 9.54 & UVOT & B & 16.50 & 0.08 \\ 
2458380.3172 & 9.67 & UVOT & B & 16.68 & 0.09 \\ 
2458380.7873 & 10.14 & UVOT & B & 16.80 & 0.07 \\ 
2458381.6447 & 11.0 & UVOT & B & 17.29 & 0.14 \\ 
2458381.7774 & 11.13 & UVOT & B & 16.80 & 0.11 \\ 
2458381.8438 & 11.2 & UVOT & B & 16.95 & 0.12 \\ 
2458383.3045 & 12.66 & UVOT & B & 17.68 & 0.15 \\ 
2458383.3705 & 12.72 & UVOT & B & 17.35 & 0.13 \\ 
2458384.3114 & 13.66 & UVOT & B & 17.44 & 0.08 \\ 
2458371.0908 & 0.44 & UVOT & U & 16.41 & 0.06 \\ 
2458371.1591 & 0.51 & UVOT & U & 16.24 & 0.05 \\ 
2458373.8834 & 3.24 & UVOT & U & 15.75 & 0.07 \\ 
2458374.0825 & 3.44 & UVOT & U & 15.68 & 0.07 \\ 
2458374.4806 & 3.83 & UVOT & U & 15.76 & 0.07 \\ 
2458375.3411 & 4.69 & UVOT & U & 15.72 & 0.06 \\ 
2458376.4794 & 5.83 & UVOT & U & 15.95 & 0.06 \\ 
2458376.5986 & 5.95 & UVOT & U & 15.83 & 0.06 \\ 
2458379.2569 & 8.61 & UVOT & U & 16.65 & 0.07 \\ 
2458380.1836 & 9.54 & UVOT & U & 17.21 & 0.10 \\ 
2458380.3168 & 9.67 & UVOT & U & 17.28 & 0.10 \\ 
2458380.7866 & 10.14 & UVOT & U & 17.34 & 0.08 \\ 
2458381.6444 & 11.0 & UVOT & U & 17.73 & 0.14 \\ 
2458381.7771 & 11.13 & UVOT & U & 17.94 & 0.15 \\ 
2458381.8435 & 11.2 & UVOT & U & 18.06 & 0.17 \\ 
2458383.3041 & 12.66 & UVOT & U & 18.82 & 0.24 \\ 
2458383.37 & 12.72 & UVOT & U & 18.36 & 0.17 \\ 
2458384.3105 & 13.66 & UVOT & U & 19.01 & 0.16 \\ 
2458371.1013 & 0.45 & UVOT & UVM2 & 15.74 & 0.05 \\ 
2458371.1669 & 0.52 & UVOT & UVM2 & 15.61 & 0.05 \\ 
2458373.8864 & 3.24 & UVOT & UVM2 & 15.19 & 0.05 \\ 
2458374.0856 & 3.44 & UVOT & UVM2 & 15.19 & 0.05 \\ 
2458374.4841 & 3.84 & UVOT & UVM2 & 15.40 & 0.05 \\ 
2458375.3466 & 4.7 & UVOT & UVM2 & 15.88 & 0.05 \\ 
2458376.4854 & 5.84 & UVOT & UVM2 & 16.66 & 0.06 \\ 
2458376.6032 & 5.96 & UVOT & UVM2 & 16.81 & 0.06 \\ 
2458379.2631 & 8.62 & UVOT & UVM2 & 19.83 & 0.21 \\ 
2458380.1881 & 9.54 & UVOT & UVM2 & 20.14 & 0.29 \\ 
2458380.3209 & 9.67 & UVOT & UVM2 & 20.31 & 0.35 \\ 
2458380.7945 & 10.15 & UVOT & UVM2 & 21.34 & 0.64 \\ 
2458381.648 & 11.0 & UVOT & UVM2 & 21.12 & 0.67 \\ 
2458381.7807 & 11.13 & UVOT & UVM2 & 20.60 & 0.44 \\ 
2458381.8472 & 11.2 & UVOT & UVM2 & 21.87 & 1.23 \\ 
2458383.3088 & 12.66 & UVOT & UVM2 & 21.45 & 0.81 \\ 
2458383.3752 & 12.73 & UVOT & UVM2 & 22.09 & 1.38 \\ 
2458384.3213 & 13.67 & UVOT & UVM2 & 26.70 & 76.24 \\ 
2458371.0893 & 0.44 & UVOT & UVW1 & 15.86 & 0.05 \\ 
2458371.1577 & 0.51 & UVOT & UVW1 & 15.75 & 0.05 \\ 
2458373.8829 & 3.24 & UVOT & UVW1 & 15.32 & 0.06 \\ 
2458374.082 & 3.43 & UVOT & UVW1 & 15.24 & 0.05 \\ 
2458374.4801 & 3.83 & UVOT & UVW1 & 15.42 & 0.05 \\ 
2458375.3402 & 4.69 & UVOT & UVW1 & 15.73 & 0.05 \\ 
2458376.4784 & 5.83 & UVOT & UVW1 & 16.45 & 0.06 \\ 
2458376.5979 & 5.95 & UVOT & UVW1 & 16.50 & 0.06 \\ 
2458379.2558 & 8.61 & UVOT & UVW1 & 17.94 & 0.09 \\ 
2458380.1828 & 9.54 & UVOT & UVW1 & 18.69 & 0.15 \\ 
2458380.3161 & 9.67 & UVOT & UVW1 & 19.08 & 0.19 \\ 
2458380.7853 & 10.14 & UVOT & UVW1 & 18.84 & 0.13 \\ 
2458381.6437 & 11.0 & UVOT & UVW1 & 19.08 & 0.20 \\ 
2458381.7765 & 11.13 & UVOT & UVW1 & 19.41 & 0.26 \\ 
2458381.8429 & 11.2 & UVOT & UVW1 & 20.01 & 0.39 \\ 
2458383.3033 & 12.66 & UVOT & UVW1 & 20.19 & 0.38 \\ 
2458383.3692 & 12.72 & UVOT & UVW1 & 20.72 & 0.57 \\ 
2458384.3086 & 13.66 & UVOT & UVW1 & 20.57 & 0.35 \\ 
2458371.0941 & 0.45 & UVOT & UVW2 & 15.51 & 0.06 \\ 
2458371.1625 & 0.52 & UVOT & UVW2 & 15.38 & 0.06 \\ 
2458373.8845 & 3.24 & UVOT & UVW2 & 15.62 & 0.06 \\ 
2458374.0835 & 3.44 & UVOT & UVW2 & 15.68 & 0.06 \\ 
2458374.4818 & 3.83 & UVOT & UVW2 & 15.88 & 0.06 \\ 
2458375.343 & 4.7 & UVOT & UVW2 & 16.37 & 0.06 \\ 
2458376.4815 & 5.83 & UVOT & UVW2 & 17.09 & 0.07 \\ 
2458376.6002 & 5.95 & UVOT & UVW2 & 17.29 & 0.08 \\ 
2458379.2591 & 8.61 & UVOT & UVW2 & 19.61 & 0.18 \\ 
2458380.1852 & 9.54 & UVOT & UVW2 & 20.11 & 0.28 \\ 
2458380.3183 & 9.67 & UVOT & UVW2 & 20.33 & 0.34 \\ 
2458380.7894 & 10.14 & UVOT & UVW2 & 20.37 & 0.28 \\ 
2458381.6457 & 11.0 & UVOT & UVW2 & 20.23 & 0.34 \\ 
2458381.7783 & 11.13 & UVOT & UVW2 & 20.86 & 0.56 \\ 
2458381.8447 & 11.2 & UVOT & UVW2 & 21.58 & 1.00 \\ 
2458383.3057 & 12.66 & UVOT & UVW2 & 21.53 & 0.85 \\ 
2458383.3718 & 12.72 & UVOT & UVW2 & 21.77 & 1.00 \\ 
2458384.3142 & 13.67 & UVOT & UVW2 & 21.17 & 0.48 \\ 
2458371.0965 & 0.45 & UVOT & V & 17.25 & 0.13 \\ 
2458371.1649 & 0.52 & UVOT & V & 16.79 & 0.10 \\ 
2458373.8852 & 3.24 & UVOT & V & 16.24 & 0.14 \\ 
2458374.0843 & 3.44 & UVOT & V & 16.13 & 0.13 \\ 
2458374.4827 & 3.84 & UVOT & V & 16.08 & 0.12 \\ 
2458375.3444 & 4.7 & UVOT & V & 16.22 & 0.10 \\ 
2458376.483 & 5.84 & UVOT & V & 16.07 & 0.09 \\ 
2458376.6013 & 5.95 & UVOT & V & 16.03 & 0.11 \\ 
2458379.2607 & 8.61 & UVOT & V & 16.41 & 0.11 \\ 
2458380.1863 & 9.54 & UVOT & V & 16.41 & 0.13 \\ 
2458380.3193 & 9.67 & UVOT & V & 16.62 & 0.15 \\ 
2458380.7914 & 10.14 & UVOT & V & 16.62 & 0.11 \\ 
2458381.6466 & 11.0 & UVOT & V & 16.61 & 0.16 \\ 
2458381.7793 & 11.13 & UVOT & V & 16.70 & 0.17 \\ 
2458381.8456 & 11.2 & UVOT & V & 16.51 & 0.16 \\ 
2458383.3069 & 12.66 & UVOT & V & 16.58 & 0.14 \\ 
2458383.3731 & 12.73 & UVOT & V & 16.74 & 0.15 \\ 
2458384.317 & 13.67 & UVOT & V & 16.86 & 0.11 \\ 
\enddata 
\end{deluxetable} 

\section{UV and Optical Spectroscopy}
\label{appendix:uv-opt-spec}

\begin{figure*}[ht]
\centering
\includegraphics[scale=0.70]{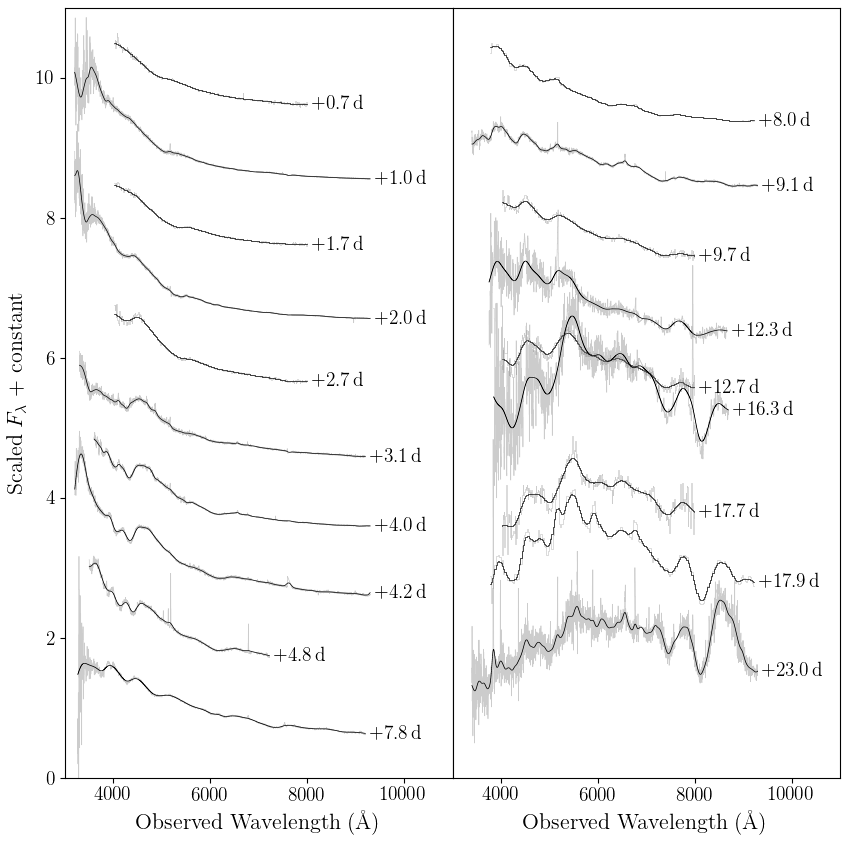}
\caption{Ground-based optical spectra of SN2018gep.
The light grey represents the observed spectrum, interpolating over host emission lines and telluric features. The black line is a Gaussian-smoothed version of the spectrum, using a Gaussian width that is several times the width of a galaxy emission line at that resolution.
For more details on the smoothing procedure, see Section 2.1 of \citet{Ho2017}.
}
\label{fig:spec-sequence}
\end{figure*}

The observation log of our UV and optical spectra is provided in 
Table \ref{tab:opt-spec}.
A plot showing the full sequence of optical spectra
is shown in Figure \ref{fig:spec-sequence}.

\startlongtable
\begin{deluxetable*}{lrrrrr} 
\tablecaption{Log of \name\ optical spectra \label{tab:opt-spec}} 
\tablewidth{0pt} 
\tablehead{ \colhead{Start Time (UTC)} & \colhead{$\Delta t$} & \colhead{Instrument} & \colhead{Exp. Time (s)} & \colhead{Airmass} } 
\rotate 
\tabletypesize{\scriptsize} 
\startdata 
 2018 Sep 09 20:30:01 & 0.7 
 & LT+SPRAT & 1200 & 1.107 \\
 2018 Sep 10 04:28:51 & 1.0 & P200+DBSP & 600 & 1.283 \\
 2018 Sep 10 21:03:42 & 1.7 
 & LT+SPRAT & 900 & 1.182 \\
 2018 Sep 11 04:59:19 & 2.0 & P200+DBSP & 600 & 1.419 \\
 2018 Sep 11 20:22:35 & 2.7 
 & LT+SPRAT & 900 & 1.107 \\
 2018 Sep 12 06:09:59 & 3.1 & P200+DBSP \\
 2018 Sep 13 03:52:58 & 4.0 & P200+DBSP & 300 & 1.209 \\
 2018 Sep 13 09:17:25 & 4.2 & Keck1+LRIS & 300 & 3.483 \\
 2018 Sep 14 02:44:24.24 & 4.8 & DCT+Deveny+LMI & 300 & 1.11 \\
 2018 Sep 17 04:38:40 & 8.0 & P60+SEDM & 1440 & 1.435 \\
 2018 Sep 17 20:40:25.750 & 8.7 & NOT+ALFOSC & 1800 & 1.19\\
 2018 Sep 18 05:21:58 & 9.1 & P200+DBSP & 600 & 1.720 \\
 2018 Sep 18 20:14:35 & 9.7 
 & LT+SPRAT & 1000 & 1.143 \\
  2018 Sep 21 11:15:10 & 12.3 & XLT+BFOSC & 3000 & 1.181 \\
 2018 Sep 21 20:58:21 & 12.7 
 & LT+SPRAT & 1000 & 1.293 \\
 2018 Sep 25 11:16:43 & 16.3 & XLT+BFOSC & 3000 & 1.225 \\
 2018 Sep 26 20:22:54 & 17.7 
 & LT+SPRAT & 1000 & 1.242 \\
 2018 Sep 27 02:42:29 & 17.9 & P60+SEDM & 1440 & 1.172 \\
 2018 Oct 02 04:34:35 & 23.0 & P200+DBSP & 600 & 1.780 \\
 2018 Nov 09 05:26:17 & 61.1 & Keck1+LRIS & 900 & 3.242 \\
\enddata 
\tablecomments{Gratings used: Wasatch600 (LT+SPRAT), Gr4 (NOT+ALFOSC), 600/4000 (P200+DBSP; blue side), 316/7500 (P200+DBSP; red side), 400/8500 (Keck1+LRIS; red side). \\
Filters used: 400nm (LT+SPRAT), open (NOT+ALFOSC), clear (Keck1+LRIS) \\ 
Wavelength range: 4020--7995\,\AA\ (LT+SPRAT),
3200--9600\,\AA\ (NOT+ALFOSC),
1759--10311\,\AA\ (Keck1+LRIS),
3777--9223\,\AA\ (P60+SEDM)\\
Resolution: 20 (LT+SPRAT), 710 (NOT+ALFOSC)}
\end{deluxetable*}

\section{Atomic data for spectral modeling}
\label{appendix:atomic-data}

The atomic data used for the spectral modelling in Section \ref{sec:spec-evolution} is the same as described in Appendix A.4 of \citet{Ergon2018}, but with the following modifications. The stage II-IV ions where (whenever possible) updated to include at least 50 levels for N, Na, Al, Ar and Ca, at least 100 levels for C, O, Ne, Mg, Si and S, and at least 300 levels for Sc, Ti, V, Cr, Mn, Fe, Co and Ni. In addition we updated the \ion{C}{2} - \ion{C}{4} and \ion{O}{2} - \ion{O}{3} ions with specific recombination rates from the online table by S. Nahar\footnote{http://www.astronomy.ohio-state.edu/\texttt{\char`\~}nahar/\texttt{\char`\_}naharradiativeatomicdata/}.

\section{Data for measuring host properties}
\label{appendix:host-data}

\begin{figure}[ht]
    \centering
    \includegraphics[width=1.0\columnwidth]{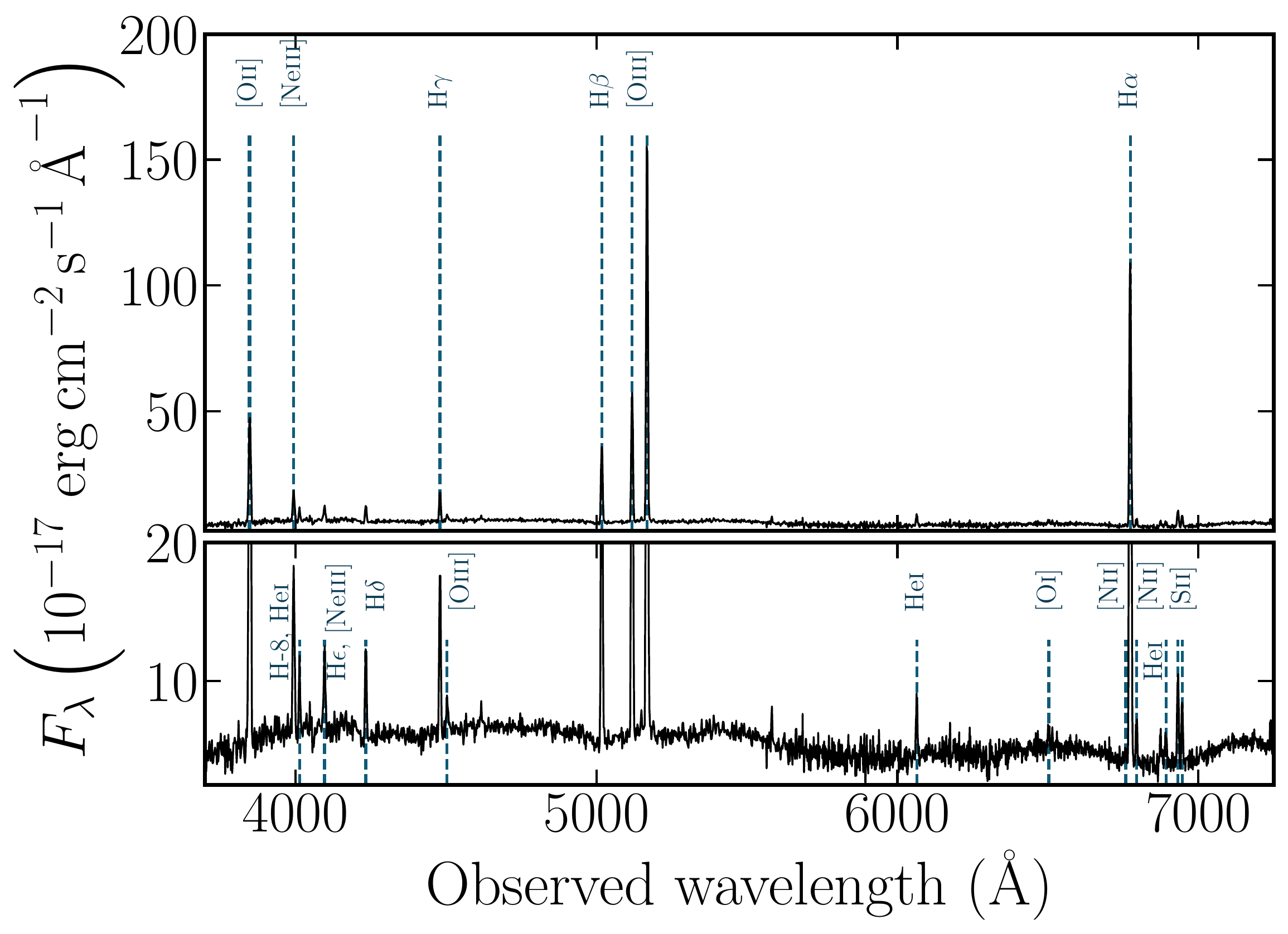}
    \caption{Host spectrum of \name\ obtained with Keck/LRIS on 9 November 2018, about two months after explosion. Strong emission lines from the host galaxy are labeled. The low host metallicity of 0.1 solar is reflected by very small \ion{N}{2}/H$\alpha$ flux ratio. The large rest-frame [\ion{O}{3}]$\lambda$5007 equivalent width of $>160$~\AA\ puts the host also in regime of extreme emission-line galaxies. These galaxy class constitute  $<2\%$ of all star-forming galaxies at $z<0.3$ in the SDSS DR15 catalogue. The undulations are due to the supernova. The spectrum is truncated at 7250~\AA\ for presentation purposes, and it is corrected for Galactic reddening. }
    \label{fig:host_spectrum}
\end{figure}

In this section we provide the data that we used to derive properties of the host galaxy of SN2018gep:
the host-galaxy spectrum (Figure \ref{fig:host_spectrum}),
line fluxes extracted from this spectrum (Table \ref{tab:eml_host}),
and host-galaxy photometry (Table \ref{tab:host_phot}).

\begin{table*}
\caption{Line fluxes from the host galaxy of \name\ extracted from the Keck/LRIS spectrum obtained on 9 November 2018. }\label{tab:eml_host}
\centering
\begin{tabular}{llcc}
\toprule
Transition			& $\lambda_{\rm obs}$& $F$	\\
					& (\AA)	& $\left(10^{-17}~{\rm erg\,cm}^{-2}\,{\rm s}^{-1}\right) 	$\\
\midrule
{[\ion{O}{2}]}$\lambda\lambda$3726,3729 &$ 3848.17 \pm 0.05	$&$	334.5	\pm	6.23	$\\
{[\ion{Ne}{3}]}$\lambda$3869			& $ 3993.50 \pm 0.16	$&$	82.34	\pm	6.18	$\\
 \ion{He}{1}$\lambda$3889,H-8			& $ 4014.49 \pm 0.16	$&$	29.01	\pm	4.73	$\\
{[\ion{Ne}{3}]}$\lambda$3968,H$\epsilon$& $ 4096.66 \pm 0.26	$&$	36.61	\pm	3.98	$\\
H$\delta$ 								& $ 4233.87 \pm 0.13	$&$	44.88	\pm	2.59	$\\
H$\gamma$								& $ 4480.20 \pm 0.10	$&$	81.95	\pm	3.74	$\\
{[\ion{O}{3}]}$\lambda$4364 			& $ 4503.68 \pm 0.10	$&$	15.01	\pm	2.69	$\\
H$\beta$								& $ 5017.87 \pm 0.08	$&$	213.41	\pm	10.53	$\\
{[\ion{O}{3}]}$\lambda$4960 			& $ 5118.61 \pm 0.04	$&$	352.42	\pm	6.50	$\\
{[\ion{O}{3}]}$\lambda$5008				& $ 5168.04 \pm 0.04	$&$	1066.70	\pm	19.50	$\\
 \ion{He}{1}$\lambda$5877				& $ 6064.21 \pm 0.20	$&$	27.04	\pm	2.30	$\\
 \ion{O}{1}$\lambda$6302				& $ 6502.18 \pm 1.08	$&$	6.72	\pm	2.94	$\\
{[\ion{N}{2}]}$\lambda$6549				& $ 6758.16 \pm 0.02	$&$	11.15	\pm	6.73	$\\
H$\alpha$								& $ 6773.40 \pm 0.02	$&$	723.85	\pm	7.65	$\\
{[\ion{N}{2}]}$\lambda$6585				& $ 6794.67 \pm 0.02	$&$	19.01	\pm	5.76	$\\
{[\ion{He}{1}]}$\lambda$6678			& $ 6890.29 \pm 0.14	$&$	7.88	\pm	2.19	$\\
{[\ion{S}{2}]}$\lambda$6718 			& $ 6931.83 \pm 0.10	$&$	41.76	\pm	2.38	$\\
{[\ion{S}{2}]}$\lambda$6732 			& $ 6946.68 \pm 0.10	$&$	28.15	\pm	2.19	$\\
\bottomrule
\end{tabular}
\tablecomments{All measurements are corrected for Galactic reddening.}
\end{table*}

\begin{table*}
\centering
\caption{Brightness of the host galaxy from UV ot IR wavelenghts}\label{tab:host_phot}
\begin{tabular}{cccccc}
\toprule
Instrument/	    & $\lambda_{\rm eff}$   & Brightness	& Instrument/   & $\lambda_{\rm eff}$   & Brightness	\\
Filter          & (\AA)                 & (mag)         & Filter        & (\AA)                 & (mag)         \\
\midrule
GALEX/FUV		& 1542.3  & $ 20.20 \pm 0.03$	& SDSS/$i'$ 		& 7439.5     &$ 18.62 \pm 0.04$\\
GALEX/NUV		& 2274.4  & $ 20.09 \pm 0.03$	& SDSS/$z'$ 		& 8897.1     &$ 18.59 \pm 0.12$\\
UVOT/$w2$ 		& 2030.5  & $ 19.91 \pm 0.12$	& PS1/$g_{\rm PS1}$	& 4775.6     &$ 18.96 \pm 0.04$\\
UVOT/$m2$		& 2228.1  & $ 20.00 \pm 0.14$	& PS1/$r_{\rm PS1}$	& 6129.5     &$ 18.82 \pm 0.04$\\
UVOT/$w1$ 		& 2589.1  & $ 20.11 \pm 0.16$	& PS1/$i_{\rm PS1}$	& 7484.6     &$ 18.88 \pm 0.04$\\
UVOT/$u$		& 3501.2  & $ 19.74 \pm 0.16$	& PS1/$z_{\rm PS1}$	& 8657.8     &$ 18.71 \pm 0.05$\\
UVOT/$b$		& 4328.6  & $ 19.45 \pm 0.20$	& WIRCam/$J$ 		& 12481.5    &$ 18.99 \pm 0.09$\\
UVOT/$v$		& 5402.1  & $ 18.45 \pm 0.21$	& 2MASS/$H$ 		& 16620.0    &$ 18.33 \pm 0.36$\\
SDSS/$u'$ 		& 3594.9  & $ 19.97 \pm 0.12$	& WISE/$W1$ 		& 33526.0    &$ 19.39 \pm 0.08$\\
SDSS/$g'$ 		& 4640.4  & $ 18.88 \pm 0.02$	& WISE/$W2$ 		& 46028.0    &$ 19.85 \pm 0.19$\\
SDSS/$r'$ 		& 6122.3  & $ 18.76 \pm 0.05$	& \\
\bottomrule
\end{tabular}
\tablecomments{All measurements are reported in the AB system and are not corrected for reddening. For guidance, we report the effective wavelengths of each filter.}
\end{table*}

\listofchanges

\end{document}